\documentclass[pre,preprintnumbers,amsmath,amssymb,nofootinbib]{revtex4-2}
\usepackage{graphicx,color}
\usepackage{bm}
\usepackage{epsfig}
\usepackage{hyperref}
\usepackage{float}
\usepackage{enumerate}  
\definecolor{amethyst}{rgb}{0.6, 0.4, 0.8}
\usepackage{amsmath}
\usepackage{stmaryrd}
\usepackage{soul}
\usepackage{bbold}
\usepackage{footmisc}

\newcommand{\red}{\normalcolor}
\newcommand{\blue}{\color{blue}}
\newcommand{\be}{\begin{equation}}
\newcommand{\ee}{\end{equation}}
\newcommand{\beq}{\begin{equation}}
\newcommand{\eeq}{\end{equation}}
\newcommand{\bea}{\begin{eqnarray}}
\newcommand{\eea}{\end{eqnarray}}
\newcommand{\nn}{\nonumber}

\hypersetup{
    colorlinks=true,       
    linkcolor=red,          
    citecolor=blue,        
    filecolor=magenta,      
    urlcolor=cyan           
}

\begin{document}

\begin{abstract}

We consider a generic one-dimensional stochastic process $x(t)$, or a random walk $X_n$, which describes the position of a particle evolving inside an interval $[a,b]$, with absorbing walls located at $a$ and $b$. In continuous time, $x(t)$ is driven by some equilibrium process ${\bm \theta}(t)$, while in discrete time, the jumps of $X_n$ follow a stationary process that obeys a time reversal property. An important observable to characterize its behaviour is the exit probability $E_b(x,t)$, which is the probability for the particle to be absorbed first at the wall $b$, before or at time $t$, given its initial position $x$. In this paper we show that the derivation of this quantity can be tackled by studying a dual process $y(t)$ very similar to $x(t)$ but with hard walls at $a$ and $b$. More precisely, we show that the quantity $E_b(x,t)$ for the process $x(t)$ is equal to the probability $\tilde \Phi(x,t|b)$ of finding the dual process inside the interval $[a,x]$ at time $t$, with $y(0) =b$. {\red This is known as Siegmund duality in mathematics. Here} we show that this duality applies to various processes which are of interest in physics, including models of active particles, diffusing diffusivity models, a large class of discrete and continuous time random walks, and even processes subjected to stochastic resetting. For all these cases, we provide an explicit construction of the dual process. We also give simple derivations of this identity both in the continuous and in the discrete time setting, as well as numerical tests for a large number of models of interest. Finally, we use simulations to show that the duality is also likely to hold for more complex processes such as fractional Brownian motion.

\end{abstract}

\title{{\red Siegmund duality for physicists: a bridge between spatial and first-passage properties of continuous and discrete time stochastic processes}}

\author{Mathis \surname{Gu\'eneau}}
\affiliation{Sorbonne Universit\'e, Laboratoire de Physique Th\'eorique et Hautes Energies, CNRS UMR 7589, 4 Place Jussieu, 75252 Paris Cedex 05, France}

\author{L\'eo \surname{Touzo}}
\affiliation{Laboratoire de Physique de l'Ecole Normale Sup\'erieure, CNRS, ENS and PSL Universit\'e, Sorbonne Universit\'e, Universit\'e Paris Cit\'e,
24 rue Lhomond, 75005 Paris, France}

\date{\today}
{
\let\clearpage\relax
\maketitle
{
  \hypersetup{hidelinks}  
  \newpage
  \tableofcontents
}
}

\section{Introduction}
The study of first-passage properties of stochastic processes has a long history, both in mathematics and in physics, with applications ranging from biology to mathematical finance. Consider for instance a random walk on the interval $[a,b]$. Finding its first-passage properties consists in addressing the following questions: what is the probability that it reaches $b$ before reaching $a$ (called the {\it exit}, {\it splitting} or {\it hitting} probability) ? What is the probability that it remains in the interval up to time $t$ (the \textit{survival} probability) ? What is the distribution of first-passage time at $b$ ? Beyond their direct applications, these very general questions also share strong connections with the study of extreme value statistics. For general surveys on the topic, see \cite{redner, Metzler_book, EVS1,handbookSM, livreSG}.

One type of models for which these questions have attracted a lot of attention recently are active particles. These systems, akin to living organisms, are capable of converting energy into directed motion, and they exhibit distinctive out-of-equilibrium stochastic dynamics \cite{activeintro1,activeintro2,activeintro3,activeintro4,activeintro5}.
At variance with a passive Brownian particle subjected to a simple white noise, active particles experience a random noise that is correlated in time -- a colored noise \cite{colorednoise} -- leading to a persistence in their motion. This distinctive property gives rise to many interesting features such as non-Boltzmann steady states \cite{DKM19, MalakarRTP, Tailleur_RTP,Wijland21, Sevilla} and also leads to emergent collective behaviors, such as motility-induced phase transitions \cite{separation1,separation2,separation3,separation4} or flocking phenomena \cite{flocking1,flocking2}, among others. Famous models of active particles include the run-and-tumble particle (RTP), inspired from the motion of bacteria such as E. Coli \cite{Tailleur_RTP, Berg2004, Cates2012} (also known in the math literature as the persistent random walk \cite{Ors1990, Kac1974}), the active Ornstein-Uhlenbeck particle (AOUP) \cite{AOUP,Wijland21} and the active Brownian particle (ABP) \cite{activeintro4, ABM}.

First-passage problems are particularly relevant in biological and chemical contexts -- for instance, for search processes where one considers a searcher (e.g. an animal) looking for a target (e.g. food) without knowing its location \cite{benichou1,benichou2}. For active particles, most of the analytical studies focus on the simplest case of a free run-and-tumble particle (RTP), for which the exit probability, survival probability and mean first-passage time (MFPT) have been computed in various settings, in one-dimension \cite{MalakarRTP, SurvivalRTPDriftDeBruyne, Singh2020, Singh2022, MFPT1DABP} or higher \cite{RTPsurvivalMori,TVB12,RBV16}. More recently, some extensions of these results in the presence of an external potential \cite{ExitProbaShort,MFPT_1D_RTP, DKM19} or partially absorbing boundaries \cite{BressloffStickyBoundaries, AngelaniGenericBC, RTPpartiallyAbsorbingTarget} have been obtained, but this can be quite challenging. 
Few attempts were made beyond the RTP case, apart from the escape time of an Active Ornstein-Uhlenbeck particle (AOUP) from a well \cite{AOUPEscapeLecomte}. 

On a seemingly unrelated note, active particles exhibit peculiar behavior under confinement. In particular their persistent motion leads them to accumulate near walls \cite{Lee2013, Yang2014, Uspal2015, Duzgun2018}. It is thus natural to study their properties in the presence of hard walls, especially the distribution of their positions. This was done in the 1d case, for RTPs \cite{AngelaniHardWalls}, as well as for AOUPs \cite{hardWallsJoanny,hardWallsCaprini} (in the second case the hard wall was modeled by a steep but finite potential).

Surprisingly, there is a strong connection between absorbing and hard wall (or reflective) boundary conditions. This was first pointed out by L\'evy in the case of Brownian motion \cite{Levy}, and by Lindley for discrete random walks \cite{Lindley}. Siegmund later generalized this relation, which became known as {\it Siegmund duality}: two processes $x(t)$ and $y(t)$ (where $t$ is either discrete or continuous) such that $x(0)=x$ and $y(0)=y$ are said to be Siegmund duals if, {\normalcolor at any time $t$,}
\be\label{siegmundintro}
\mathbb{P}(x(t) \geq y | x(0)=x) = \mathbb{\tilde P}(y(t) \leq x | y(0)=y) \;,
\ee 
{\normalcolor where for the sake of clarity we will denote with a tilde all quantities associated to the dual}. Siegmund showed the existence of a Siegmund dual for any one-dimensional stochastically monotone Markov process (meaning that $\mathbb{P}(x(t) \geq y | x(0)=x)$ is a non-decreasing function of $x$) \cite{Siegmund}. Since then, extensions to more general Markov processes have been considered \cite{SiegmundDualityClifford, KolokoltsovDuality, partiallyOrdered, DualityZhao}. Depending on the setting it is not always obvious how to explicitly construct this dual. {\normalcolor However, if $x(t)$ is a Brownian motion or a random walk with i.i.d steps, with absorbing boundary conditions, then $y(t)$ has the exact same dynamics as $x(t)$ but with hard walls.}

\begin{figure}[t]
\centering
    \includegraphics[width=0.9\linewidth]{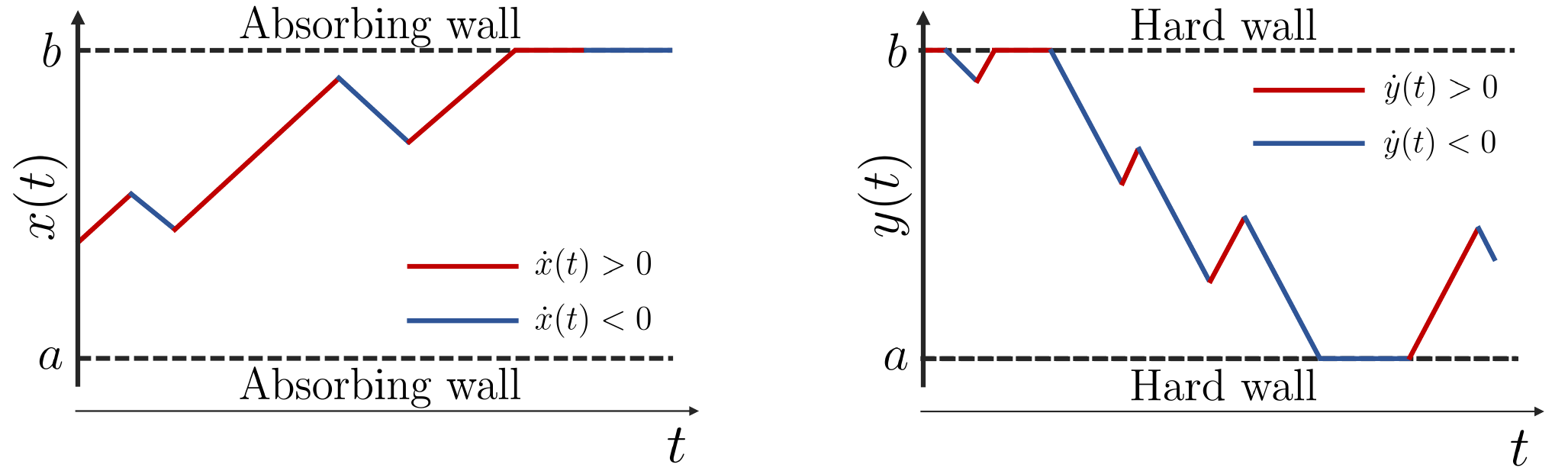}
    \caption{In this figure, we show the two types of boundaries that we consider throughout the whole paper, and how particles behave around them. The trajectories shown are schematics and could be for instance sampled from a run-and-tumble dynamics. \textbf{Left}:~The process $x(t)$ is surrounded by two absorbing walls at $a$ and $b$. If the particle reaches one of the two walls it will be absorbed and stay stuck to the wall forever. \textbf{Right}: The process $y(t)$ is surrounded by two hard walls at $a$ and $b$. In this case, if the particle is at a wall, and if its velocity $\dot {y}(t)$ is directed toward the wall, it will stay at the wall until $\dot {y}(t)$ changes sign.}
\label{figure_walls}
\end{figure}

In this paper, $x(t)$ takes values in an interval $[a,b]$ and has absorbing walls at $a$ and $b$ (thus $y(t)$ has hard walls at $a$ and $b$). By {\it absorbing boundary conditions} we mean that, if the particle reaches $a$ or $b$, it ``sticks'' to the wall and remains there afterwards {\red (i.e. it does not disappear, as it is sometimes the case in the physics literature \cite{redner, Metzler_book, EVS1})}. On the other hand, {\it hard walls} are considered as an infinite potential step. Thus, if the particle has a persistent motion, it will stay at the wall until its velocity changes sign (see \cite{ExitProbaShort, AngelaniHardWalls}). In Figure \ref{figure_walls}, we show a schematic description of these boundary conditions. Our main quantity of interest is the finite time exit probability at $b$, which we define as the probability to be absorbed at $b$ before or at time $t$, when $x(0)=x$,
\be 
E_b(x,t) = \mathbb{P}(x(t) = b | x(0)=x) \;.
\ee
Here we mostly focus on the case where $y=b$ in \eqref{siegmundintro}. The identity (\ref{siegmundintro}) then gives us a relation between $E_b(x,t)$ and the cumulative distribution of the dual $y(t)$ {\normalcolor denoted $\tilde \Phi(x,t|b)$ (since with our definition of absorbing walls, $x(t)\geq b$ is simply equivalent to $x(t)=b$)}
\be\label{siegmundExitIntro}
E_b(x,t) = \tilde \Phi(x,t|b) \quad , \quad \text{ where } \quad \tilde \Phi(x,t|b)=\mathbb{\tilde P}(y(t) \leq x | y(0)=b) \;.
\ee 
A simple example is when $x(t)$ is a Brownian motion (in which case $y(t)$ is a Brownian motion with hard walls). In that case $E_b(x,t)$ and $\tilde \Phi(x,t|b)$ both obey the diffusion equation $\partial_t E_b(x,t) = T \partial^2_x E_b(x,t)$,
with the same boundary conditions {\normalcolor $E_b(a,t) = {\tilde \Phi}(a,t|b) = 0$ and $E_b(b,t) = {\tilde \Phi}(b,t|b) = 1$, and the same initial condition $E_b(b,0) = {\tilde \Phi}(b,0|b) = 1$, or $E_b(x,0) = {\tilde \Phi}(x,0|b) = 0$ if $x< b$.} This is a direct consequence of \eqref{siegmundExitIntro}. 

Note that if $x(t)$ now diffuses in a potential $V(x)$, it is well known that in the stationary state, the exit probability reads $\int_a^x dz\, e^{\frac{V(z)}{T}} / \int_a^b dz\, e^{\frac{V(z)}{T}}$. This is the same expression as the cumulative distribution of a Brownian motion with hard walls at $a$ and $b$, in the presence of a reversed potential $-V(x)$ (this has been noticed for instance in \cite{hittingProbaAnomalous}). This is a general result: if $x(t)$ is subjected to an external force, the sign of the force is reversed in the case of its dual $y(t)$. In Appendix \ref{ProofsSimple}, we give a simple derivation of \eqref{siegmundExitIntro} for a Brownian motion with an external force, which illustrates this statement beyond the stationary state.

Another class of processes for which one can apply (\ref{siegmundintro}) is L\'evy flights. In the continuum limit, {\normalcolor when $x(t)$ is} a L\'evy flight whose increments obey a L\'evy stable symmetric law of index $0<\mu\leq 2$, the probability that it exits the interval $[a,b]$ at $b$ {\normalcolor after an infinite time} reads \cite{WidomExitprobaLevyflight, Levy3, hittingProbaAnomalous}
\begin{equation} \label{exitprobalevy}
    E_b(x,{\normalcolor t\to +\infty}) = \frac{\Gamma(2\phi)}{\Gamma(\phi)^2} (b-a)^{1-2\phi} \int_a^x du\, \left[(u-a)(b-u)\right]^{\phi - 1}\, ,
\end{equation}
where $\phi = \mu/2$. {\normalcolor If $x(t)$ is a Lévy flight, then its dual $y(t)$ is also a Lévy flight with the same increments. This is confirmed by a more recent and completely independent derivation of the stationary distribution of a L\'evy flight between two hard walls, $\tilde P_{st}(x)=\partial_x \tilde \Phi(x,t\to+\infty|b)$, in \cite{DenisovLevy},}
\begin{equation}
    \tilde P_{st}(x) =\frac{\Gamma(2\phi)}{\Gamma(\phi)^2} (b-a)^{1-2\phi} \left[(x-a)(b-x)\right]^{\phi - 1} = \frac{dE_b(x,{\normalcolor t\to \infty})}{dx} \, .
\end{equation}
As we can see, $\tilde P_{st}(x)$ is simply the derivative of $E_b(x)$, which is a direct consequence of the duality (\ref{siegmundExitIntro}) at infinite time\footnote{The derivation in Sec.~\ref{DiscreteProof}, as well as the original results by Lindley, apply in particular to Lévy flights in discrete time. Taking the continuous time limit would require a rigorous analysis on its own, but it is reasonable to assume that the duality still holds in that case.}. {\red The possibility of deriving one quantity directly from the other is a strong analytical motivation for the usefulness of the duality presented here.}

The original results by Siegmund were derived for one-dimensional Markov processes. Active particle models are by definition non-Markovian when one only considers the position of the particle, but they are Markovian when considering both the position and the driving velocity.
In fact, the existence of a dual process was also shown for processes driven by a stationary process, in \cite{AsmussenDiscrete} for discrete time and in \cite{SigmanContinuous} for continuous time (it has even been extended to higher dimensions in \cite{DualityMultidimensions}). In this setting, inspired by applications to finance, the exit probability (called ruin probability in this context) is related to the cumulative distribution of some dual process, defined in very general terms.

Siegmund duality can be seen as a particular case of Markov duality (see \cite{JansenDualityReview, DualityGenerators, CoxEntranceExitLaws, PathwiseDualSturm} for reviews and other examples). Such duality relations are frequently used by mathematicians and have been applied in various contexts, including queuing theory, finance and population genetics, as well as interacting particle systems and systems with a reservoir of particles \cite{Kolotsovkthorder, DualityLevyProcessesGoffard, MohleDualityGenetics, DualityBranchingFoucart, LiggettInteractingParticles, DualityBoundaryDriven}. Yet, they seem to be less well-known among the physics community{\red ,} although similar relations have been sometimes pointed out \cite{hittingProbaAnomalous, Comtet2011, Comtet2020, ThibautDual}. 
{\red Other connections between different types of boundary conditions have however been studied for various Markov processes in the context of physics. For instance, in \cite{Szabo, Spouge}, the authors showed that the propagator of a diffusion process with partially absorbing boundary conditions can be obtained from the propagator of the same process with reflective boundary conditions, an approach which was later extended to other situations \cite{Scher1, Scher2, Guerin}. Let us also mention the defect technique \cite{defect1, defect2}, which allows for example to relate the first passage distribution of a diffusing particle in the presence of an absorbing wall to the cumulative distribution without walls (in Laplace space) \cite{defect3}.}

In \cite{ExitProbaShort}, we highlighted the connection between absorbing boundaries and hard walls for a specific model of active particle: a run-and-tumble particle inside an arbitrary external potential. For this system, we gave an explicit formulation of the Siegmund dual and derived the relation between the exit probability and the distribution of positions of the particle at any time. In the stationary state, both quantities were computed explicitly. Here, we generalize the relation presented in \cite{ExitProbaShort} for RTPs to a broader class of systems, including the most well-known models of active particles.

In this paper we consider two types of models. The first is a one-dimensional continuous stochastic process subjected to a force and to a Gaussian white noise, with absorbing boundary conditions. Both the force and the temperature may depend on a parameter which is itself a Markov process at equilibrium. This definition was inspired by active particle models, but it also includes other models of interest such as diffusing diffusivity models \cite{DiffDiffChubynsky, DiffDiffChechkin, DiffDiffJain, DiffDiffFPTSposini}. The latter were introduced recently as an attempt to understand the ``non-Gaussian normal diffusion" observed in several soft matter systems \cite{DDsoftmatter1,DDsoftmatter2,DDsoftmatter3} (meaning that the mean squared displacement is linear in time but the distribution of position is not Gaussian). The second model is a 1d random walk with stationary increments (and again absorbing boundary conditions). In both settings, we construct explicitly a dual process, with hard wall boundary conditions, such that \eqref{siegmundintro} is satisfied. {\red Although some general mathematical results already exist (see, e.g. \cite{AsmussenDiscrete, SigmanContinuous}), to our knowledge, this is the first time that a fully explicit construction of the dual process is proposed for such a wide range of physically relevant models (including active particles).} We provide {\red original and} intuitive derivations {\red of the duality relation \eqref{siegmundintro} for both the continuous time and the discrete time settings}, as well as analytical and numerical illustrations for some well-known models. Our goal is not to provide fully rigorous proofs, but rather to extend the use of Siegmund duality to new fields and to convince physicists of its usefulness.

{\red There are several situations in which one may want to use such a relation. First of all, analytically computing the distribution of positions and the first passage properties of a stochastic process can be quite cumbersome, and knowing that it is possible to derive one from the other could save a lot of effort in some situations (see e.g. the example of L\'evy flights mentioned above). Additionally, in numerical simulations as well as in experiments, it is often much simpler to compute spatial properties than first-passage properties. In particular, if one is interested in the stationary state of an ergodic system, the former can be computed from a single, long enough time series of data, contrary to the latter.}

The precise setting and the main results of this paper are presented in Sec.~\ref{mainresults}. In Sec.~\ref{FPderivationsection}, we derive these results in the case of continuous stochastic processes, using the Fokker-Planck formalism. We also provide some numerical results which illustrate this duality for common models of active particles and other processes. Similar results exist for discrete time random walks (as well as continuous time random walks). We derive them in Sec.~\ref{discrete_case} using a completely different approach from the continuous case, based on a mapping between individual trajectories of the model and its dual. As an illustration we discuss the case of the persistent random walk, a discrete time version of the RTP \cite{PRWWeiss, PRWMasoliverReview, PRWSurvivalLacroixMori}. Finally, in Sec.~\ref{sec:resetting} we extend the duality to processes subjected to stochastic resetting \cite{resettingPRL,resettingReview, resettingBriefReview}. These systems restart their dynamics at a given point at random times, which makes them out-of-equilibrium. An important application of such models are search processes, which makes the study of their first-passage properties particularly relevant \cite{resettingPRL, resettingRTP, Besga20, Faisant21, resettingInInterval}.
We conclude this paper by discussing some limitations and possible extensions of our results in Sec.~\ref{sec:generalize}.

\section{Main results}
\label{mainresults}

We consider a one-dimensional stochastic process $x(t)$ which evolves according to the following stochastic differential equation (SDE),
\begin{equation}
    \dot{x}(t) = f\left(x(t),\bm{\theta}(t)\right) + \sqrt{2\mathcal{T}\left(x(t),\bm{\theta}(t)\right)}\, \xi(t)\, ,
\label{LangevinIntroduction}
\end{equation}
where $f\left(x,\bm{\theta}\right)$ and $\mathcal{T}\left(x,\bm{\theta}\right)$ are arbitrary functions which act respectively as a force and a temperature. In Eq.~(\ref{LangevinIntroduction}), $\xi(t)$ represents a Gaussian white noise with zero mean and unit variance, and $\bm{\theta}(t)$ is a vector of arbitrary dimension with stochastic components governed by the Markovian dynamics (independent of $x(t)$)
\begin{equation}
    \bm{\dot{\theta}}(t) = \bm{g}\left(\bm{\theta}(t)\right) + \left[2\underline{\mathcal{D}}(\bm{\theta}(t))\right]^{1/2} \cdot \bm{\eta}(t)\, ,
\label{SDEtheta_Introduction}
\end{equation}
where $\underline{\mathcal{D}}$ is a positive matrix\footnote{ Throughout the paper, we use bold letters to denote vectors and we underline matrices.}. The $\eta_i(t)$'s are again independent Gaussian white noises with zero mean and unit variance. In the whole paper we use the It\=o prescription for multiplicative noise \cite{handbookSM,riskenBook}. Additionally, we allow $\bm{\theta}(t)$ to jump from a value $\bm{\theta}$ to $\bm{\theta}'$ with a transition kernel $\mathcal{W}(\bm{\theta}'|\bm{\theta})$. {\normalcolor More precisely, during a time interval $dt$, ${\bm \theta}(t)$ evolves according to \eqref{SDEtheta_Introduction} with probability $1-dt \int d{\bm \theta}' \mathcal{W}(\bm{\theta}'|\bm{\theta})$, or jumps to some value $\bm{\theta}'$ with probability $\mathcal{W}(\bm{\theta}'|\bm{\theta}) d\bm{\theta}' \, dt$.} We assume that $\bm \theta(t)$ admits an equilibrium distribution $p_{eq}(\bm{\theta})$ which satisfies the local detailed balance conditions\footnote{When initialized in its equilibrium distribution $p_{eq}({\bm \theta})$, ${\bm \theta}(t)$ satisfies $P({\bm \theta}(t_1), ..., {\bm \theta}(t_n)) = P({\bm \theta}(t_1+\tau), ..., {\bm \theta}(t_n+\tau))$ for any times $t_1,...,t_n$ and any time-shift $\tau$, and is called a stationary process.}
\begin{eqnarray}\label{detailed_balance_Introduction1}
&&- g_i(\bm{\theta}) p_{eq}(\bm{\theta}) + \sum_{j} \partial_{\theta_j}[\mathcal{D}_{ij}(\bm{\theta}) p_{eq}(\bm{\theta})]=0 \,, \, \forall \ i, \\
&& \mathcal{W}(\bm{\theta}|\bm{\theta}')p_{eq}(\bm{\theta}') = \mathcal{W}(\bm{\theta}'|\bm{\theta})p_{eq}(\bm{\theta}) \,. 
\label{detailed_balance_Introduction2}
\end{eqnarray}
{\normalcolor The first equation \eqref{detailed_balance_Introduction1} corresponds to the vanishing of the probability current in \eqref{SDEtheta_Introduction}. Since discrete jumps are allowed, we also need a separate detailed balance condition given by \eqref{detailed_balance_Introduction2}.}

This definition encompasses a wide variety of stochastic processes which are relevant to physics. At the end of this section we will show how it can be specialized to active particles, leading to the results of our previous paper in the case of RTPs \cite{ExitProbaShort}. {\normalcolor To put it simply, for these models ${\bm \theta}(t)=v(t)$ corresponds to the intrinsic velocity of the active particle. In models such as the AOUP, the evolution of $v(t)$ follows a Langevin equation similar to \eqref{SDEtheta_Introduction} (an Ornstein-Uhlenbeck process in that case), while for other models such as the RTP, $v(t)$ takes discrete values, hence the transition kernel $\mathcal{W}(\bm{\theta}'|\bm{\theta})$. A combination of these two types of evolution is possible, as it is the case for the direction reversing active Brownian particle \cite{DRABP1,DRABP2}.} Another important class of models included in this definition are diffusing diffusivity models \cite{DiffDiffChubynsky, DiffDiffChechkin, DiffDiffJain, DiffDiffFPTSposini}: in that case, $\bm \theta$ is a $d$-dimensional Ornstein-Uhlenbeck process ($\underline{\mathcal{D}}$ is a constant, $\bm g$ is a harmonic force and $\mathcal{W}=0$), $f\left(x,\bm{\theta}\right)=F(x)$ and $\mathcal{T}\left(x,\bm{\theta}\right)=\bm \theta^2$. {\normalcolor In some situations, one may also want to consider a force $f(x,{\bm \theta},t)$ with an explicit time dependence. We will mention this case in Sec.~\ref{FPderivationsection}.}

We are interested in a situation where two absorbing walls are present at the positions $x=a$ and $x=b$, with $b>a$. By absorbing wall we mean that if the particle reaches one of these walls at any point in time, it remains there indefinitely (see left panel of Fig.~\ref{figure_walls}). It is important to stress that, contrary to another common definition of absorbing boundary conditions, the particle does not disappear, such that the integral of the particle density over the interval $[a,b]$, including walls, remains equal to $1$ at all times. The particle is initially placed inside the box such that $x(0)\in [a,b]$. The quantity that we want to compute is the probability that the particle reaches the wall located at $b$ before or at time $t$ (without touching the wall at $a$). This is called the \textit{exit probability} at $b$. In the current setting it can be formally defined as 
\begin{equation}\label{defExitProba}
    E_b(x,\bm{\theta},t) = \mathbb{P}(x(t)=b|x(0)=x,\bm{\theta}(0)=\bm{\theta}) \,.
\end{equation}
Here and in the rest of the paper, we use the notation $\mathbb{P}(x(t)\in...)$ to refer to the probability mass function that $x(t)$ belongs to a given set, while the notation $P$ or $p$ refers to a probability density. Being able to compute this quantity for a certain process gives us a lot of information about the first passage properties of this process. In particular, it contains the information on the survival probability, i.e. the probability that the particle is still in the interval $[a,b]$\footnote{{\normalcolor Eq.~\eqref{survivalmainresult} is mostly interesting for $x\in [a,b]$, but with the definition \eqref{defExitProba} it remains valid for $x=a$ and $x=b$ since $E_a(a,{\bm \theta},t)=E_b(b,{\bm \theta},t)=1$.}} after a duration $t$,
\begin{equation}
    Q_{[a,b]}(x,\bm{\theta},t) =1-E_a(x,\bm{\theta},t)-E_b(x,\bm{\theta},t) \; .
\label{survivalmainresult}
\end{equation}
One may also take the limit $a\to -\infty$, in which case it describes the probability that the particle is still below the value $x=b$ after a duration $t$. The survival probability can in turn be used to obtain the first passage time distribution of the process. It also plays an important role in the computation of extreme value statistics \cite{EVS1,EVS2,reviewEVSPal}.
\

\begin{figure}[t]
\centering
    \includegraphics[width=0.9\linewidth]{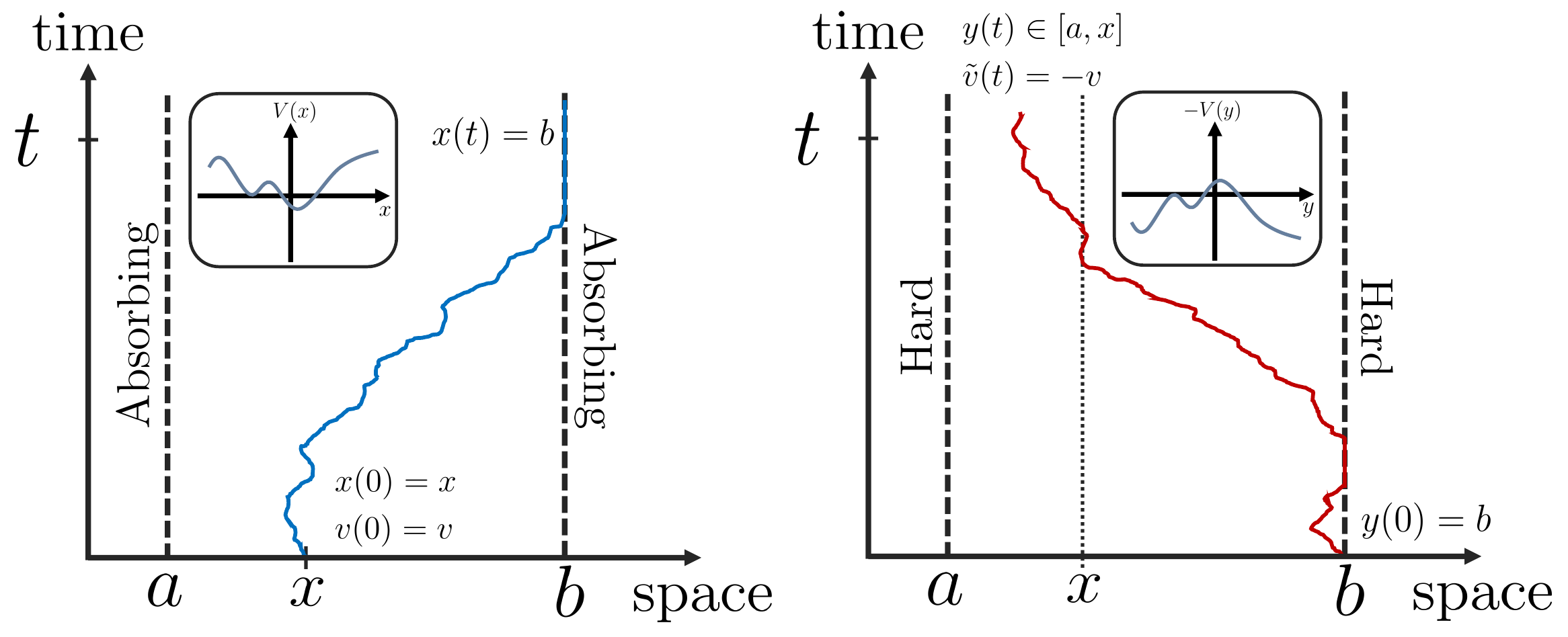} 
    \caption{Schematic representation of typical trajectories which contribute to the probabilities $E_b(x,v,t)$ and $\tilde \Phi(x,t|-v;b)$. \textbf{Left}:~The process $x(t)$ following the Langevin dynamics~(\ref{Langevin_Active_Introduction}) initiates its motion at position $x(t=0)=x$ with velocity $v(t=0)=v$. Its motion is eventually influenced by the presence of an external potential $V(x)$. Two absorbing walls are located at $a$ and $b$. The trajectory shown is absorbed at $b$ before time $t$ and hence contributes to the exit probability $E_b(x,v,t)$. \textbf{Right}:~The dual process of $x(t)$, namely $y(t)$ (with Langevin dynamics~(\ref{Langevin_Active_Introduction_Dual})), featuring hard walls at $x=a$ and $x=b$. The dual particle initiates its motion at $y(t=0)=b$ with an initial velocity drawn from the equilibrium distribution of the process $v(t)$, and in the presence of the reversed potential $-V(y)$. The trajectory shown contributes to $\Phi(x,t|\tilde v(t)=-v;b)$, i.e. the probability of locating the dual particle within the interval $[a,x]$ with a velocity $-v$ at time $t$. In this paper we show that $E_b(x,v,t) = \tilde \Phi(x,t|-v;b)$.
    }
\label{figureIntro_duality}
\end{figure}

The main results of this paper rely on the definition of a dual process of $x(t)$, which we denote $y(t)$. The evolution of $y(t)$ is governed by the equation
\begin{equation}
    \dot{y}(t) = \tilde f\left(y(t),\bm{\theta}(t)\right) + \sqrt{2\mathcal{T}\left(y(t),\bm{\theta}(t)\right)}\, \xi(t) \quad , \quad \tilde f(x,{\bm \theta})=-f\left(x,\bm{\theta}\right) + \partial_x \mathcal{T}\left(x,\bm{\theta}\right) \;.
\label{LangevinIntroductionDUAL}
\end{equation}
If the temperature is independent of the position, $y(t)$ has the same dynamics as $x(t)$ but with the sign of $f$ reversed, while if the temperature is space dependent the force also has an additional term $\partial_x \mathcal{T}$. Note that the transformation $f \to -f+\partial_x \mathcal{T}$ is its own inverse. Thus if one wants to find the process $x(t)$ which is dual to $y(t)$, the force applied to $x(t)$ is also given by $f=-\tilde f +\partial_x \mathcal{T}$. Here $\bm{\theta}(t)$ and $\xi(t)$ denote different realisations of the same processes as the ones appearing in \eqref{LangevinIntroduction}. In addition, we assume the presence of hard walls at $a$ and $b$ instead of absorbing walls. This means that if the particle is at $y=b$ with a positive velocity, it remains at $y=b$ until its velocity changes sign (and symmetrically for $y=a$ -- see right panel of Fig. \ref{figure_walls}).

Let us now assume that the dual process is initialized at position $y(0)=b$ and that $\bm{\theta}(0)$ is drawn from the equilibrium distribution $p_{eq}(\bm{\theta})$. Given these initial conditions, we are interested in the probability of finding the dual particle in an interval $[a,y]$ at time $t$, conditioned on the value of $\bm{\theta} (t)$, i.e. the cumulative distribution of the dual process $y(t)$ given $\bm{\theta}(t)$. We write it as follows
\begin{equation}
    \tilde \Phi(y,t|\bm{\theta};b) = \tilde{ \mathbb{P}}(y(t)\leq y | \bm{\theta}(t)=\bm{\theta};y(0)=b, \bm{\theta}(0)^{eq})\, ,
    \label{Phi_def_MainResults1}
\end{equation}
where the tildes indicate that the observable concerns the dual process. We also use a semicolon to separate the events at time $t$ (the conditioning on $\bm{\theta}(t)$, on the left in Eq.~(\ref{Phi_def_MainResults1})) from the initial condition (on the right). In the whole paper, a conditioning on $\bm{\theta}(0)^{eq}$ means that $\bm{\theta}(0)$ is drawn from $p_{eq}(\bm{\theta})$.

The primary objective of this paper is to demonstrate that, based on the above assumptions, it is possible to establish the following duality relation at all time:
\begin{equation} \label{mainRelation}
    E_b(x,\bm{\theta},t) = \tilde \Phi(x,t|\bm{\theta};b) \;.
\end{equation}
In simpler terms, this relation implies that the probability that the process $x(t)$ reaches the boundary $b$ before time $t$ when initiated with $\bm{\theta}(0)=\bm{\theta}$, is equal to the probability of finding the dual process $y(t)$ - where $\bm{\theta}(t)=\bm{\theta}$ - within the interval $[a,x]$ at time $t$, where $x=x(t=0)$ and $y(t=0)=b$. {\red This is a new, explicit formulation of Siegmund duality, adapted for the study of many relevant physical models.} Of course, the relation \eqref{mainRelation} has an equivalent for the exit probability at $x=a$, $E_a(x,\bm{\theta},t)$, which can be immediately deduced by symmetry (one simply needs to revert the inequality in the definition of $\tilde \Phi$ and to choose $y(0)=a$). The relation (\ref{mainRelation}) also remains valid in the limit where there is only one absorbing wall, for instance when $a\to - \infty$ (see appendix \ref{suvivalBM} for an illustration in the case of Brownian motion). Note that $E_b(x,\bm{\theta},t)$ and $\tilde \Phi(x,t|\bm{\theta};b)$ are both increasing functions of $x$ at fixed $t$ and increasing functions of $t$ at fixed $x$ (as long as $f$ does not depend explicitly on time).

A particular situation where the relation \eqref{mainRelation} can be useful is when the process $y(t)$ admits a stationary distribution, in which case one may consider the infinite time limit. If the system is ergodic, the stationary distribution is unique and independent of the initial condition. In this case \eqref{mainRelation} becomes at large times
\begin{equation} \label{mainRelationStationary}
    E_b^{st}(x,\bm{\theta})= \tilde{\Phi}^{st}( x|\bm{\theta})\, ,
\end{equation}
where the notation $``st"$ means that the quantities are drawn from there stationary distribution ($E_b^{st}(x,\bm{\theta}) = \lim_{t\to+\infty} E_b(x,\bm{\theta},t)$ and similarly for $\tilde \Phi$). In that case one has $E_b^{st}(x,\bm{\theta})+E_a^{st}(x,\bm{\theta})=1$.

One can also imagine situations where the initial value of $\bm{\theta}$ is unknown. In this case we assume that it is drawn from its equilibrium distribution $p_{eq}(\bm{\theta})$ and we define
\begin{equation}
    E_b(x,t) = \int d\bm{\theta} \, p_{eq}(\bm{\theta}) E_b(x,\bm{\theta},t) \;,
\end{equation}
as well as
\begin{equation}
    \tilde \Phi(y,t|b) = \tilde{ \mathbb{P}}(y(t)\leq y , t | y(0)=b, \bm{\theta}(0)^{eq}) = \int d\bm{\theta} \, p_{eq}(\bm{\theta}) \tilde \Phi(y,t|\bm{\theta};b) \, .
    \label{Phi_def_MainResults1}
\end{equation}
Averaging both sides of \eqref{mainRelation} over $p_{eq}(\bm{\theta})$ we obtain
\begin{equation} \label{mainRelation_averaged}
    E_b(x,t) = \tilde \Phi(x,t|b) \;.
\end{equation}

Finally, it is possible to derive the equivalent of \eqref{siegmundintro} for the Langevin dynamics (\ref{LangevinIntroduction}) (see Appendix~\ref{ContinuousProof_duality} for the complete derivation). The duality relation reads in this case
\begin{equation} \label{Siegmund}
    \mathbb{P}(x(t)\geq y|x(0)=x,\bm{\theta}(0)^{eq}) = \tilde{\mathbb{P}}(y(t)\leq x|y(0)=y,\bm{\theta}(0)^{eq})\, .
\end{equation}
This relation connects the full probability density of $x(t)$ (in the presence of absorbing walls) at any finite time with the probability density of $y(t)$ (in the presence of hard walls). The relation~(\ref{mainRelation_averaged}) can be recovered from Eq.~(\ref{Siegmund}) by taking $y=b$. It is worth mentioning that Eq.~(\ref{Siegmund}) still holds when there are no walls.

{\red The duality relation \eqref{mainRelation} is illustrated schematically in Fig.~\ref{figureIntro_duality}, where typical trajectories contributing to the probabilities $E_b(x,\bm{\theta},t)$ and $\tilde \Phi(x,t|\bm{\theta};b)$ are shown. It can be intuitively understood as a form of time reversal symmetry (hence the minus sign that appears in front of the force for the dual, while the additional term $\partial_x{\mathcal T}$ compensates for the flux of probability generated by the gradient of diffusion coefficient). The detailed balance conditions \eqref{detailed_balance_Introduction1}-\eqref{detailed_balance_Introduction2} for the driving process $\bm{\theta}(t)$ play a crucial role in this symmetry. This interpretation will become clearer with the derivation of the discrete time variant of this identity in Sec.~\ref{DiscreteProof}, which relies on a one-to-one mapping between the trajectories of the process and its dual.}
\\

{\bf Specialization to active particle models.} A particular class of models for which this duality relation can be interesting is active particles models. {\red The application of Siegmund duality to such systems was introduced in our previous work \cite{ExitProbaShort}.} Here we explain how to recover the results of \cite{ExitProbaShort} from the more general results derived here. The most well-known active particle models can all be described as follows: $\bm{\theta}(t)=v(t)$ is a scalar which represents the random velocity, and the dynamics typically involve $f(x,v) = F(x) + \alpha(x)\, v(t)$, with $F(x)$ denoting an external force, and $\alpha(x)$ a (generally positive) function modulating the particle's intrinsic speed, while we choose the temperature $\mathcal{T}(x,v) =T(x)$ to be independent of $v$. The equation of motion is then
\begin{equation}
    \dot{x}(t) = F(x(t)) + \alpha(x(t))\, v(t) + \sqrt{2 T(x(t))}\, \xi(t)\, .
\label{Langevin_Active_Introduction}
\end{equation}
This setting includes in particular:

- the run-and-tumble particle (RTP), which corresponds to $g=0$, $\mathcal{D}=0$, and $\mathcal{W}(v'|v)=\gamma\, \delta(v+v')$ in Eq. \eqref{SDEtheta_Introduction}, having for initial condition $v(0)=\pm v_0$, resulting in a constant velocity of $\pm v_0$ throughout. 

- the active Ornstein-Uhlenbeck particle (AOUP), represented by $g(v)=-\frac{v}{\tau}$ and $\mathcal{D}=\frac{D}{\tau^2}$ (constant), with $\mathcal{W}=0$. 

- for a two-dimensional active Brownian particle (ABP) projected into one dimension, it is simpler to choose $\bm{\theta}$ in \eqref{LangevinIntroduction} as the angle $\varphi$ between the particle's orientation and the $x$-axis. Here, the dynamics involve $f(x,\varphi) = F(x)+\alpha(x)\cos \varphi$ (we assume that the external force along the $x$-direction does not vary along the other space direction), with parameters $g=0$, $\mathcal{D}$ a constant and $\mathcal{W}=0$. However this can be easily rewritten under the form \eqref{Langevin_Active_Introduction} by writing $v(t)=\cos\varphi$.

For an active particle, the dual process of Eq.~(\ref{Langevin_Active_Introduction}) is equivalent to
\begin{equation}
    \dot{y}(t) = \tilde F(y(t)) + \alpha(y(t))\, \tilde{v}(t) + \sqrt{2T(y(t))}\, \xi(t) \quad , \quad \tilde F(x) = -F(x) + \partial_x T(x) \;,
\label{Langevin_Active_Introduction_Dual}
\end{equation}
where $\tilde{v}(t)$ has the same dynamics as $-v(t)$. This definition is more in line with the interpretation of $v$ as a velocity (changing the sign of $\alpha(x)$ would lead to a counter-inuitive situation where a positive velocity, i.e. positive value of $v(t)$, would push the particle in the $-$ direction). For all the examples considered in this paper (RTP, AOUP and ABP), the equation of motion is invariant under the change $v \to -v$, so that $\tilde v(t)$ evolves as $v(t)$.

In this setting, Eq.~(\ref{mainRelation}) reads
\begin{equation}  \label{mainRelationActive}
    E_b(x,v,t) = \tilde \Phi(x,t|\tilde v(t)=-v;b) \;.
\end{equation}
Here, it relates the exit probability of a particle with initial velocity $v$ {\red to} the cumulative distribution of the dual particle with velocity $-v$ at time $t$. Similarly, the results (\ref{mainRelationStationary}), (\ref{mainRelation_averaged}) and ({\ref{Siegmund}}) can be rewritten in terms of $v(t)$ and $\tilde v(t)$.
\\

{\bf Discrete time random walks.}
All the results above have an equivalent for discrete time random walks of the form
\be
    X_{n} = X_{n-1} + W_{n} \;,
\label{discreteevolutionmainresults}
\ee
where $W_n$ is a stationary stochastic process whose stationary distribution $p_{st}(w)$ obeys the time reversal property:
\begin{equation}
    p_{st}(w_1) P(W_2=w_2,...,W_T=w_T|W_1=w_1) = p_{st}(w_T) P(W_2=w_{T-1},...,W_T=w_1|W_1=w_T) \, .
    \label{timereversalmainresults}
\end{equation}
The process $W_n$ can take either discrete or continuous values, but its evolution does not depend on the position $X_n$. If $X_n$ has absorbing boundary conditions, we show in Sec.~\ref{discrete_case} how to construct its dual $Y_n$, with hard wall boundary conditions, such that the results above hold. 
We also propose an extension to continuous time random walks. An interesting example is the persistent random walk (which can be considered as a discrete model of active particles), which we use to illustrate the discrete time result in Sec.~\ref{sec:PRW}.
\\

{\normalcolor \textbf{Stochastic resetting.} All the models considered here can be extended by adding Poissonian resetting of the position. During a time interval $[t,t+dt]$, the process $x(t)$ evolves according to \eqref{LangevinIntroduction} with probability $1-\sum_{i=1}^n r_i dt$, or restarts it dynamics at position $X_r^i \in [a,b]$ with probability $r_i dt$, where $r_i$ ($i=1,...,n$) is the resetting rate associated with the resetting position $X_r^i$. In that case, the definition of the dual is a bit more subtle. Similarly to $x(t)$, the dual process $y(t)$ ``resets'' with rates $r_i$, $i=1,...,N$, but instead of resetting to a fixed position, the position $y(t^+)$ after the reset now depends on the position $y(t^-)$ just before the reset: if $y(t^-) \leq X_r^i$, then $y(t^+)=a$, but if $y(t^-) > X_r^i$, one has instead $y(t^+)=b$. Resetting can also be considered for discrete random walks. The details are given in Sec.~\ref{sec:resetting}. {\red The construction of a Siegmund dual for stochastic resetting is another important new result of this paper.}}
\\

{\red \textbf{$N$ independent stochastic processes.} An interesting and non-trivial new result for extreme value statistics can be derived as a corollary of the duality relation (\ref{Siegmund}). Consider $N$ independent processes $x_i(t)$ ($i \in \{1,2,\ldots N\}$) following the Langevin dynamics (\ref{LangevinIntroduction}) with absorbing walls located at $a$ and $b$, all starting at the same position $x_i(0)=x\in [a,b]$. Consider also $N$ independent realisations $y_i(t)$ of the dual process, which evolve according to (\ref{LangevinIntroductionDUAL}) with hard walls at $a$ and $b$, and initial condition $y_i(0)=y\in [a,b]$ for all $i$. The driven equilibrium process $\bm{\theta}(t)$ is initialized in its equilibrium distribution in both cases. In Appendix \ref{NparticlesAppendix}, we show that there exists a relation between the cumulative distributions of the maximum (resp. minimum) of the $x_i$'s, and the minimum (resp. maximum) of the $y_i$'s,
\be\label{}
\mathbb{P}\left(\max_{0\leq i\leq N} x_i(t) \geq y\, \Big|\, x_i(0)=x\right) = \mathbb{P}\left(\min_{0\leq i\leq N} y_i(t) \leq x\, \Big|\, y_i(0)=y \right)\;,
\ee 
and
\be\label{}
\mathbb{P}\left(\min_{0\leq i\leq N} x_i(t) \geq y\, \Big|\, x_i(0)=x\right) = \mathbb{P}\left(\max_{0\leq i\leq N} y_i(t) \leq x\, \Big|\, y_i(0)=y\right)\;.
\ee 
When specializing to $y=b$, this relates the exit probability at time $t$ of the maximum (resp. minimum) of $x_i$ at $b$, to the cumulative distribution of the minimum (resp. maximum) of $N$ independent dual processes starting at $b$.
\\
}

Since the equations used to compute analytically the two quantities $E_b$ and $\tilde{\Phi}$ are quite similar (indeed in Sec.~\ref{ContinuousProof} we show that in the continuous case they satisfy the same differential equation with the same boundary and initial conditions), one could expect that the difficulty associated with both computations will be comparable in most cases. However, it is useful to know that they are connected by such a simple relation, so that one can always obtain the second quantity as a byproduct of the other. In addition, some approximations or solution methods which might appear natural in one case might be less intuitive in the other (e.g. the replacement of the hard wall by a steep harmonic potential as done in \cite{hardWallsJoanny, hardWallsCaprini}). If one wants to compute these quantities using numerical simulations however, there are cases where one quantity is clearly simpler to obtain than the other. In particular if one is interested in the large time limit, if the system is ergodic, the stationary density in the presence of hard walls can be obtained using a single simulation and averaging over time. In contrast, computing the exit probability requires a large number of simulations starting from every position $x$ and waiting for the particle to reach one of the walls every time, which requires much more computation time. This could also be useful in experimental settings, where it is often much simpler to obtain long time series of data from a single particle trajectory \cite{singletraj}. One could therefore imagine an experimental protocol to indirectly measure the exit probability of a particle via the measure of the distribution of positions of a ``dual'' particle. On the other hand, the probability density conditioned on the value of ${\bm\theta}$ can be challenging to compute numerically, while computing the exit probability for a given initial value of ${\bm \theta}$ is quite straightforward.

\section{Duality for continuous stochastic processes}\label{FPderivationsection}

In this section we derive the duality relation \eqref{mainRelation} for continuous stochastic processes using the Fokker-Planck formalism. This derivation is a generalization of the one presented in Appendix \ref{ProofsSimple} for a Brownian motion with an external force and in \cite{ExitProbaShort} for a run-and-tumble particle. We then illustrate this relation on several examples through numerical simulations.

\subsection{Derivation of the duality relation from the Fokker-Planck equation} \label{ContinuousProof}

In this section we will show the main result \eqref{mainRelation} for continuous stochastic processes, using the Fokker-Planck equation. We consider a general continuous stochastic process of the form:
\begin{equation}
    \dot{x}(t) = f\left(x(t),\bm{\theta}(t)\right) + \sqrt{2\mathcal{T}\left(x(t),\bm{\theta}(t)\right)}\, \xi(t)\, ,
\label{SDEgeneral}
\end{equation}
where $\xi(t)$ is Gaussian white noise with zero mean and unit variance. $\bm{\theta}(t)$ is a vector of parameters which follows a stochastic evolution independent of $x$, of the form 
\begin{equation}
    \bm{\dot{\theta}}(t) = \bm{g}\left(\bm{\theta}(t)\right) + \left[2\underline{\mathcal{D}}(\bm{\theta}(t))\right]^{1/2} \cdot \bm{\eta}(t)\, .
\label{SDEtheta}
\end{equation}
where $\underline{\mathcal{D}}$ is a positive matrix and the $\eta_i(t)$'s are independent Gaussian white noises with zero mean and unit variance. In addition, it can jump from the value $\bm{\theta}$ to $\bm{\theta}'$ with a transition kernel $\mathcal{W}(\bm{\theta}'|\bm{\theta})$.

The Fokker-Planck equation for the joint probability density $P(x,\bm{\theta},t)$ reads (with the It\=o convention)
\begin{equation}
\partial_t P = - \partial_x [f(x,\bm{\theta}) P] + \partial_x^2 [\mathcal{T}(x,\bm{\theta}) P] - \sum_i \partial_{\theta_i}[g_i(\bm{\theta}) P] + \sum_{i,j} \partial_{\theta_i\theta_j}^2 [\mathcal{D}_{ij}(\bm{\theta}) P] + \int d\bm{\theta}' \left[\mathcal{W}(\bm{\theta}|\bm{\theta}')P(x,\bm{\theta}',t) - \mathcal{W}(\bm{\theta}'|\bm{\theta})P(x,\bm{\theta},t)\right] \,.
\label{FPgeneral}
\end{equation}
We will denote $P_{\bm{\theta}}(x,t)=P(x,t|\bm{\theta})$ the probability density of the positions conditioned on $\bm{\theta}$ and $p(\bm{\theta},t)$ the density probability of $\bm{\theta}$, such that $P(x,\bm{\theta},t) = p(\bm{\theta},t) P_{\bm{\theta}}(x,t)$. Integrating \eqref{FPgeneral} over $x$ we obtain the Fokker-Planck equation for $p(\bm{\theta},t)$ (we will consider densities with a finite support $[a,b]$, thus the boundary terms vanish)
\begin{equation}
\partial_t p = - \sum_i \partial_{\theta_i}[g_i(\bm{\theta}) p] + \sum_{i,j} \partial_{\theta_i\theta_j}^2 [\mathcal{D}_{ij}(\bm{\theta}) p] + \int d\bm{\theta}' \, [\mathcal{W}(\bm{\theta}|\bm{\theta}')p(\bm{\theta}',t) - \mathcal{W}(\bm{\theta}'|\bm{\theta})p(\bm{\theta},t)] \, .
\label{FPtheta}
\end{equation}
Note that we can also derive Eq.~(\ref{FPtheta}) directly from Eq.~(\ref{SDEtheta}).
We now assume that equation \eqref{FPtheta} has an equilibrium solution $p_{eq}(\bm{\theta})$, i.e. a stationary solution which satisfies the local detailed balance conditions
\begin{eqnarray}
&&0 = - g_i(\bm{\theta}) p_{eq}(\bm{\theta}) + \sum_{j} \partial_{\theta_j}[\mathcal{D}_{ij}(\bm{\theta}) p_{eq}(\bm{\theta})] \,, \, \forall \ i,\label{detailed_balance1_proof_continuous} \\
&& \mathcal{W}(\bm{\theta}|\bm{\theta}')p_{eq}(\bm{\theta}') = \mathcal{W}(\bm{\theta}'|\bm{\theta})p_{eq}(\bm{\theta}) \,. 
\label{detailed_balance2_proof_continuous}
\end{eqnarray}

Let us first consider the dynamics \eqref{SDEgeneral}-\eqref{SDEtheta} on an interval $[a,b]$ with absorbing boundary conditions at $x=a$ and $x=b$. In this section we assume that the whole interval is accessible to the particle\footnote{This may not be the case{\blue ,} e.g. when the noise terms have a finite amplitude (for instance in the case of RTPs) and the external force is too strong. See \cite{ExitProbaShort} to see how this can be dealt with in the particular case of RTPs at infinite times.}. We are interested in the probability that a particle starting at position $x$ at time $t=0$, with a certain initialisation value of $\bm{\theta}$, is absorbed at $x=b$ before time $t$, denoted $E_b(x,\bm{\theta},t)$. Since the joint process $(x,\bm{\theta})$ is Markovian, one has
\begin{eqnarray}
E_b(x,\bm{\theta},t+dt) &=& \mathbb{E}_{\xi,\bm \eta}\left[E_b\left(x+dt[f(x,\bm{\theta})+\sqrt{2\mathcal{T}(x,\bm{\theta})} \, \xi(t)],\bm{\theta}+dt[\bm{g}(\bm{\theta}) + (2\underline{\mathcal{D}}(\bm{\theta}))^{1/2} \cdot \bm{\eta}(t)],t\right)\right] \nonumber \\
&& +\, dt \int d\bm{\theta}' \, \mathcal{W}(\bm{\theta}'|\bm{\theta})[E_b(x,\bm{\theta}',t) - E_b(x,\bm{\theta},t)] \;,
\end{eqnarray}
which leads to the backward Fokker-Planck equation for $E_b(x,\bm{\theta},t)$,
\begin{equation}
\partial_t E_b = f(x,\bm{\theta}) \partial_x E_b + \mathcal{T}(x,\bm{\theta}) \partial_x^2 E_b + \sum_i g_i(\bm{\theta}) \partial_{\theta_i} E_b + \sum_{i,j} \mathcal{D}_{ij}(\bm{\theta}) \partial_{\theta_i\theta_j}^2 E_b + \int d\bm{\theta}'\, \mathcal{W}(\bm{\theta}'|\bm{\theta})[E_b(x,\bm{\theta}',t) - E_b(x,\bm{\theta},t)] \;,
\label{backwardFPgeneral}
\end{equation}
which is complemented by the boundary conditions\footnote{In Eq.~(\ref{exit_bc}), by ``$\mathcal{T}(x,\bm{\theta})>0 \text{ in the vicinity of } a$'' we mean that there does not exist $\epsilon>0$ such that $\mathcal{T}(x,\bm{\theta})=0$ for every $x\in[a,a+\epsilon]$.} 
\begin{eqnarray} \label{exit_bc}
&&E_b(a^+, \bm{\theta},t) = 0 \quad \text{for all } f(a^+,\bm{\theta})<0, \text{ or if } \mathcal{T}(x,\bm{\theta})>0 \text{ in the vicinity of } a, \\
&&E_b(b^-, \bm{\theta},t) = 1 \quad \text{for all } f(b^-,\bm{\theta})>0, \text{ or if } \mathcal{T}(x,\bm{\theta})>0 \text{ in the vicinity of } b, \nonumber
\end{eqnarray}
and the initial conditions
\begin{eqnarray}\label{exit_ic}
&&E_b(x,\bm{\theta},0) = 0 \text{ for } x<b, {\color{red}}\\
&&E_b(b,\bm{\theta},0) = 1. \nonumber
\end{eqnarray}
The boundary conditions \eqref{exit_bc} require some explanation. Although the values of $E_b(x,\bm{\theta},t)$ exactly at $x=a$ and $x=b$ are fixed by the absorbing conditions, there can be discontinuities in some cases. Indeed if $\mathcal{T}(x,\bm{\theta})=0$ and the force $f(x,\bm{\theta})$ is driving the particle away from the wall at time $t=0$, then a particle starting at an infinitesimal distance from the wall will not be absorbed immediately (in fact it may even escape and reach the opposite wall). However if the force is driving the particle towards the wall, then it will be absorbed with probability 1 and there will be no discontinuity. Additionally, if a Brownian term is present, i.e. $\mathcal{T}(x,\bm{\theta})>0$, then there can be no discontinuity independently of the sign of $f$. Indeed at infinitely small times, only the Brownian term is relevant in this case. Since a Brownian motion always goes back to its starting point infinitely many times before moving away, the particle will be absorbed.

Let us now consider the probability density in the presence of hard walls at $x=a$ and $x=b$. In addition, we replace the force $f(x,{\bm \theta})$ by some $\tilde f(x,{\bm \theta})$. All the probabilities and probability densities related to this new process will be denoted with a tilde (however the distribution $p_{eq}({\bm \theta})$ remains the same). In this case, if $\mathcal{T}=0$, the density may have some delta peaks at $x=a$ and $x=b$ (see examples in section \ref{numerical_evidence_section}). Here we assume that at $t=0$ the parameter $\bm{\theta}$ is initialised in its equilibrium distribution $p_{eq}(\bm{\theta})$ (and thus it keeps the same distribution at all times). Starting from \eqref{FPgeneral}, we derive an equation for the conditional density $\tilde P_{\bm{\theta}}(x,t)$ (using $\tilde P(x,\bm{\theta},t) = p_{eq}(\bm{\theta}) \tilde P_{\bm{\theta}}(x,t)$),
\begin{eqnarray}
p_{eq}(\bm{\theta}) \partial_t \tilde P_{\bm{\theta}} &=& - p_{eq}(\bm{\theta}) \partial_x \left[\tilde f(x,\bm{\theta}) \tilde P_{\bm{\theta}}\right] + p_{eq}(\bm{\theta}) \partial_x^2 [\mathcal{T}(x,\bm{\theta}) \tilde P_{\bm{\theta}}] + \tilde P_{\bm{\theta}} \sum_i \partial_{\theta_i}\Big\{-g_i(\bm{\theta}) p_{eq}(\bm{\theta}) + \sum_j \partial_{\theta_j}[\mathcal{D}_{ij}(\bm{\theta}) p_{eq}(\bm{\theta})]\Big\} \nonumber \\
&& + \sum_{i} \Big\{-g_i(\bm{\theta}) p_{eq}(\bm{\theta}) + 2 \sum_j \partial_{\theta_j} [\mathcal{D}_{ij}(\bm{\theta}) p_{eq}(\bm{\theta})]\Big\} \partial_{\theta_i} \tilde P_{\bm{\theta}} + \sum_{i,j} \mathcal{D}_{ij}(\bm{\theta}) p_{eq}(\bm{\theta}) \partial_{\theta_i\theta_j}^2 \tilde P_{\bm{\theta}} \\
&&+ \int d\bm{\theta}' \left[\mathcal{W}(\bm{\theta}|\bm{\theta}') p_{eq}(\bm{\theta}') \tilde P_{\bm{\theta}'} - \mathcal{W}(\bm{\theta}'|\bm{\theta}) p_{eq}(\bm{\theta}) \tilde P_{\bm{\theta}}\right] \nonumber \\
=  p_{eq}(\bm{\theta}) \hspace{-0.2cm} &\Big\{& \hspace{-0.2cm} -\partial_x [\tilde f(x,\bm{\theta}) \tilde P_{\bm{\theta}}] + \partial_x^2 [\mathcal{T}(x,\bm{\theta}) \tilde P_{\bm{\theta}}] + \sum_{i} g_i(\bm{\theta}) \partial_{\theta_i} \tilde P_{\bm{\theta}} + \sum_{i,j} \mathcal{D}_{ij}(\bm{\theta}) \partial_{\theta_i\theta_j}^2 \tilde P_{\bm{\theta}} + \int d\bm{\theta}' \mathcal{W}(\bm{\theta}'|\bm{\theta}) [\tilde P_{\bm{\theta}'} - \tilde P_{\bm{\theta}}] \Big\} \nonumber
\end{eqnarray}
where we have made an extensive use of the detailed balance conditions \eqref{detailed_balance1_proof_continuous} and \eqref{detailed_balance2_proof_continuous} to obtain the second identity. We can then eliminate $p_{eq}(\bm{\theta})$ to obtain
\begin{equation}
\partial_t \tilde P_{\bm{\theta}} = -\partial_x [\tilde f(x,\bm{\theta}) \tilde P_{\bm{\theta}}] + \partial_x^2 [\mathcal{T}(x,\bm{\theta}) \tilde P_{\bm{\theta}}] + \sum_{i} g_i(\bm{\theta}) \partial_{\theta_i} \tilde P_{\bm{\theta}} + \sum_{i,j} \mathcal{D}_{ij}(\bm{\theta}) \partial_{\theta_i\theta_j}^2 \tilde P_{\bm{\theta}} + \int d\bm{\theta}' \mathcal{W}(\bm{\theta}'|\bm{\theta}) [\tilde P_{\bm{\theta}'} - \tilde P_{\bm{\theta}}] \;.
\label{FPconditional}
\end{equation}

We now introduce the cumulative distribution of the conditional density $\tilde P_{\bm{\theta}}$,
\begin{equation}
    \tilde \Phi(x, t|\bm{\theta}) = \int_{a^-}^x dy \, \tilde P_{\bm{\theta}}(y,t) = \int_{a^-}^x dy \, \tilde P(y,t|\bm{\theta}) \quad , \quad \tilde P_{\bm{\theta}}(x,t) = \partial_x \tilde \Phi(x, \bm{\theta}, t)
\end{equation}
where the integral starts at $a^-$ to include a potential delta peak at $x=a$. Writing \eqref{FPconditional} in terms of $\tilde \Phi$ yields
\begin{equation}
\partial_x \partial_t \tilde \Phi = \partial_x \Big\{-\tilde f(x,\bm{\theta}) \partial_x \tilde \Phi + \partial_x [\mathcal{T}(x,\bm{\theta}) \partial_x \tilde \Phi] + \sum_{i} g_i(\bm{\theta}) \partial_{\theta_i} \tilde \Phi + \sum_{i,j} \mathcal{D}_{ij}(\bm{\theta}) \partial_{\theta_i\theta_j}^2 \tilde \Phi + \int d\bm{\theta}' \mathcal{W}(\bm{\theta}'|\bm{\theta}) [\tilde \Phi(x, t|\bm{\theta}') - \tilde \Phi(x, t|\bm{\theta})] \Big\} \;.
\label{preFPphi}
\end{equation}
We can then integrate this equation between $x=-\infty$ and $x$, using that $\tilde \Phi(-\infty,t|\bm{\theta})=0$ as well as its derivatives. Using that $\partial_x [\mathcal{T}(x,\bm{\theta}) \partial_x \tilde \Phi] = \partial_x \mathcal{T}(x,\bm{\theta}) \partial_x \tilde \Phi + \mathcal{T}(x,\bm{\theta}) \partial_x^2 \tilde \Phi$, this finally yields the differential equation satisfied by $\tilde \Phi(x, t|\bm{\theta})$,
\begin{equation}
\partial_t \tilde \Phi = [-\tilde f(x,\bm{\theta}) + \partial_x \mathcal{T}(x,\bm{\theta})] \partial_x \tilde \Phi + \mathcal{T}(x,\bm{\theta}) \partial_x^2 \tilde \Phi + \sum_{i} g_i(\bm{\theta}) \partial_{\theta_i} \tilde \Phi + \sum_{i,j} \mathcal{D}_{ij}(\bm{\theta}) \partial_{\theta_i\theta_j}^2 \tilde \Phi + \int d\bm{\theta}' \mathcal{W}(\bm{\theta}'|\bm{\theta}) [\tilde \Phi(x, t|\bm{\theta}') - \tilde \Phi(x, t|\bm{\theta})] \;.
\label{FPphi}
\end{equation}

Let us now fix $\tilde f(x,{\bm \theta}) = -f(x,{\bm \theta}) + \partial_x \mathcal{T}(x,\bm{\theta})$. Then this is exactly the same as the equation \eqref{backwardFPgeneral} for the exit probability at $E_b(x,\bm{\theta},t)$. Note that to go from Eq.~(\ref{FPconditional}) to Eq.~(\ref{FPphi}), it is essential that $\bm g$ and $\underline{\mathcal{D}}$ do not depend on $x$. The boundary conditions are
\begin{eqnarray}\label{cumul_bc}
&&\tilde \Phi(a^+,t|\bm{\theta}) = 0 \quad \text{for all } \tilde f(a^+,\bm{\theta})>0, \text{ or if } \mathcal{T}(x,\bm{\theta})>0 \text{ in the vicinity of } a, \\
&&\tilde \Phi(b^-,t|\bm{\theta}) = 1 \quad \text{for all } \tilde f(b^-,\bm{\theta})<0, \text{ or if } \mathcal{T}(x,\bm{\theta})>0 \text{ in the vicinity of } b, \nonumber
\end{eqnarray}
which are also the same as \eqref{exit_bc} when writing $\tilde f(x,{\bm \theta}) = -f(x,{\bm \theta}) + \partial_x \mathcal{T}(x,\bm{\theta})=0$ (indeed if $\mathcal{T}(x,\bm{\theta})=0$ near the wall then one simply has $\tilde f(x,{\bm \theta}) = -f(x,{\bm \theta})$ in this region). These conditions translate the fact that there can be an accumulation of particles at the boundaries (i.e. a delta in the density at $a$ or $b$), but only if the velocity is oriented towards the wall, and if there is no Brownian noise. Finally, one can choose an initial condition which matches \eqref{exit_ic} 
by assuming that at $t=0$ all particles are at $x=b$,
\begin{eqnarray}\label{cumul_ic}
&& \tilde \Phi(x,0|\bm{\theta};b) = 0 \text{ for } x<b, \\
&& \tilde \Phi(b,0|\bm{\theta};b) = 1 \nonumber
\end{eqnarray}
(remember that $\bm{\theta}$ is initialised at equilibrium). With this choice of initial condition, the two quantities $E_b(x,\bm{\theta},t)$ and $\tilde \Phi(x, t|\bm{\theta};b)$ 
follow the same differential equation with the same boundary and initial condition. One can thus reasonably assume that\footnote{In theory, one would need to show the unicity of the solution of the PDE with these initial and boundary conditions. Here we choose to leave aside these considerations.}
\begin{equation}
    E_b(x,\bm{\theta},t) = \tilde \Phi(x, t|\bm{\theta};b) \;.
\label{identity_general}
\end{equation}

In Appendix \ref{ContinuousProof_duality}, using the same derivation procedure, we show the more general result
\begin{equation}
     \mathbb{P}(x(t)\geq y|x,{\bm \theta}) = \tilde{\mathbb{P}}(y(t)\leq x|{\bm \theta}(t)={\bm \theta};y,{\bm \theta}(0)^{eq})  \, ,
\end{equation}
from which \eqref{Siegmund} can be deduced by averaging over $p_{eq}({\bm \theta})$. This relates the cumulative distribution of the process with absorbing wall with the cumulative of its dual initialized at position $y$. It yields back \eqref{identity_general} when taking $y=b$.

To conclude this section, let us add that the result can be extended to a time-dependent force $f(x,{\bm \theta}, t)$. However, in this case, the time-dependence of the force should be reversed in the dual process. More precisely, if one wants to compute the exit probability after some time $t_f$, the force applied in the dual process should be $\tilde f(x,{\bm \theta},t) = -f(x,{\bm \theta},t_f-t) + \partial_x \mathcal{T}(x,\bm{\theta})$. We will not expand upon this case in the present article, but an example, with some numerical results, is given in Sec.~\ref{sec:timedepf}.

\subsection{Examples and numerical results}\label{numerical_evidence_section}

The duality relation (\ref{mainRelation}), derived in Sec.~\ref{ContinuousProof}, extends to a broad spectrum of continuous stochastic processes, including the most well-known models of active particles and diffusing diffusivity models. In this section, we consider several examples of frequently studied continuous processes, and show numerical evidence that this relation holds in all these cases. To do this, we performed direct simulations of the stochastic differential equation for each model and computed the two quantities of interest as follows:

(i) For given initial values of $x$ and ${\bm \theta}$, we compute $E_b(x,{\bm \theta},t)$ by simulating independently the trajectories of $N$ particles over a time interval $t$ and counting how many of them escape at $b$.

(ii) For the cumulative distribution, we simply simulate independently the trajectories of $N$ particles over a time interval $t$, starting from $x=b$ and with the initial value of ${\bm \theta}$ drawn from $p_{eq}({\bm \theta})$, and compute a histogram of positions at time $t$, for each value of ${\bm \theta}$ (if ${\bm \theta}$ varies continuously we need to discretize the ${\bm \theta}$ space).

\begin{figure}[t]
\centering
    \begin{minipage}[c]{.32\linewidth}
        \centering
        \includegraphics[width=1.\linewidth]{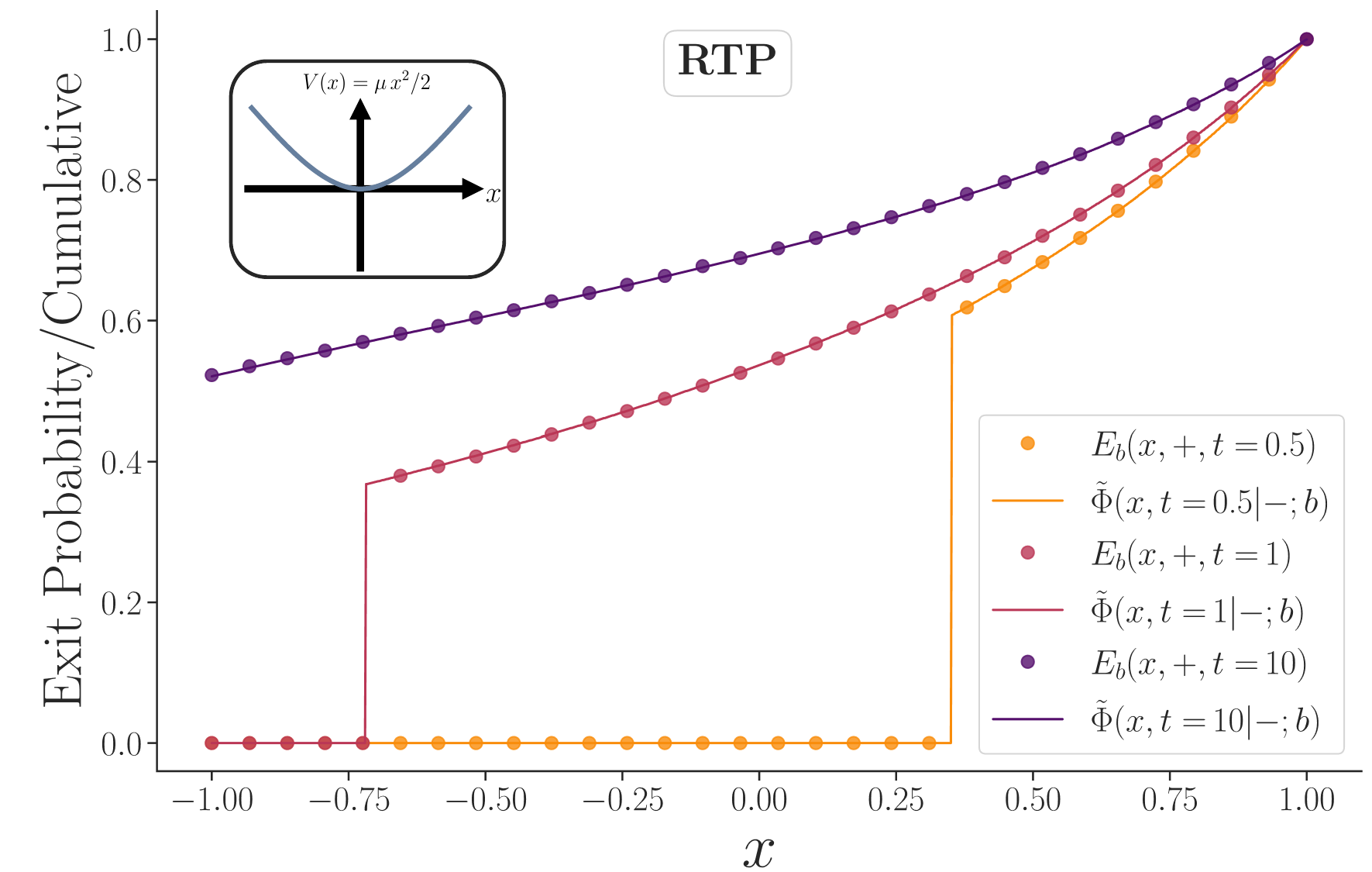}
    \end{minipage}
    \hfill%
    \begin{minipage}[c]{.32\linewidth}
        \centering
        \includegraphics[width=1.\linewidth]{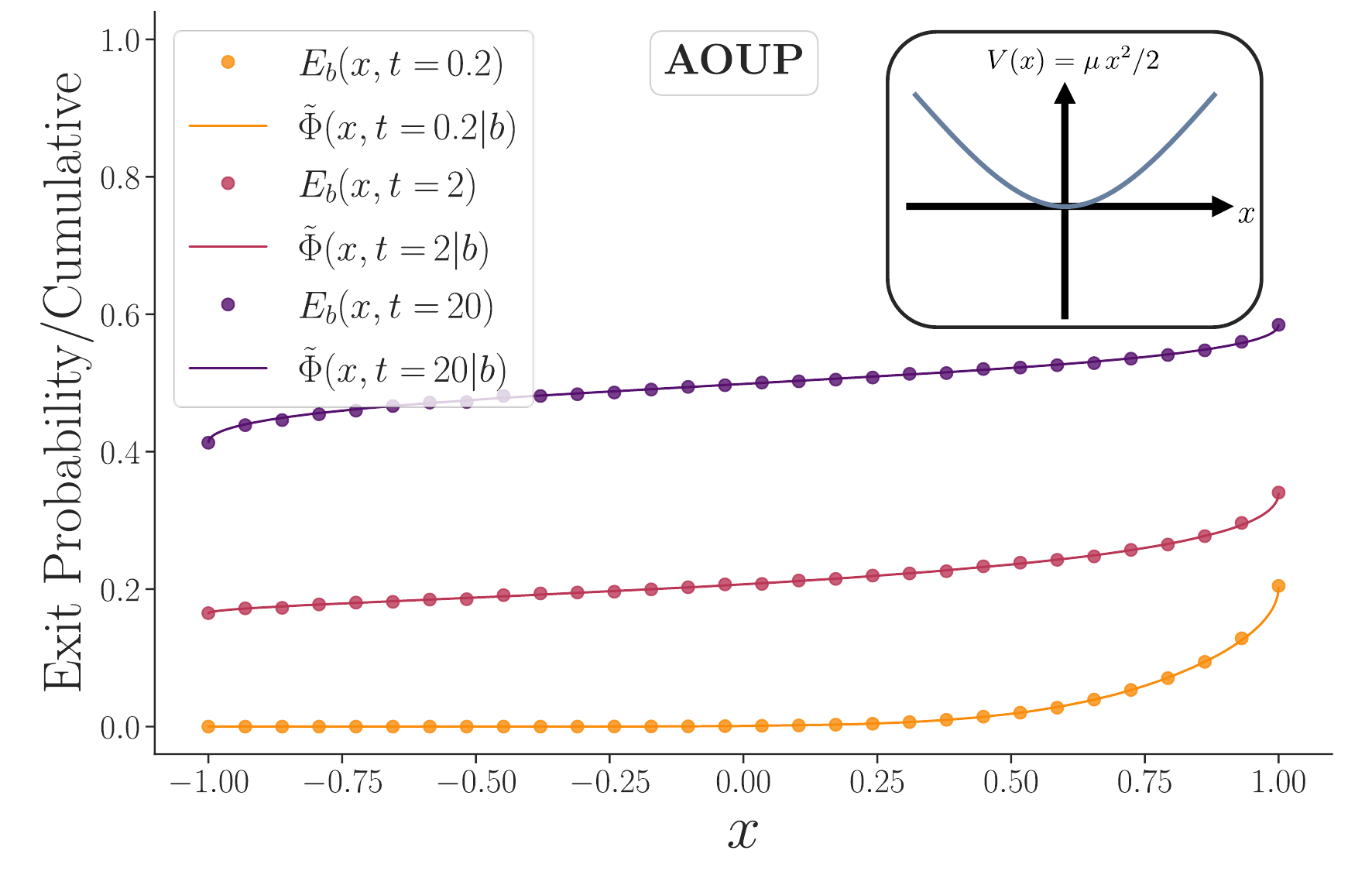}
    \end{minipage}
    \hfill%
    \begin{minipage}[c]{.32\linewidth}
        \centering
        \includegraphics[width=1.\linewidth]{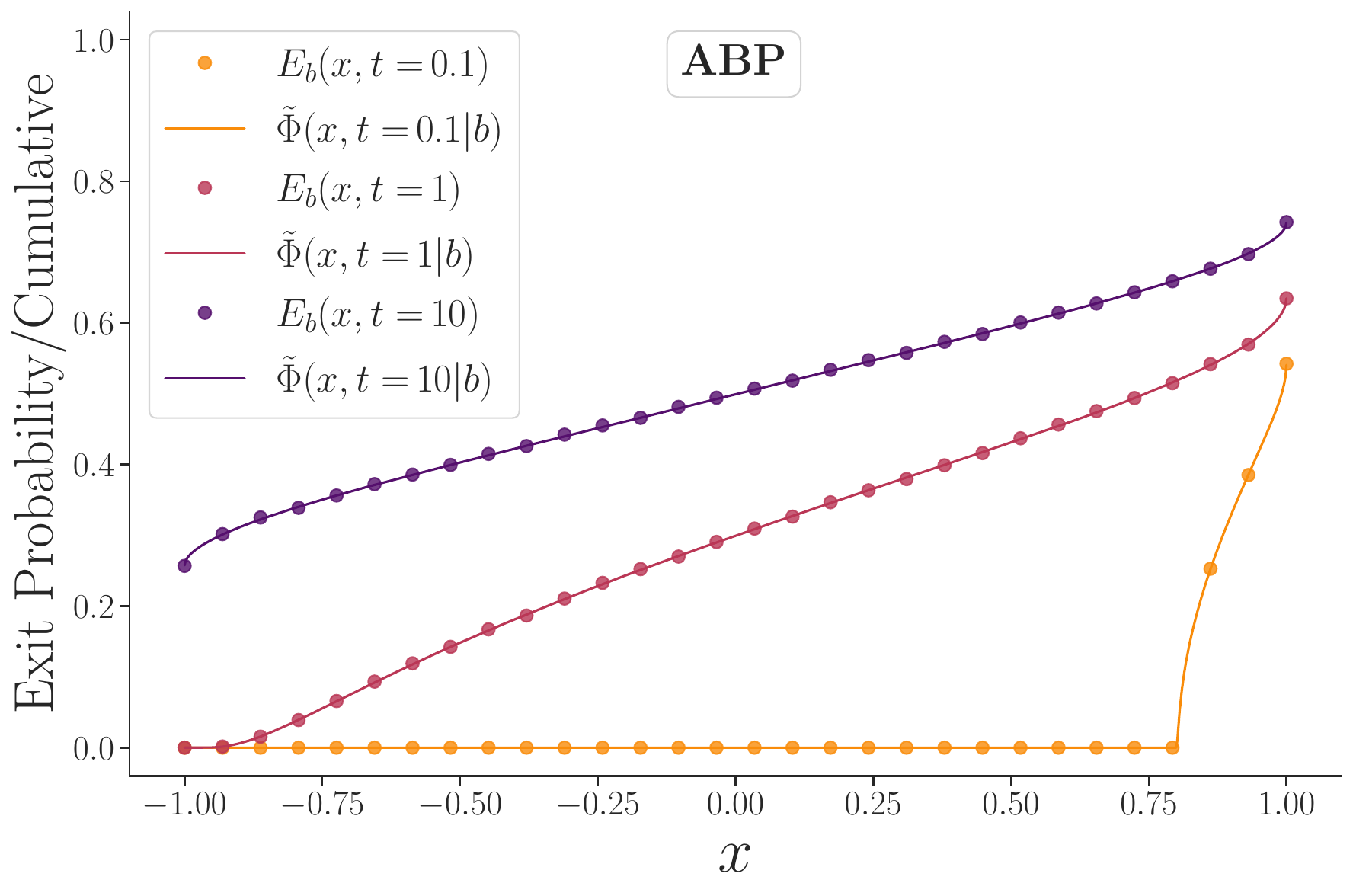}
    \end{minipage}
    \hfill%

    \caption{We validate the duality relations~\eqref{mainRelationActive} and~\eqref{mainRelation_averaged} numerically considering various models of active particles at finite times. The dots represent the exit probability, while the solid lines show the empirical cumulative distribution of the dual process, both being computed numerically through averages over many trajectories. \textbf{Left}: run-and-tumble particle (RTP) with $v_0=2$ and $\gamma=1$. The particle is initialized in the $+$ state for the exit probability and in the presence of a harmonic potential $V(x)=\frac{\mu}{2}x^2$ with $\mu=1$ for $E_b$. The dual process is in the presence of the reversed potential $-V(x)$. \textbf{Center}: active Ornstein-Uhlenbeck particle (AOUP) with $D=4$, $\tau=1$, in the presence of a harmonic potential with $\mu=1$ for $E_b$ and $-V(x)$ for $\tilde \Phi$. \textbf{Right}: active Brownian particle (ABP) with velocity $v_0=2$ and diffusion coefficient $D=1${\red , in the absence of external potential}. In each case, the exit probability overlaps the empirical cumulative of the dual process perfectly. The discontinuities observed in the yellow curve can be attributed to the fact that particles starting on the left side of these discontinuities do not have enough time to exit the interval, even if they remain in the positive state throughout the entire simulation.}
\label{FiniteTimeFig}
\end{figure}

In practice we only tested the complete relation \eqref{mainRelation} in the RTP case (see the left panel of Figure \ref{FiniteTimeFig}). For models where ${\bm \theta}$ takes continuous values we only tested the integrated version \eqref{mainRelation_averaged}. We used $N\sim 10^5 \hspace{-0.05cm}-\hspace{-0.05cm}10^7$ depending on the model\footnote{If we only wanted to compute the cumulative distribution in the stationary state, it would be more efficient to perform the simulation for a single particle and compute the histogram of positions over time once it has reached stationarity, as in \cite{ExitProbaShort}.}.

The first three models considered in this section are well-known models of active particles, which can all be written under the form \eqref{Langevin_Active_Introduction} (i.e. ${\bm \theta}$ plays the role of a driving velocity), and for which the dual process and duality relation take the forms \eqref{Langevin_Active_Introduction_Dual} and \eqref{mainRelationActive} respectively. We then consider a diffusing diffusivity model, for which the parameter ${\bm \theta}$ now modifies the temperature of the process $x(t)$. Finally we consider two simple cases with a space-dependent temperature and a time-dependent external force.

\subsubsection{RTP}
We start with the run-and-tumble particle (RTP) model {\red \cite{ Tailleur_RTP,DKM19, Cates2012}}, defined as
\begin{equation}\label{defAOUP}
\frac{dx}{dt} = F(x) + v_0\, \sigma(t) \quad , \quad
\sigma(t+dt) = \begin{cases}
\sigma(t) &\text{, with probability } (1 - \gamma \, dt)\\
-\sigma(t) &\text{, with probability } \gamma\, dt 
\end{cases}\, ,
\end{equation}
where $\sigma(0)=\pm 1$, and $F(x)=-V'(x)$ is an arbitrary external force. This case has been studied extensively in \cite{ExitProbaShort}, where we have explicitly computed both sides of \eqref{mainRelationActive} in the stationary state. In Fig.~\ref{FiniteTimeFig} (left panel) we illustrate this relation at finite time:
\begin{equation} \label{RTPfinitietimeDual}
    E_b(x,\pm,t) = \tilde \Phi(x,t|\tilde \sigma(t)=\mp;b) \;.
\end{equation}
When computing the r.h.s the initial value of $\tilde \sigma$ should be taken as $\pm 1$ with equal probability (since $p_{eq}(\sigma=\pm 1)=\frac{1}{2}$).
\subsubsection{Active Orstein-Uhlenbeck particle (AOUP)}

We then consider an active Orstein-Uhlenbeck particle (AOUP) {\red \cite{AOUP,Wijland21}},
for which the equation of motion is
\begin{equation}\label{defAOUP}
\frac{dx}{dt} = F(x) + v(t) \quad , \quad
\tau \frac{dv}{dt} = -v(t) + \sqrt{2D} \ \eta(t)\, ,
\end{equation}
where $F(x)=-V'(x)$ is an external force, $\tau$ is the persistence time, $D$ is a {\red diffusion coefficient}, and $\eta$ a Gaussian white noise. For the dual, the velocity $v$ should again be initialised in the equilibrium distribution, now given by
\be \label{eqAOUP}
p_{eq}(v)= \sqrt{\frac{\tau}{2\pi D}} \, e^{-\frac{\tau v^2}{2D}}\, .
\ee
Since conditioning on a continuous variable is more challenging numerically, in this case and the next ones we instead test the integrated identity (\ref{mainRelation_averaged}). Thus the velocity is also initialised in the distribution \eqref{eqAOUP} for the exit probability. The central panel of Fig.~\ref{FiniteTimeFig} shows a perfect agreement between the two quantities, for an AOUP in a harmonic potential.

\subsubsection{2d active Brownian particle (ABP) between two parallel walls}

Another common model of active particles is the active Brownian particle (ABP) {\red \cite{activeintro4, ABM}}. It can only be defined in two dimensions or higher (in 1d it coincides with the RTP). Here we consider the 2d case in the particular situation where it can effectively be described in 1d. More precisely, we assume that the particle is confined between two parallel walls perpendicular to the $x$-direction, at $x=a$ and $x=b$, and that the external force in the $x$-direction is independent of the second coordinate $z$, so that the evolution along the $x$-axis is completely independent of the $z$ coordinate,
\begin{equation}\label{defABP}
\frac{dx}{dt} = F_x(x) + v_0 \cos \varphi(t) \quad , \quad
\frac{dz}{dt} = F_z(x,z) + v_0 \sin \varphi(t) \quad , \quad
\frac{d\varphi}{dt} = \sqrt{2D} \ \eta(t),
\end{equation}
where $F_x(x)$ and $F_y(x,z)$ are the projections of the external force, $v_0$ is a constant speed, $D$ is again a {\red diffusion coefficient}, and $\eta(t)$ a Gaussian white noise. Forgetting about the $z$ coordinate, this problem is very similar to the AOUP case. The equilibrium distribution of $\varphi(t)$ is simply the uniform distribution on $[0,2\pi)$. The right panel of Figure \ref{FiniteTimeFig} illustrates Eq.~(\ref{mainRelation_averaged}) in this case.

\subsubsection{Diffusing diffusivity}\label{numerics_diffusingdiffusivity}
\begin{figure}[t]
\centering
    \begin{minipage}[c]{.32\linewidth}
        \centering
        \includegraphics[width=1.\linewidth]{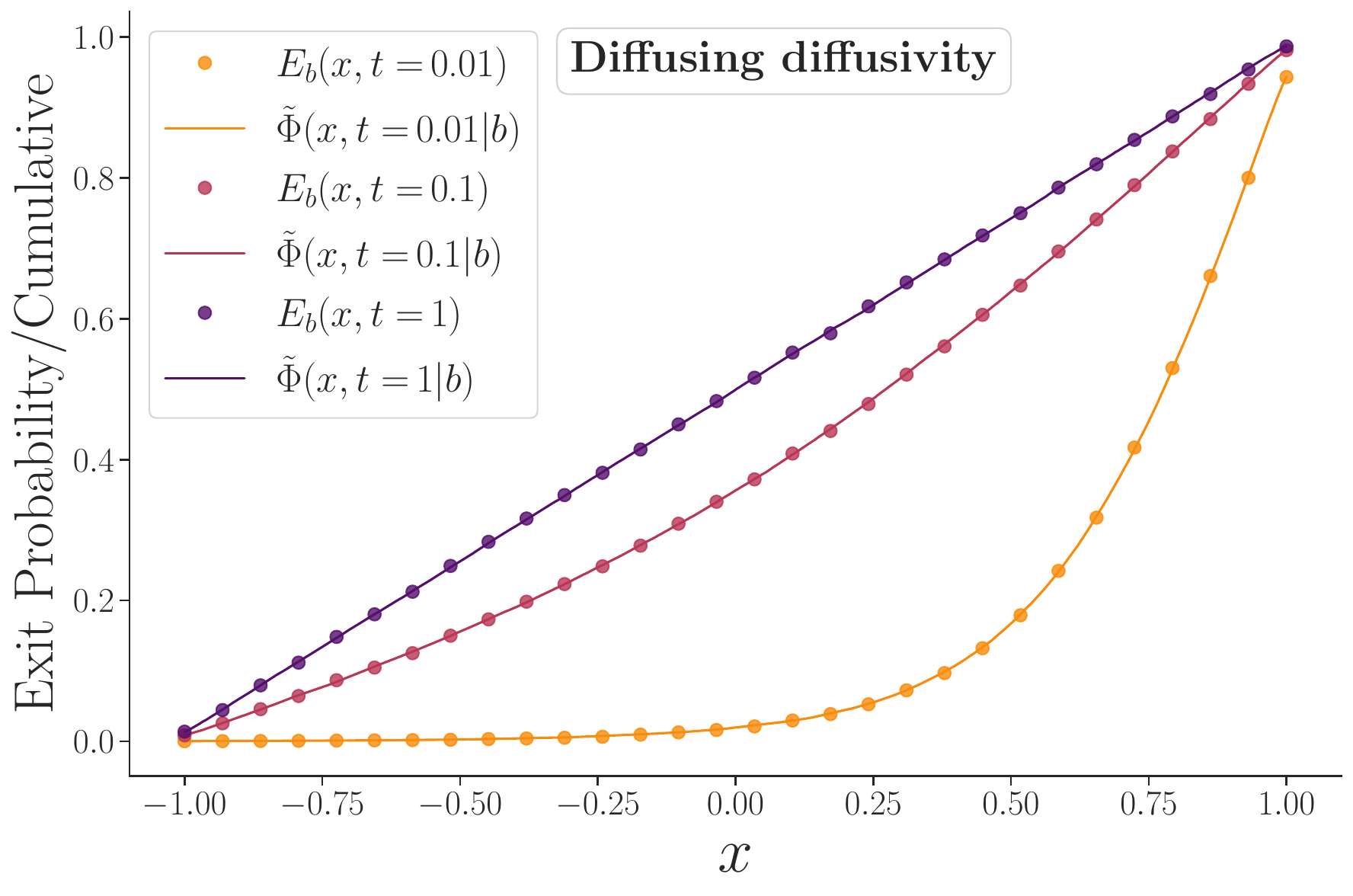}
    \end{minipage}
    \hfill%
    \begin{minipage}[c]{.32\linewidth}
        \centering
        \includegraphics[width=1.\linewidth]{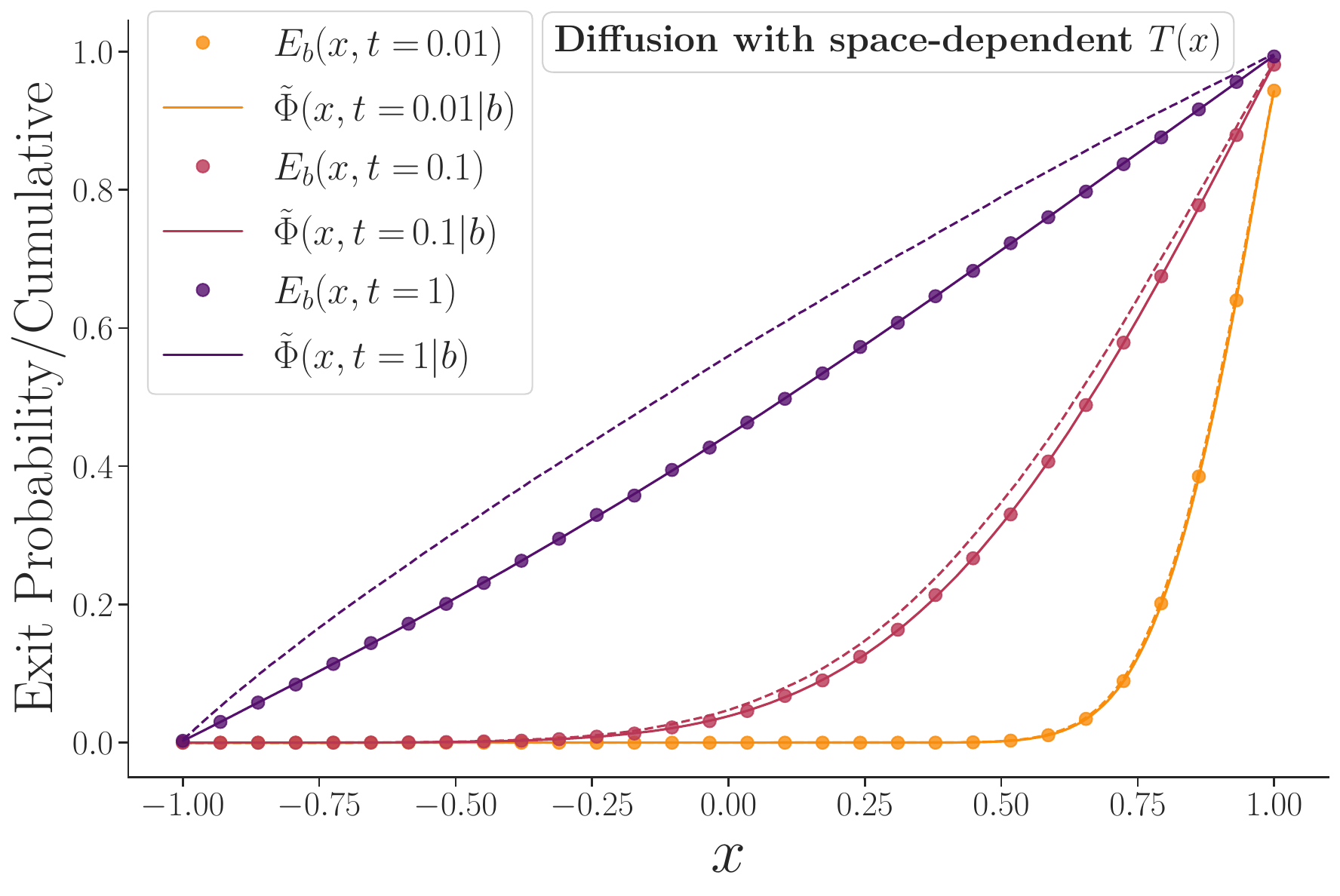}
    \end{minipage}
    \hfill%
    \begin{minipage}[c]{.32\linewidth}
        \centering
        \includegraphics[width=1.\linewidth]{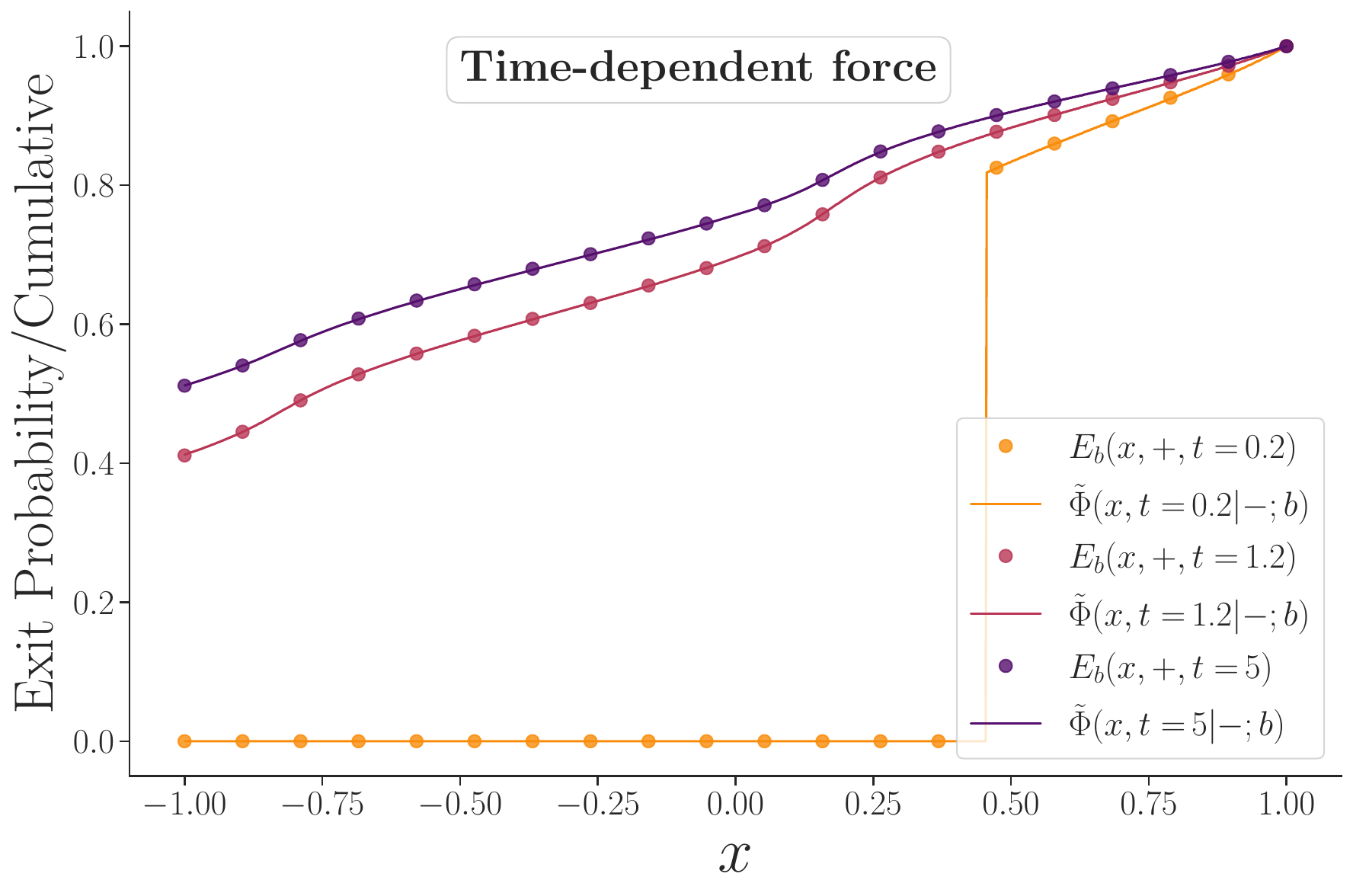}
    \end{minipage}
    \hfill%

    \caption{Here, we show numerically the validity of Eq.~\eqref{mainRelation_averaged} and \eqref{mainRelationActive} at finite time in three different cases. Dots show the value of the exit probability while the solid lines are for the cumulative distribution for the dual process. \textbf{Left}: diffusing diffusivity model with diffusion coefficient $D=4$ and relaxation time $\tau=1$. \textbf{Center}: Brownian particle with space-dependent diffusion coefficient $T(x)=1+\frac{x}{2}$ (and no external potential). To compensate the variations in temperature, the dual process is subjected to a constant force $\tilde f=\partial_x T=\frac{1}{2}$. The dashed line shows the cumulative distribution in the absence of this force. \textbf{Right}:~RTP initialized in the $+$ state, subjected to a time dependent force $f(t)=a\sin(2\pi \frac{t}{\tau})$. The dual process is in the presence of hard walls, with a force $\tilde f(t)=-a\sin(2\pi \frac{t_f-t}{\tau})$ where $t_f$ is the time at which we measure the exit probability. The cumulative distribution is conditioned on $\sigma(t)=-1$.  Here $a=1$, $\tau=0.5$, $v_0=2$ and $\gamma=1$.}
\label{FiniteTimeFig2}
\end{figure}

We now consider a family of models which is not connected to active particles, but which has also attracted a lot of attention in recent year{\red s}: diffusing diffusivity models {\red \cite{DiffDiffChubynsky, DiffDiffChechkin, DiffDiffJain, DiffDiffFPTSposini}}. One example is a Brownian particle where the diffusion coefficient is itself the square norm of an Ornstein-Uhlenbeck process in arbitrary dimension $d$,
\begin{equation}\label{defDiffDiff}
\frac{dx}{dt} = F(x) +  \sqrt{2T(t)} \ \xi(t) \quad , \quad T(t)=\bm{\theta}^2(t) \quad , \quad \tau \frac{d\bm{\theta}}{dt} = -\bm{\theta}(t) + \sqrt{2 D} \ \bm{\eta}(t) \;,
\end{equation}
where $F(x)$ is the external force, $D$ is a {\red diffusion coefficient}, and $\xi$ and $\eta$ are Gaussian white noises. For the simulations, we restrict ourselves to the simplest case $d=1$, although the results are valid for any $d$, as we have shown in Sec.~\ref{ContinuousProof}. For $d=1$, the equilibrium distribution of $\theta$ is the same as for $v$ in the AOUP case, i.e. \eqref{eqAOUP}. In the left panel of Fig.~\ref{FiniteTimeFig2} we confirm the validity of the relation (\ref{mainRelation_averaged}).

\subsubsection{Multiplicative noise}
In the central panel of Fig.~\ref{FiniteTimeFig2} we illustrate the effect of a multiplicative noise on the duality relation 
with a simple example. We consider a Brownian particle, with no external force, but with a space-dependent diffusion coefficient $T(x)=1+\frac{x}{2}$. For the dual, this means that we need to add a constant force $\tilde f=\partial_x T=\frac{1}{2}$. In the infinite time limit, the exit probability is $E_b^{st}(x)=\frac{x-a}{b-a}=\frac{1}{2}(x+1)$ (when $a=-b=1$ as in Fig.~\ref{FiniteTimeFig2}), exactly like for a constant $T$. This is true for any $T(x)$ as long as there is no external force, since the stationary equation is $T(x)\partial_x^2 E_b^{st}=0$, leading to a linear solution independent of $T(x)$. On the other hand, the equilibrium probability density with hard walls is the solution of $0=-\tilde f(x) P_{\rm eq} + \partial_x[T(x)P_{\rm eq}]$. In the absence of $\tilde f$, the solution is $P_{\rm eq}(x)=\frac{1}{T(x)}/\int_a^b \frac{1}{T(x)}$, i.e. the particle density is higher where the diffusion coefficient is smaller. For $\tilde f=\partial_x T$, the equation simplifies to $\partial_x P_{\rm eq}=0$, leading to a constant density compatible with \eqref{mainRelationStationary}.

\subsubsection{Time-dependent force} \label{sec:timedepf}
We end this section with an example of a time-dependent external force. We consider a RTP subjected to a force $f(t)=a\sin(2\pi \frac{t}{\tau})$. The duality relation \eqref{mainRelationActive} still applies, but now the dual process is subjected to a force $\tilde f(t)=-a\sin(2\pi \frac{t_f-t}{\tau})$ which depends on the time $t_f$ at which the exit probability is measured. In terms of numerical simulations, this means that if we want to obtain the exit probability at different times from the cumulative of the dual process, we need to run a different simulation of the dual for each time. Taking this into account, the left panel of Fig.~\ref{FiniteTimeFig2} shows that \eqref{mainRelationActive} holds once again.

\section{Duality for discrete time random walks} \label{discrete_case}

In this section we show the main result Eq.~(\ref{mainRelation}) for discrete time random walks. The route taken here is completely different from the previous section, and provides a more pictorial way to understand the duality relation. {\red It relies on a one-to-one mapping between the trajectories of the process and its dual. This mapping can be understood as a form of time-reversal symmetry, where the boundaries play a non-trivial role.} We then extend the derivation to continuous time random walks. We also explain briefly how the result for continuous stochastic processes derived in the previous section can be recovered as a limit of the discrete time results, although in a slightly less general form. Finally we show how this duality applies to a particular model of discrete time random walk connected to active particles: the persistent random walker.

\subsection{Derivation of the {\red duality} relation {\red in discrete time from a mapping between trajectories}}
\label{DiscreteProof}
{\red {\it General setting.}} Consider a discrete-time random walk whose position at time $n$ is denoted  $X_n$. The walker starts at position $X_0$ inside the interval $[a,b]$ and there are absorbing boundary conditions at $x=a^-$, and $x=b$. Here, $a^-$ should be understood as $a-\epsilon$ with $\epsilon \to 0$, i.e. we consider $x=a^-$ as a point which is distinct from $x=a$, with $a^-<a$, but any negative jump starting from $x=a$ will lead to $x < a^-$. {\red The reason for this will be clarified below.} For convenience we define the notation
\begin{eqnarray}
    && (x)_{[a,b]} = \begin{cases} a \quad {\rm if}\  x \leq a \\
    x \quad {\rm if} \ a < x < b \\
    b \quad {\rm if} \ x \geq b \end{cases}\, .
\end{eqnarray}
The dynamics is as follows,
\begin{eqnarray}
    && X_{n} = \begin{cases} (X_{n-1} + W_{n})_{[a^-,b]} \quad {\rm if} \ X_n \in ]a^-,b[ \\
    X_{n-1} \quad {\rm if} \ X_{n-1} = a^- \ {\rm or} \ b \end{cases}\, ,
\label{defXt}
\end{eqnarray}
where $W_n$ is a stationary stochastic process with stationary distribution $p_{st}(w)$, which satisfies the following time reversal property:
\begin{equation}
    p_{st}(w_1) P(W_2=w_2,...,W_T=w_T|W_1=w_1) = p_{st}(w_T) P(W_2=w_{T-1},...,W_T=w_1|W_1=w_T) \label{DBstatement1} \, .
\end{equation}
The process $W_n$ can take either discrete or continuous values, but its evolution should not depend on the position $X_n$ (note that on this last point, the discrete case is more restrictive than the continuous one, since in the previous section $f(x,\bm{\theta})$ and $\mathcal{T}(x,\bm{\theta})$ could depend on the position $x$). Note that a realisation of the process $\{X_n,W_n\}=(X_0,...,X_T,W_1,...,W_T)$ is completely determined by the initial position $X_0$ and the set of $W_n$'s.

This framework is very general and can be used to describe a large variety of processes, including discrete versions of active particle models, such as the persistent random walk discussed in Sec. \ref{sec:PRW}. Let us note that, in the case where the distribution of jumps $W_n$ is continuous, the absorbing wall can be placed at $x=a$ instead of $a^-$, since the probability that the process reaches exactly $x=a$ at some time $n$ is zero. On the other hand, if one considers a lattice model where the position of the particle can take only integer values $a, a+1, ..., b$, then placing the boundary condition at $a^-$ instead of $a$ amounts to adding a state $a-1$, the absorbing states of $X_n$ being $a-1$ and $b$ (as in Sec.~\ref{sec:PRW}).

A particular case which is closer to the hypotheses we have used for continuous stochastic processes is when $W_n=W(\bm{\Theta}_n)$, where $W(\bm{\Theta})$ is an arbitrary function from $\mathbb{R}^d$ to $\mathbb{R}$ and $\bm{\Theta}_n$ is a Markov process on $\mathbb{R}^d$ with transition probability $\pi(\bm{\Theta}_n|\bm{\Theta}_{n-1})$. In addition we require that $\bm{\Theta}_n$ admits an equilibrium distribution $p_{eq}(\bm{\Theta})$, which satisfies the detailed balance condition
\begin{equation}
\pi(\bm{\Theta}_n|\bm{\Theta}_{n-1}) p_{eq}(\bm{\Theta}_{n-1}) = \pi(\bm{\Theta}_{n-1}|\bm{\Theta}_n) p_{eq}(\bm{\Theta}_n)\; .
\label{detailed_balance}
\end{equation}
Applying \eqref{detailed_balance} recursively straightforwardly leads to \eqref{DBstatement1}. Note that in the following we will condition the exit probability on $W_1$, but we might as well condition it on ${\bm\Theta}_1$ in order to obtain a result closer to the continuous case. More details on the connection with continuous stochastic processes will be given in Sec. \ref{sec:DtoC}.

\begin{figure}
  \centering
    \includegraphics[width=0.49\textwidth]{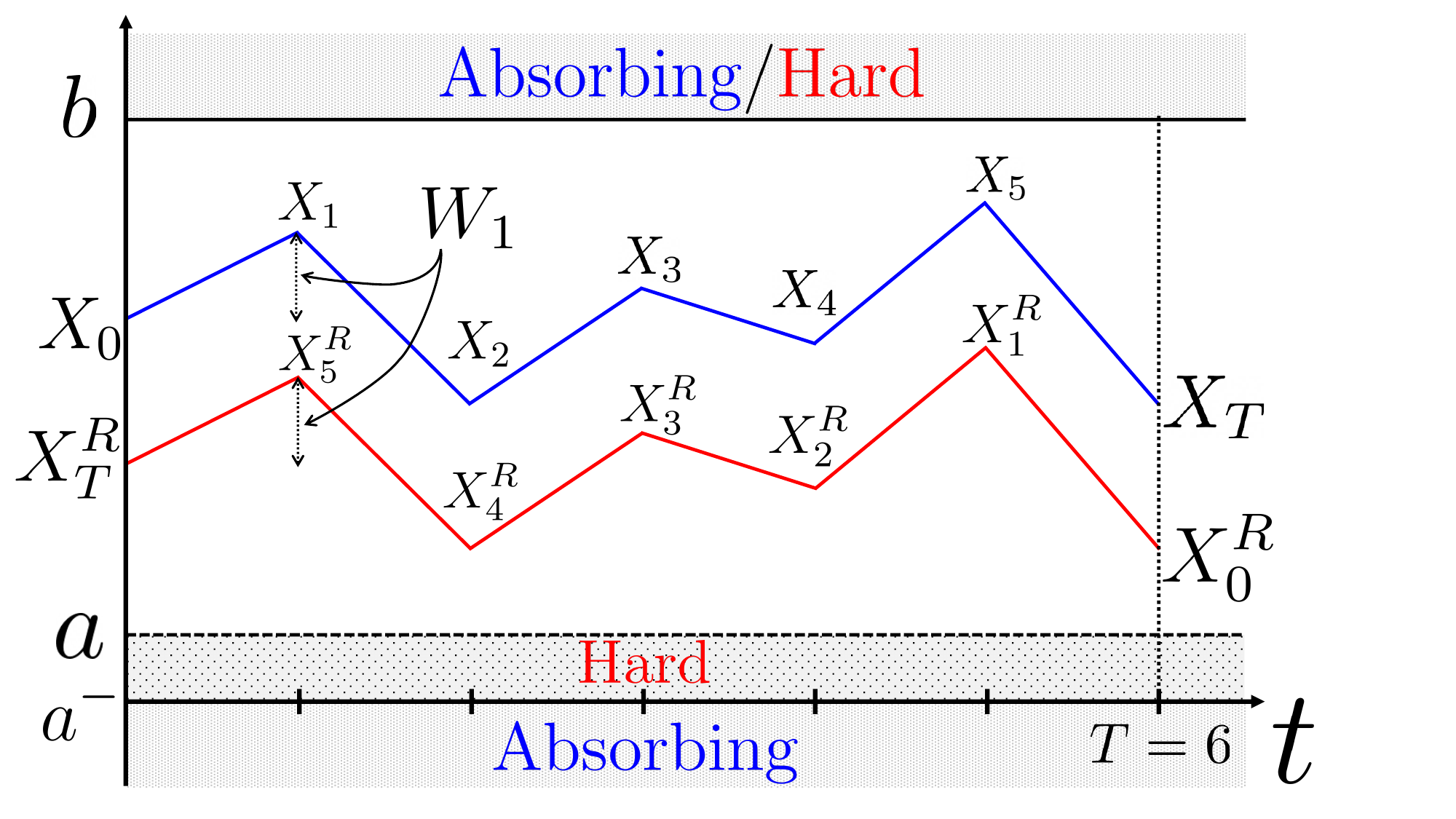}
    \label{fig:subfig1_process}
    \includegraphics[width=0.49\textwidth]{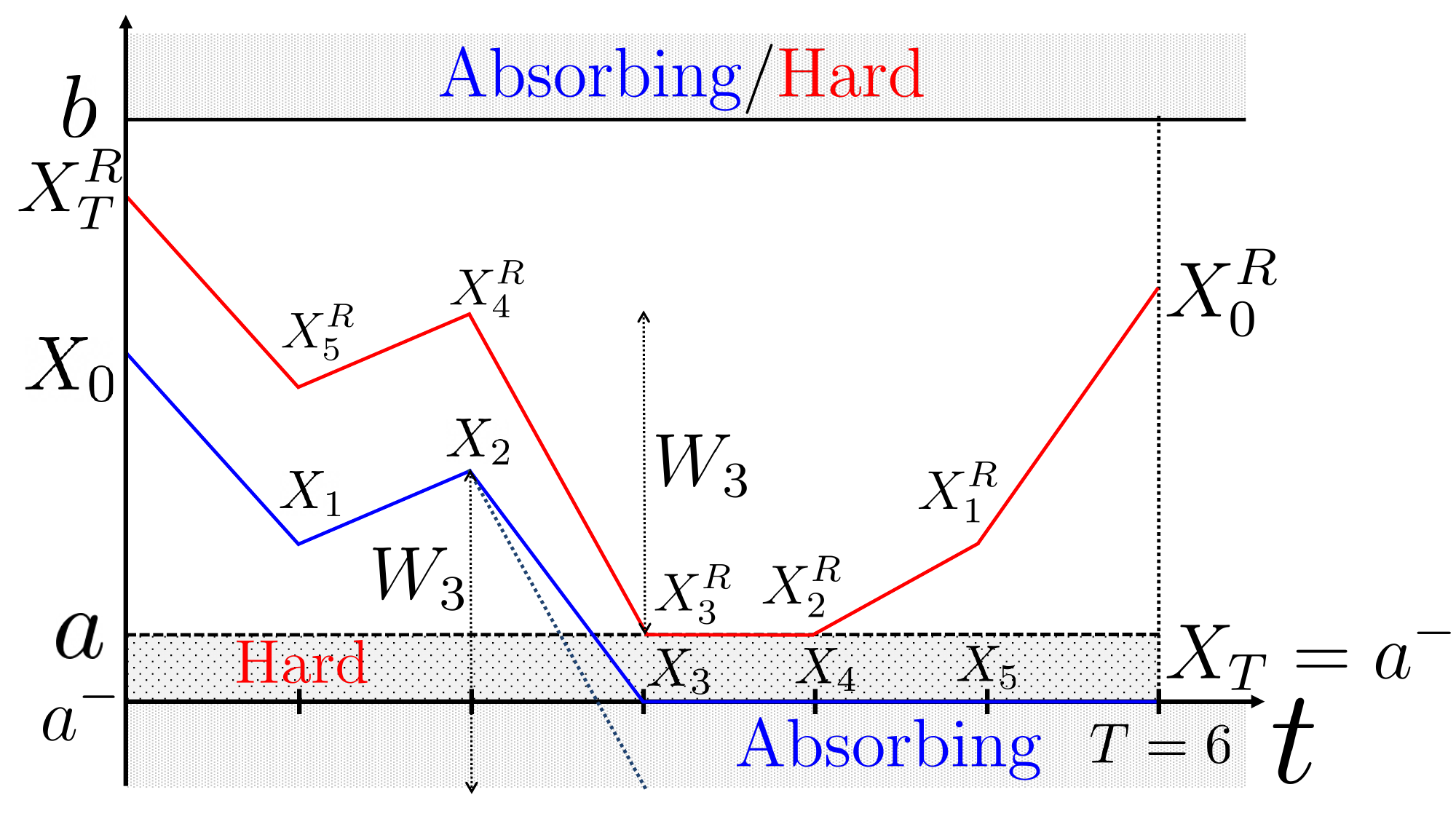}
    \label{fig:subfig2_process}
  \caption{We show in blue two examples of trajectories of a process $X_n$ (defined in Eq. (\ref{defXt})), and in red we show $\hat{X}^R_n= X^R_{T-n}$, where $X_n^R$ is the dual trajectory. 
  \textbf{Left}: We illustrate the fact that when  $X_n$ never reaches a wall, $\hat{X}_n^R$ is parallel to $X_n$ and never crosses it. \textbf{Right}: $X_n$ reaches a wall at time $t_0=3$ and is absorbed at $x=a^-$. On the other hand, for $\hat{X}^R_n$, $a$ and $b$ are hard walls and it can still move inside $[a,b]$. In both cases, the equivalence (\ref{equivalence}) holds.}
  \label{fig:two_subfigs_process}
\end{figure}

Here, we study the process \eqref{defXt} on a finite time interval $\llbracket 0,T \rrbracket$. On this interval we define the {\it dual process} of $X_n$, which we denote $Y_n$. It is initialised at some value $Y_0 \in[a,b]$ and then follows the dynamics
\begin{equation}
Y_n =  (Y_{n-1} - \tilde W_{n})_{[a,b]} \;,
\label{defdual}
\end{equation}
where the $\tilde W_n$'s follow the same stochastic dynamics as the $W_n$'s. This can be seen as a time-reversed version of {\red the process} $X_n$ but with an arbitrary initial position $Y_0$ and with hard walls at $a$ and $b$ instead of absorbing ones (as shown in Fig. \ref{fig:two_subfigs_process}). Note that the left wall is at $x=a$ and not $a^-$ in the dual.

{\red In addition, for a fixed value of $X_0^R\in [a,b]$, we define the {\it dual trajectory} (or time-reversed trajectory) $X^R_n$ of a given realisation of $X_n$ with jump sequence $(W_1,W_2,\ldots,W_T)$, as the trajectory which starts at the initial position $X_0^R$, and whose jump sequence is $(-W_T,-W_{T-1},\ldots,-W_0)$. Such a trajectory can be seen as a particular realisation of $Y_n$ defined in (\ref{defdual}), which satisfies $Y_0=X_0^R$ and for all $n$, $\tilde W_n = W_{T+1-n}$.} We also define $\hat{X}^R_n=X^R_{T-n}${\red , which is easier to interpret visually.} Examples of trajectories along with their dual trajectories are shown in Fig.~\ref{fig:two_subfigs_process}.

As for the continuous case, we are interested in making a connection between the exit probability at $b$ for the process $X_n$,
\begin{equation}
    E_b(x,w,T) = \mathbb{P}(X_T=b|X_0=x, W_1=w)\, , 
    \label{exit_def_discrete}
\end{equation}
and the conditional cumulative of its dual $Y_n$, where $\tilde W_1$ is drawn from $p_{st}(w)$
\begin{equation}
    \tilde{\Phi}(x,T|w;y) = \mathbb{\tilde P}(Y_n\leq x|\tilde W_T=w, Y_0=y,\tilde W_1^{eq}) \equiv \int dw \ p_{st}(w_1) \mathbb{\tilde P}(Y_T\leq x|\tilde W_T=w, Y_0=y,\tilde W_1=w_1)\; .
    \label{cumul_def}
\end{equation}
Here and in the whole section, the tilde means that we refer to the distribution of the dual $Y_t$. Note that $E_b(x,w,T)$ is conditioned on the next step {\em after} time $n=0$ (i.e. the jump that occurs between $X_0$ and $X_1$), while the cumulative is conditioned on the last step {\em before} time $T$ (i.e. between $Y_{T-1}$ and $Y_T$).
\\

{\red {\it Sketch of the proof.} The proof of the relation \eqref{mainRelation} for discrete random walks relies on a mapping between the trajectories of the process $X_n$ and its dual $Y_n$, through the notion of dual trajectory $X_n^R$.} We will first show that, under the hypotheses defined above, the following equivalence holds:
\begin{equation}
    X_0 \geq X^R_T \Leftrightarrow X_T \geq X^R_0 \, .
    \label{equivalence}
\end{equation}
{\red We will then use the time reversal property \eqref{DBstatement1}, and integrate it over all the realisations where the events in \eqref{equivalence} are realised (seeing $X^R_T$ as a realisation of $Y_n$). Using the definition of conditional probabilities, this will lead us to an intermediate result \eqref{Siegmund_precise}, from which both the relation for the exit probability \eqref{mainRelation} for the present setting, and the Siegmund duality \eqref{Siegmund} can be derived.}

{\red Before proving the equivalence \eqref{equivalence}, let us justify why the absorbing wall for the process $X_n$ should be at $a^-$ instead of $a$. This is simply due to the fact that the negation of $\geq$ is $<$, and hence we need this ``shift'' of the boundary, in particular to prove the point (iii) below. Without this, it would be possible to have simultaneously $X_0=X_T^R$ and $X_T=a<X_0^R$ (if $X_n$ reaches $a$ before time $T$), hence the equivalence \eqref{equivalence} would not hold. Instead if the absorbing wall is at $a^-$, point (iii) holds, i.e. $X_T=a^-<X_0^R$ implies $X_0<X_T^R$, in agreement with \eqref{equivalence}.}
\\

{\em Proof of \eqref{equivalence}.}
We first need to prove the following 3 properties:
\vspace{0.15cm}

(i) If $X_n \in [a,b{\red[}$ and $\hat X^R_n \in {\red]}a,b{\red[}$ for all $n\in \llbracket n_1,n_2\rrbracket$, then the trajectories of $X_n$ and $\hat X^R_n$ are parallel on this time interval.

(ii) If $X_T=b$, then $X_0 \geq X^R_T$.

(iii) If $X_T=a^-$, then $X_0<X^R_T$.
\vspace{0.15cm}

The property (i) simply comes from the fact that the $W_n$'s that appear in the definition of the processes $X_n$ and $X^R_n$ are the same. Let us now assume that $X_n$ reaches $x=b$ for the first time at time $n_0$. Then one has $\hat X^R_{n_0} \leq b=X_{n_0}$. Furthermore, $X_{n_0-1} \geq X_{n_0} - W_{n_0}=b  - W_{n_0}$ (here $W_{n_0}>0$) and $\hat X^R_{n_0-1}=\hat X^R_{n_0} - W_{n_0} \leq b - W_{n_0}$. Thus $X_{n_0-1} \geq \hat X^R_{n_0-1}$, and since the two trajectories are parallel up to time $n_0-1$, one necessarily has $X_0 \geq \hat{X^R}_0=X^R_T$, which proves (ii). Similarly, let us assume that $X_n$ reaches $x=a^-$ for the first time at time $n_0$. Then one has $\hat X^R_{n_0} \geq a > a^-=X_{n_0}$. Furthermore, $X_{n_0-1} \leq X_{n_0} - W_{n_0}=a^-  - W_{n_0}$ (here $W_{n_0}<0$) and $\hat X^R_{n_0-1}=\hat X^R_{n_0} - W_{n_0} \geq a - W_{n_0}$. Thus $X_{n_0-1}<\hat X^R_{n_0-1}$, and since the two trajectories are parallel up to time $n_0-1$, one necessarily has $X_0 < \hat{X}^R_0=X^R_T$, which completes the proof of (iii).

To prove the equivalence \eqref{equivalence}, we need to distinguish 3 cases. In each case, we will show that either both inequalities in (\ref{equivalence}) hold or none of them holds (i.e. $X_0<X^R_T \Leftrightarrow X_T<X^R_0$). If $X_T=b$, then one necessarily has $X_T \geq \hat X^R_T=X^R_0$, and according to (ii), $X_0 \geq X^R_T$. Similarly if $X_T=a^-$, then $X_T < X^R_0${\red , and (iii) implies} $X_0<X^R_T$. Thus the equivalence is satisfied in both of these cases. The last case to consider is when $X_n \in [a,b{\red [}$ for all $n$. In this case, $\hat X^R_n$ cannot reach both $a$ and $b$ during the time interval $\llbracket 0,T \rrbracket$. Indeed, consider {\red for instance} the case where $\hat X^R_n$ leaves the wall $b$ at time $n_1$ and reaches $a$ at time $n_2$. One then has $\hat X^R_{n_1}-\hat X^R_{n_1+1} \leq X_{n_1} - X_{n_1+1}$ and the trajectories of $X_n$ and $\hat X^R_n$ are parallel on the interval $\llbracket n_1+1,n_2 \rrbracket$ {\red (property (i))}. Since $X_{n_1} < b$, this implies that $X_{n_2} < a$, which contradicts the assumption. Therefore there are only 3 possible sub-cases: either $\hat X^R_n$ never reaches a wall, in which case {\red (i) implies that} the trajectories of $X_n$ and $X^R_n$ are parallel on $\llbracket 0,T \rrbracket$ and the equivalence holds, or $\hat X^R_n$ reaches only one of the two walls. By a similar reasoning to what we have done to show (ii) and (iii), one shows that if this wall is $a$, one has $X_0\geq X^R_T$ and $X_T \geq X^R_0$, while if it is $b$, $X_0<X^R_T$ and $X_T<X^R_0$. Thus the equivalence \eqref{equivalence} again holds in this case, which completes its proof. 

Note that this reasoning actually gives us a stronger result, namely for any $n_1, n_2 \in \llbracket 0,T \rrbracket$,
\begin{equation}
    X_{n_1} \geq X^R_{T-n_1} \Leftrightarrow X_{n_2} \geq X^R_{T-n_2} \, .
    \label{equivgeneral}
\end{equation}
We will use the equivalence \eqref{equivgeneral} in Sec. \ref{sec:resetting} when we will include stochastic resetting.
\\

{\red{\it Proof of the duality relation.}} Let us now fix $X_0=x \in [a,b]$ and $Y_0=y \in [a,b]$. We choose a given realisation of $X_n$ and we fix $Y_n=X_n^R$, such that for all $n$, $W_n=\tilde W_{T+1-n}=w_n$. Keeping $w_1$ and $w_T$ fixed and integrating Eq. (\ref{DBstatement1}) over all values of $w_2, ..., w_{T-1}$ such that the events in \eqref{equivalence} are realized (using that $W_n$ and $\tilde W_n$ have the same law), we obtain
\bea
\int dw_2 ... dw_{T-1} \, \mathbb{1}_{X_T\geq Y_0} \, p_{st}(w_1) P(W_2=w_2,...,W_T=w_T|W_1=w_1) \quad \quad \quad \quad \nn \\
\quad \quad \quad \quad = \int dw_2 ... dw_{T-1} \, \mathbb{1}_{X_0\geq Y_T} \, p_{st}(w_T) \tilde P(\tilde W_2=w_2,...,\tilde W_T=w_T|\tilde W_1=w_1)\, .
\eea
Performing the integrals leads to
\begin{equation}
p_{st}(w_1) \ P(X_T \geq y, W_T=w_T|X_0=x, W_1=w_1)=
    p_{st}(w_T) \ \tilde{P}(Y_T \leq x, \tilde W_T=w_1|Y_0=y, \tilde W_1=w_T)  \; .
\label{firststepproof}
\end{equation}
$P$ and $\tilde P$ refer to the density of the $W$'s and $\tilde W$'s while $\mathbb{P}$ and $\mathbb{\tilde P}$ refer to the probability of the event $X_T \geq y$ and $Y_T \leq x$ respectively. Integrating over $w_T$, fixing $w_1=w$ and dividing by $p_{st}(w)$, we obtain
\begin{equation}
\mathbb{P}(X_T \geq y|X_0=x, W_1=w) = \frac{\tilde{P}(Y_T \leq x, \tilde W_T=w|Y_0=y, \tilde W_1^{st})}{p_{st}(w)}
\label{generalidentityconditionalNEW}
\end{equation}
where we recall that the conditioning on $\tilde W_1^{st}$ means that $\tilde W_1$ is drawn from the distribution $p_{st}(\tilde w_1)$, i.e.
\begin{eqnarray}
   \tilde{P}(Y_T \leq x, \tilde W_T=w|Y_0=y, \tilde W_1^{st}) = \int d\tilde w_1 \ p_{st}(\tilde w_1) \tilde{P}(Y_T \leq x, \tilde W_T=w|Y_0=y, \tilde W_1=\tilde w_1)\; .
\end{eqnarray}
Since the process $\tilde W_n$ is initialized in its stationary distribution, and since its evolution is independent of the position, it will remain stationary at all times. Thus the probability density of $\tilde W_T$ is  $p_{st}(w)$. Therefore the right-hand side of Eq. (\ref{generalidentityconditionalNEW}) can be rewritten using the definition of conditional probabilities
\begin{equation}
\mathbb{P}(X_T \geq y|X_0=x, W_1=w) = \tilde{\mathbb{P}}(Y_T \leq x|\tilde W_T=w, Y_0=y, \tilde W_1^{st}) \; .
\label{Siegmund_precise}
\end{equation} 
This is the most general form of the result presented in this paper in the case of discrete time random walks, and it is valid at any time $T\geq 0$. {\red It is the discrete time equivalent of the identity \eqref{identity_general_app1} derived in App.~\ref{ContinuousProof_duality}.} It can be made symmetric by multiplying both sides by $p_{st}(w)$ and integrating over $w$, leading to 
\begin{equation}
\mathbb{P}(X_T \geq y|X_0=x,W_1^{st}) = \tilde{\mathbb{P}}(Y_T \leq x|Y_0=y, \tilde W_1^{st}) \; .
\label{Siegmund_symmetric}
\end{equation}
which is an instance of a Siegmund duality.
Specializing \eqref{Siegmund_precise} to $y=b$, we obtain the desired relation
\begin{equation}
E_b(x,w,T) = \tilde{\Phi}(x,T|w;b) \; .
\label{ExitRW_precise}
\end{equation}

Let us now make a few important remarks. First, let us mention that the reasoning above also holds if one places the absorbing walls at $a$ and $b^+$, reverts all the inequalities and exchanges $a$ and $b$. In this case \eqref{Siegmund_precise} becomes
\begin{equation}
E_a(x,w,T) = \mathbb{P}(Y_t\geq x|\tilde W_T=w, Y_0=a,\tilde W_1^{st}) \;.
\end{equation}
Second, if we instead take $y=b$ in Eq.~\eqref{Siegmund_symmetric}, we obtain the integrated version of \eqref{ExitRW_precise},
\begin{eqnarray}
    E_b(x,T) &=& \tilde{\Phi}(x,T|b)\, , \\
    {\rm with}\, \,  \quad E_b(x,t) &=& \int dw \ p_{st}(w) \mathbb{P}(X_t=b|X_0=x, W_1=w) \, ,\\
   {\rm and} \quad \tilde{\Phi}(x,T|b) &=& \tilde{\mathbb{P}}(Y_T \leq x|Y_0=b, \tilde W_1^{st}) = \int d\tilde w_1 \ p_{st}(\tilde w_1) \tilde{P}(Y_T \leq x|Y_0=y, \tilde W_1=\tilde w_1) \, .
\label{Siegmund_averaged}
\end{eqnarray}
This relation is simpler to use since one does not need to condition the cumulative distribution of $Y_n$ on the value of the final jump.

\subsection{Continuous time random walks} \label{sec:CTRW}

\begin{figure}
  \centering
    \includegraphics[width=0.6\textwidth]{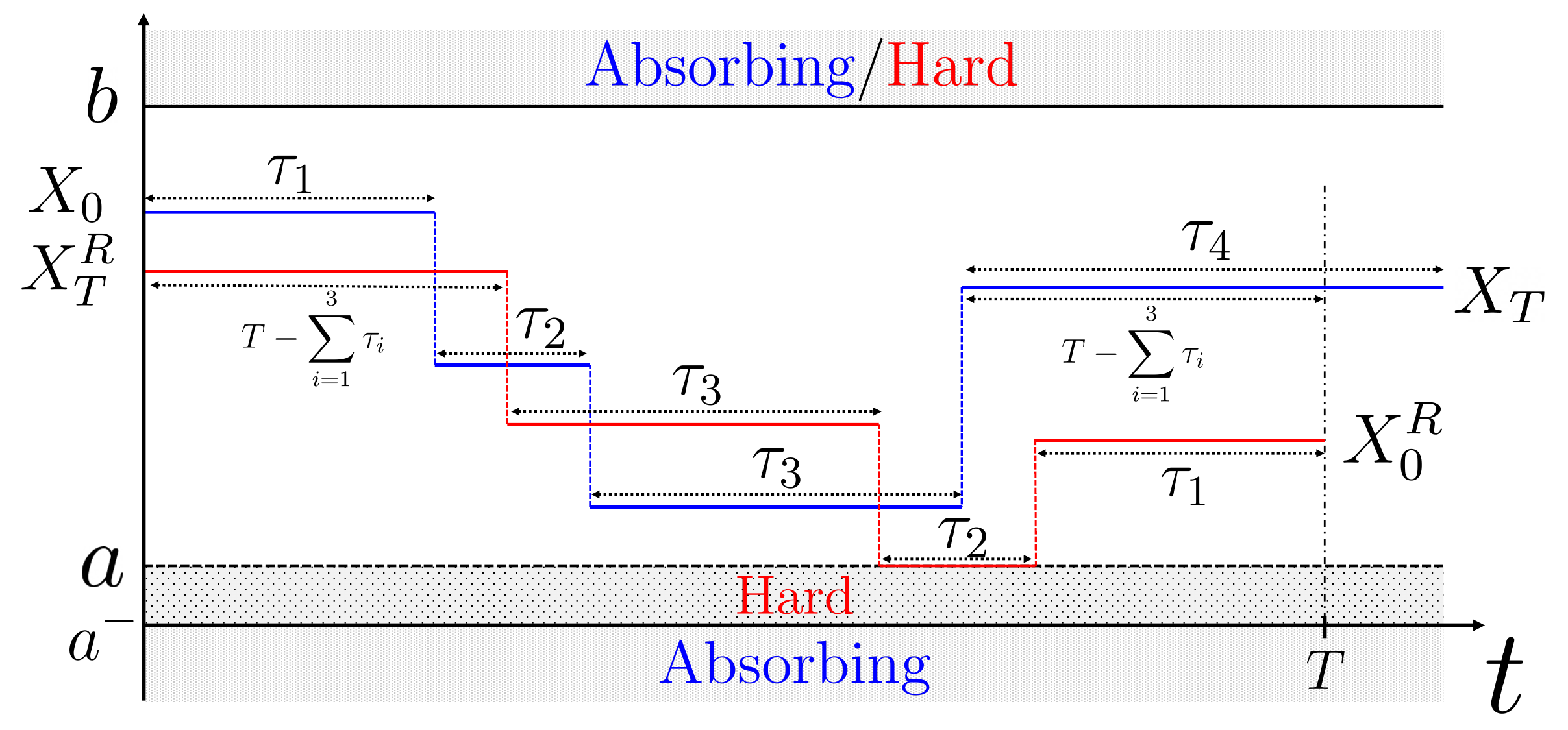}
  \caption{We show in blue a trajectory of a continuous time random walk $X_t$ starting at $X_0$ and with last position $X_T$. The jumps are separated by time intervals $\tau_i$ with distribution $w(\tau_i)$. In red, we show $\hat{X}^R_t= X^R_{T-t}$ where $X^R_t$ is the dual trajectory of $X_t$ with initial position $X_0^R$. The process $X^R_t$ jumps at the same times as $X_t$ but performs the jumps in the reversed order as illustrated. In this example, it is clear that the equivalence (\ref{equivalence}) holds.}
  \label{fig:CTRWFig}
\end{figure}

The reasoning above can be generalized to continuous-time random walks (CTRW) \cite{CTRW1,CTRW2, CTRWA, CTRWB, CTRWC}. Time is now a continuous variable, and the random walk $X_t$ performs jumps at random times. The time intervals $\tau_1, \tau_2 ...$ between successive jumps are drawn from an arbitrary probability distribution $\omega(\tau)$ until $\tau_1 + ... + \tau_N<T$ and $\tau_1 + ... + \tau_{N+1} \geq T$. Denoting $t_i=\sum_{j=1}^{i} \tau_j$ (with $t_0=0$) we then define
\begin{eqnarray}
    && X_{t_{i}} = \begin{cases} (X_{t_{i-1}} + W_{i})_{[a^-,b]} \quad {\rm if} \ X_{t_{i-1}} \in ]a^-,b[ \\
    X_{t_{i-1}} \quad {\rm if} \ X_{t_{i-1}} = a^- \ {\rm or} \ b \end{cases}\, ,
\label{defXtCTRW}
\end{eqnarray}
where the $W_i$ satisfy the same hypotheses as in Sec. \ref{DiscreteProof}, and $X_t$ remains fixed between $t_{i-1}$ and $t_i$. We now define the dual as
\begin{equation}
Y_{\tilde t_{i}} =  (Y_{\tilde t_{i-1}} - \tilde W_{i})_{[a,b]} \;.
\label{defdualCTRW}
\end{equation}
where the $\tilde W_i$'s and the $\tilde t_i$'s follow the same laws as the $W_i$'s and the $t_i$'s. The dual trajectory of $X_t$ (given $X_0^R=Y_0$) is
\begin{equation}
X^R_{t_{i}} =  (X^R_{t_{i-1}} - W_{n+1-i})_{[a,b]} \;.
\label{defdualCTRW}
\end{equation}
An example of a trajectory along with its dual is shown in Fig.~\ref{fig:CTRWFig}. The important point to notice is that, in the reasoning of Sec. \ref{DiscreteProof}, only the total number $N$ of jumps in the time interval $[0,T]$ and the values of the $W_i$'s are relevant. Although the times of the jumps are now different between the trajectories $X_t$ and $\hat X^R_t =X^R_{T-t}$, they will perform exactly the same jumps $W_i$ in the same order, and thus the equivalence \eqref{equivalence} still holds. Since the $W_i$'s have the same properties as before, the rest of the derivation naturally follows. In the left panel of Figure \ref{PRWfig}, we illustrate the duality for a CTRW with Gaussian jumps, and a Cauchy distribution for the time intervals.

\subsection{Recovering the continuous case} \label{sec:DtoC}

We will now discuss the connection between the result \eqref{ExitRW_precise} derived above for discrete random walks and the one obtained in Sec. \ref{ContinuousProof} for continuous stochastic processes. Consider a discrete random walk defined as above, where the jumps take the form $W_n=W(\bm{\Theta}_n)$ with $\bm{\Theta}_n=(\bm{\bar\theta}_n,\bar{\xi}_n)$, and a function $W(\bm{\Theta})$ such that
\begin{eqnarray}
    X_{n} = X_{n-1} + f(\bm{\bar\theta}_n) \Delta t + \bar{\xi}_n \sqrt{2\mathcal{T}(\bm{\bar\theta}_n)\Delta t} \, .
\end{eqnarray}
Here the $\bar{\xi}_n$'s are i.i.d. random variables with distribution $\mu(\bar \xi)$ (with mean zero and variance 1), and $\bm{\bar\theta}_n$ is a vector of parameters which is Markovian with jump distribution $\bar \pi(\bm{\bar\theta}_{n}|\bm{\bar\theta}_{n-1})$, independent of the position, which satisfies detailed balance. 

We then introduce the variable $t=n\Delta t$ and the process $x(t)$ defined as
\be
x(t=n\Delta t)=X_n \;,
\ee
as well as ${\bm\theta}_t={\bm\bar\theta}_{n}$ and ${\xi}_t=\bar{\xi}_n$.
By choosing the correct form for $\bar \pi(\bm{\bar\theta}_{n}|\bm{\bar\theta}_{n-1})$, this process will converge in the limit $\Delta t \to 0$ to \eqref{LangevinIntroduction}-\eqref{SDEtheta_Introduction}, with the important restriction that $f$ and $\mathcal{T}$ cannot depend on $x$ in this case. In addition, this process satisfies the hypotheses of Sec. \ref{DiscreteProof} for any value of $\Delta t$. Thus the results above apply for any $\Delta t$, however small. Furthermore, since in the continuous time limit there is zero probability that $x(t)$ reaches exactly $a$ without crossing the line $x=a$, one may place the absorbing wall at $a$ instead of $a^-$.  Finally, since the $\bar{\xi}_n$'s are i.i.d., we only need to condition the probabilities on $\bm{\bar\theta}_n$. Thus \eqref{mainRelation} holds in this limit. 

Let us conclude this section by briefly describing what happens when the function $f$ depends on $X_n$. In this case the trajectories of $X_n$ and $\hat{X}_n^R$ are no longer parallel even away from the walls. Hence nothing prevents their trajectories from crossing, and thus the results above are invalid for any finite $\Delta t$. However, if the function $f(x,\bm{\theta})$ is smooth enough as a function of $x$, then these trajectories become more and more parallel as they get closer to each other. Therefore, although at finite $\Delta t$ the trajectories may cross with finite probability, this probability will go to zero as $\Delta t \to 0$. Hence it is the continuity of the trajectories which makes the results valid even in the presence of an $x$-dependent $f$.

\subsection{An example: the persistent random walk} \label{sec:PRW}

In this section we show how Eq.~\eqref{mainRelationStationary} applies for a persistent random walk \cite{PRWWeiss, PRWMasoliverReview, PRWSurvivalLacroixMori} on a 1d lattice with $L+2$ sites $(i=0,...,L+1)$. The motion follows the rule
\begin{equation} \label{PRWx}
    x_{n} = x_{n-1} + \sigma_n\, ,
\end{equation}
where the steps $\sigma_n = \pm 1$ follow a Markov dynamics defined by
\begin{equation}
\begin{aligned} \label{PRWsigma}
\sigma_n = \begin{cases}
\sigma_{n-1} &\text{with probability } p \\
-\sigma_{n-1} &\text{with probability } 1-p
\end{cases} \, .
\end{aligned}
\end{equation}
The parameter $p \in [0,1]$ controls the ``persistence" of the random walk. When $p=1/2$, we retrieve the well-known symmetric random walk with uncorrelated steps. For $p>\frac{1}{2}$, the steps are correlated positively, leading to persistence in the motion of the walker, while for $p<\frac{1}{2}$ the steps are negatively correlated.

The first quantity of interest is the exit probability in the presence of absorbing boundary conditions at sites $0$ and $L+1$ (i.e. if the particle reaches one of those two sites, it stays there forever),
\begin{equation} \label{PRWEdef}
    E_i^\pm(n) = \mathbb{P}(\text{particle exits at } L+1 \text{ before or at time } n| \text{ particle starts at site } i \,\& \text{ first jump is } \pm 1) \;.
\end{equation}

Since the jump distribution is symmetric (i.e. we have $q(\sigma_{n+1}=\sigma_{n}|\sigma_{n}) = q(\sigma_{n+1}=-\sigma_{n}|-\sigma_{n}) = p$), the dual process of $x_n$, that we denote $y_n$, is naturally defined by the same recursive relation (see Eq.~(\ref{defdual}))
\begin{equation} \label{PRWy}
    y_{n} = y_{n-1} + \sigma_n\, ,
\end{equation}
with $y_0=L+1$. In this case, we consider hard walls at sites $1$ and $L+1$, i.e. if the particle is on site 1 and jumps to the left it stays at 1, and similarly at the other end of the lattice (as explained in Sec. \ref{DiscreteProof}, the hard wall at $a$ for the dual is shifted by 1 compared to the absorbing wall of the original process in the case of lattice random walks). Here we are interested in the cumulative of the distribution of positions of $y_n$. Let us denote $p_i^\pm(n)$ the probability that the particle is on site $i$ at time $n$ and that the previous jump was $\pm 1$. The cumulative distribution reads
\be
\tilde \Phi_i^{\pm}(n) = \sum_{j=1}^{i} P(j,n|\pm) = 2 \sum_{j=1}^{i} p_j^\pm(n)
\ee
where $P(i,n|\pm)$ is the probability that the particle is on site $i$ at time $n$ given that it has sign $\pm$ (we indeed have $P(\pm)P(i,n|\pm) = p_i^\pm(n)$, with $P(\pm) = 1/2$, the probability that $\sigma = \pm 1$). Then, according to \eqref{ExitRW_precise}, one has
\begin{equation} \label{dualityPRW}
    E_i^\pm(n) = \tilde \Phi_i^{\mp}(n)  \, .
\end{equation}

\begin{figure}
\centering
    \begin{minipage}[c]{.32\linewidth}
        \centering
        \includegraphics[width=1.\linewidth]{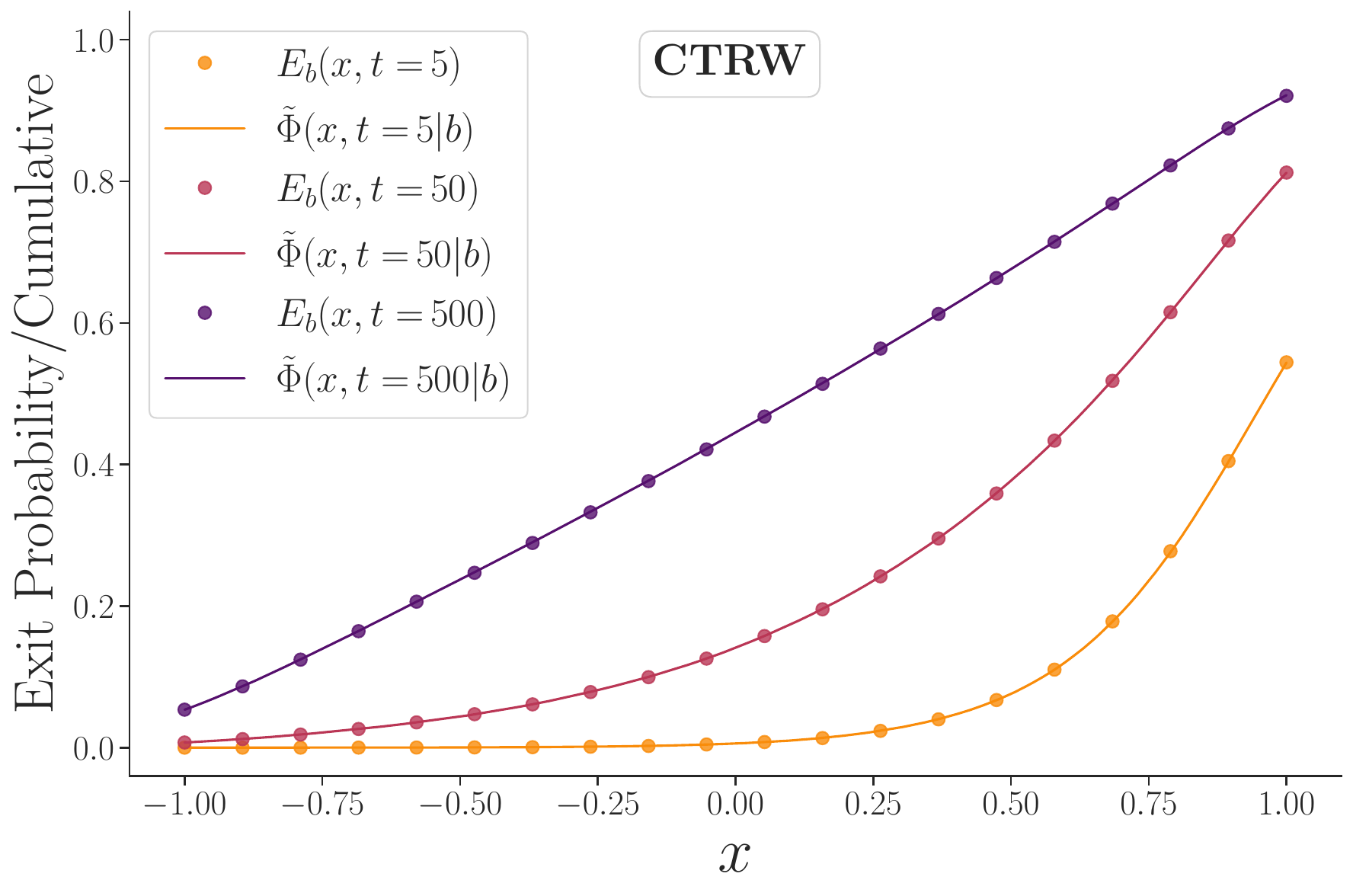}
    \end{minipage}
    \begin{minipage}[c]{.32\linewidth}
        \centering
        \includegraphics[width=1.\linewidth]{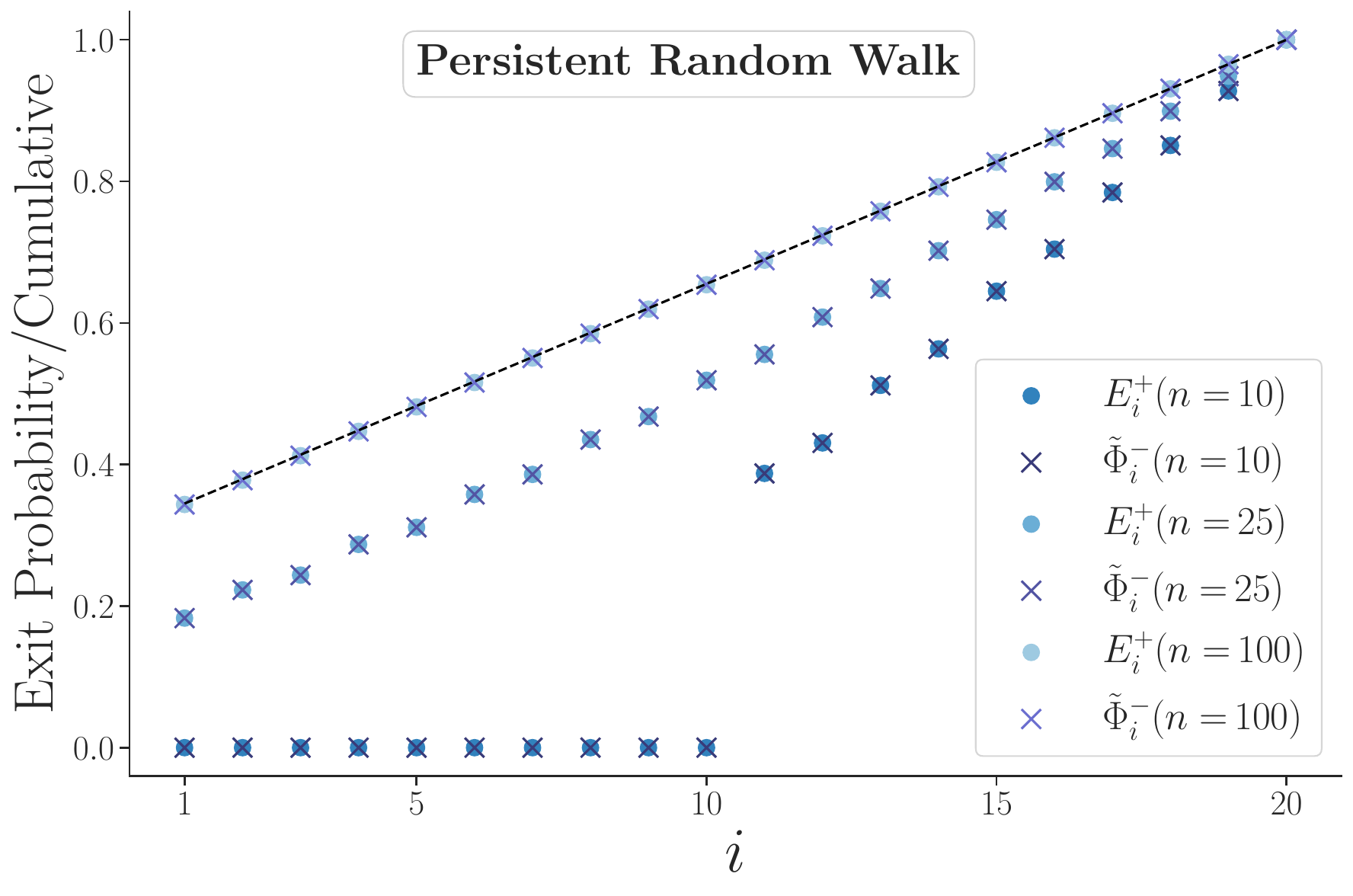}
    \end{minipage}
    \hfill%
    \begin{minipage}[c]{.32\linewidth}
        \centering
        \includegraphics[width=1.\linewidth]{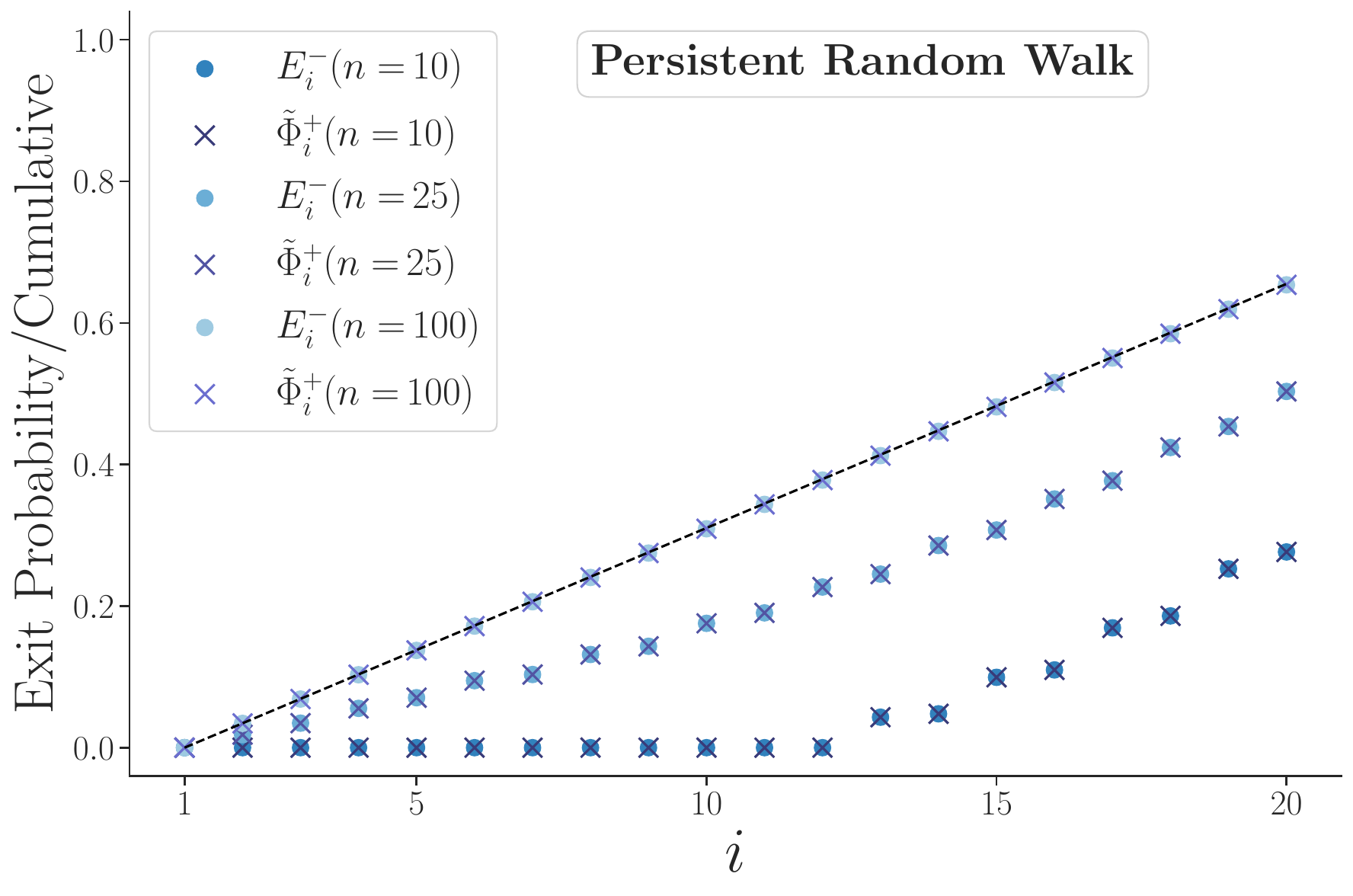}
    \end{minipage}
     \hfill
    \caption{\textbf{Left}: Numerical evaluation of the exit probability and the cumulative distribution (in the presence of hard walls) of a continuous time random walk with centered Gaussian jump distribution with variance $0.4$ and a distribution of time between jumps $p(\tau)=\frac{1}{(1+\tau)^2}$.
    \textbf{Center}:~For a persistent random walk, we compute numerically the exit probability $E_i^+(n)$ and the cumulative distribution with hard walls $\tilde \Phi_i^-(n)$ using transfer matrices, as described in App. \ref{PRW}, for $p=0.9$ and $L=20$. The dashed black line corresponds to the analytical result for the stationary state \eqref{PRWEsol} and \eqref{PRWpsol}. \textbf{Right}: Same plot for $E_i^-(n)$ and $\tilde \Phi_i^+(n)$.}
    \label{PRWfig}
\end{figure}

We have computed these two quantities at different times using transfer matrices and we show the results in Figure~\ref{PRWfig}. There is indeed a perfect overlap. The details of the numerical computations are presented in App.~\ref{PRW}. There we also compute both quantities in the stationary state. The stationary solution for the exit probability is (with $E_i^{\pm}=E_i^{\pm}(n\to+\infty)$ and similarly for $p_i^{\pm}$ below)
\begin{eqnarray} \label{PRWEsol}
    && E_i^+ = \frac{1-p}{(1-p)L+p} (i-1) + \frac{1}{(1-p)L+p} \;, \nonumber\\
    && E_i^- = \frac{1-p}{(1-p)L+p} (i-1) \;,
\end{eqnarray}
for $1 \leq i \leq L$. The stationary distribution of positions of the dual process $y_n$ is given by
\begin{eqnarray} \label{PRWpsol}
    && p_i^+ = p_i^- = \frac{1}{2} \frac{1-p}{(1-p)L+p} \quad {\rm for} \quad  2 \leq i \leq L \, , \nonumber\\
    && p_1^- = p_{L+1}^+ = \frac{1}{2} \frac{1}{(1-p)L+p} \, , \nonumber\\
    && p_1^+ = p_{L+1}^- = 0 \, ,
\end{eqnarray}
and one can indeed verify that \eqref{dualityPRW} holds in the stationary state.

\section{Duality for stochastic resetting} \label{sec:resetting}

A particular class of stochastic processes has generated a lot of interest in recent years: stochastic processes under resetting \cite{resettingPRL,resettingReview, resettingBriefReview}. The resetting procedure is the following: a random process $x(t)$ starts its dynamics at $x(0)$ and evolves, e.g. through the Langevin dynamics \eqref{LangevinIntroduction}, but at random times, it is reset at a given position $X_r$ where it restarts its dynamics. A celebrated example is the resetting Brownian motion (rBm) where the dynamics is a simple Brownian motion, and the random times are exponentially distributed. One can also consider a resetting RTP as in \cite{resettingRTP, resettingRTP2}, or any dynamics studied in the above sections. These systems are convenient to study analytically using for instance renewal theory and are also studied in experiments where periodic resetting is easier to achieve \cite{Besga20, Faisant21, resettingNoise}, and resets are non-instantaneous \cite{Besga20, Faisant21,IntermittentReset}. The interest of such models is twofold. First, the discontinuity in the position of the particle violates detailed balance, leading to out-of-equilibrium dynamics. Second, from a first-passage perspective, tuning the rate $r$ of the resets gives a way to optimise search processes \cite{resettingPRL, resettingRTP, Besga20, Faisant21, resettingInInterval}.

\begin{figure}[t]
\centering
    \includegraphics[width=0.7\linewidth]{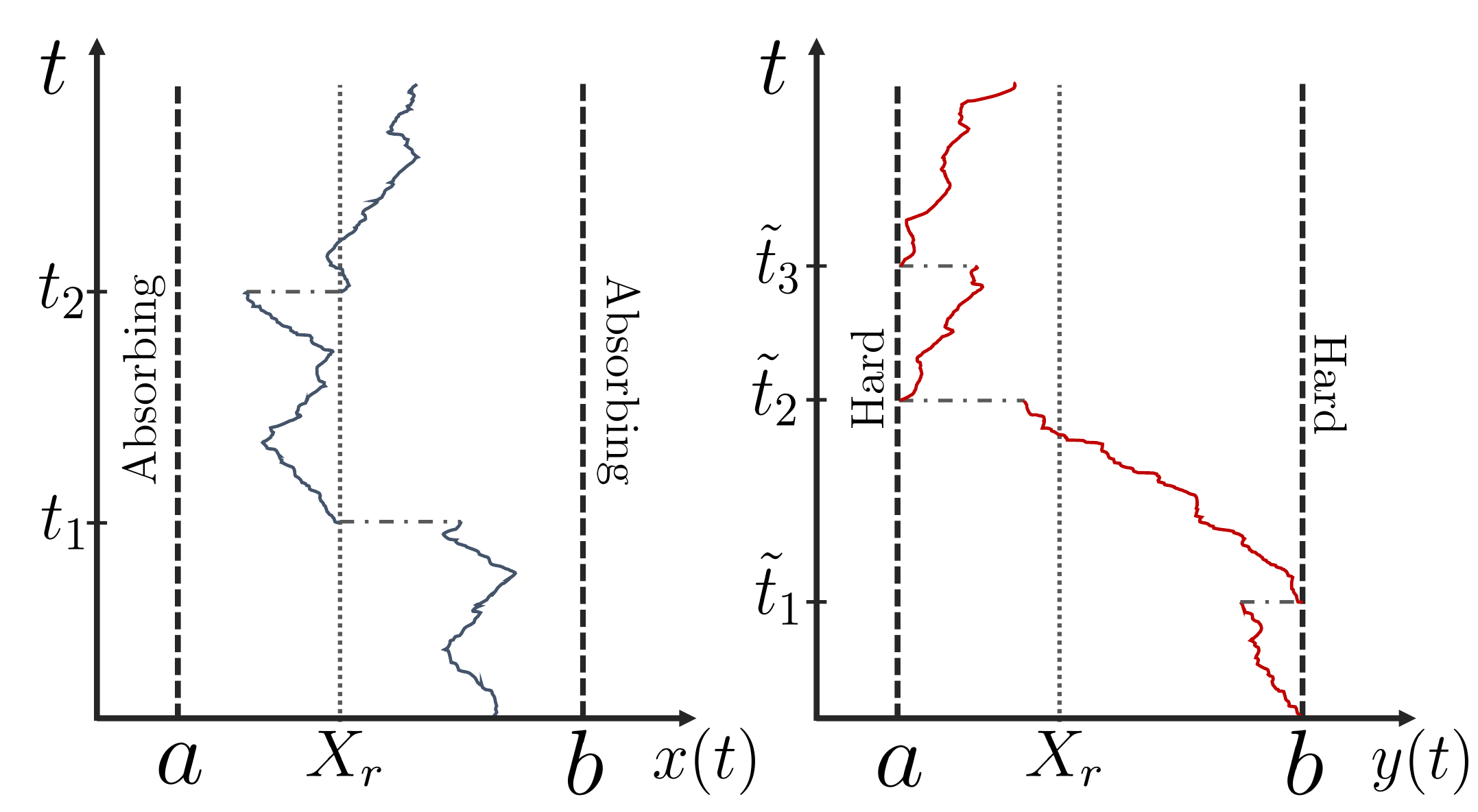}
    \caption{\textbf{Left}: Schematic of a trajectory of the process $x(t)$. At random times $t_i$ ($i \in \{1,2,3,...\}$), $x(t)$ restarts its dynamics at $X_r$. \textbf{Right:} On the other hand, at random times $\tilde t_i$ (with the same distribution as the $t_i$'s), the dual process of $x(t)$ that we denote $y(t)$, restarts its motion at $a$ if $y(t)< X_r$ just before the resetting, or at $b$ if $y(t)> X_r$.}
\label{figure_resetting_duality}
\end{figure}

The duality presented in this paper can be extended to processes undergoing resetting. The construction of the dual process is intuitive when considering non-instantaneous resetting. Instead of teleporting the particle to $X_r$ at random times, one switches on a force for a finite duration $\Delta t$ that attracts the particle toward $X_r$, for example, $f(x)=-\mu (x-X_r)$. For the dual procedure, we expect this force to be reversed, i.e. $-f(x)$, and it will now drive the particle away from $X_r$. Taking the double limit $\mu\to\infty$ and $\Delta t \to 0$ gives the dual procedure of instantaneous resetting: the dual of a process $x(t)$ with instantaneous resetting to $X_r$ and absorbing walls at $a$ and $b$ is a process $y(t)$ which is reset at random times either to $a$ if $y(t)<X_r$, or to $b$ if $y(t)>X_r$, and where walls are now hard walls (of course $X_r \in [a,b]$). For a schematic description of the dual procedure, see Figure \ref{figure_resetting_duality}. 

This duality transformation for the resetting events is actually independent of the boundary conditions. Indeed, a process $y(t)$ which resets at $X_r$ with fixed rate $r$, and hard walls located at $a$ and $b$, has a dual process $x(t)$ with absorbing walls at $a$ and $b$ which resets at $a$ if $x(t)<X_r$, or at $b$ if $x(t)>X_r$. Therefore, it is absorbed either at $a$ or $b$ after the first reset.

In this section we show that the main results of this paper still hold when processes are subjected to stochastic resetting if the dual is constructed as we explained above. We start by illustrating the duality with the resetting Brownian motion, for which explicit expressions can even be obtained in the stationary state. We then briefly explain how this duality can be extended to other continuous stochastic processes, as well as discrete time random walks.

\subsection{Duality for the resetting Brownian motion}

We consider a resetting Brownian motion in the presence of a force $F(x)$. With a fixed rate $r$, it restarts its dynamics at position $X_r$, such that it evolves through
\begin{equation}
x(t+dt) = \left\{
    \begin{array}{ll}
    x(t) + \left(F(x{\red (t)}) + \sqrt{2T}\, \xi(t)\right) dt \mbox{ with probability } (1-r\, dt)\\
        X_r \mbox{ with probability } r\, dt 
    \end{array}
\right.
,\label{rbmlangevin}
\end{equation}
where $T$ is a diffusion constant,  and $\xi(t)$ is a Gaussian white noise with zero mean and correlation function given by $\langle \xi(t)\xi(t')\rangle=\delta(t-t')$. In the presence of two absorbing walls located at $a$ and $b$, we want to compute the exit probability at wall $b$ and time $t$ of the rBm. We can write the backward Fokker-Planck equation by considering the evolution of the process between times $0$ and $dt$. Starting from $x$, the exit probability at time $t+dt$ reads\footnote{In this section we write the resetting rate as an exponent and the resetting position as an additional argument in the functions which depend on these parameters.}
\begin{equation}
    E_b^r(x,t+dt,X_r)=(1-rdt)\,  \mathbb{E}_{\xi}\left[E_b^r\left(x+\left(F(x) + \sqrt{2T}\, \xi(t)\right) dt,t,X_r\right)\right] + r dt\,  E_b^r(X_r,t,X_r)\, ,
\end{equation}
 The first term comes from the fact that with probability $1-r dt$ there is no reset, and the process diffuses from $0$ to $dt$, and one has to average over the noise. When a reset happens, with probability $r dt$, the particle restarts its dynamics at $X_r$ at time $dt$. Expanding the first term on the right hand-side and averaging over $\xi(t)$ gives the backward Fokker-Planck equation corresponding to (\ref{rbmlangevin})
\begin{equation}\label{FPrBmexit}
    \partial_t E_b^r(x,t,X_r) = T\, \partial^2_x E_b^r(x,t,X_r) + F(x)\, \partial_x E_b^r(x,t,X_r) - r E_b^r(x,t,X_r) + r E_b^r(X_r,t,X_r)\, .
\end{equation}
The boundary conditions are $E_b^r(a,t,X_r)=0$, and $E_b^r(b,t,X_r)=1$, while the initial condition reads $E_b^r(x,t=0,X_r)=0$ for $x<b$, and $E_b^r(b,t=0,X_r)=1$.

On the other hand, the dual of $x(t)$ is denoted $y(t)$ and evolves through
\begin{equation}\label{dualrbm}
y(t+dt) = \left\{
    \begin{array}{ll}
   
        y(t) + \left(\tilde F(y{\red (t)}) + \sqrt{2T}\, \xi(t)\right)\, dt \mbox{ with probability } (1-r\, dt)\\
        a \mbox{ with probability } r\, dt \mbox{ if } y(t)<X_r\\
              b \mbox{ with probability } r\, dt \mbox{ if } y(t)>X_r 
    \end{array}
\right.
,
\end{equation}
and $y(0)=b$. We do not consider the case $y(t)=X_r$ in Eq.~(\ref{dualrbm}) as it has zero measure in continuous space. For discrete time random walks, one needs to be more careful (see Sec.~\ref{discretereset}). Let us write $\tilde p(y,t)$ the probability distribution of the position of the process (\ref{dualrbm}). It obeys the following forward Fokker-Planck equation

\begin{equation}\label{FPdensitydualrbm}
    \partial_t \tilde p(y,t) = T\, \partial_y^2 \tilde p(y,t) - \partial_y \left(\tilde F(y) \tilde p(y,t)\right) -r\tilde p(y,t) + r \left[\int_a^{X_r}dz\, \tilde p(z,t)\right]\delta(y-a)+ r\left[\int_{X_r}^{b}dz\, \tilde p(z,t)\right]\delta(y-b)\, .
\end{equation}
Now we want to relate the Fokker-Planck equation (\ref{FPrBmexit}) on the exit probability to the one satisfied by the cumulative distribution of the dual process (the probability to find the dual particle in $[a,x]$ at time $t$), defined as
\begin{equation}
    \tilde \Phi^r(x,t|b,X_r) = \int_{a^-}^x dy\, \tilde p(y,t)\, .
\end{equation}
Integrating (\ref{FPdensitydualrbm}) between $a^-$ and $x$ leads to
\begin{equation} \label{FPrBmcumu}
    \partial_t \tilde \Phi^r(x,t|b,X_r) = T\, \partial^2_x \tilde \Phi^r(x,t|b,X_r) -\tilde F(x)\, \partial_x \tilde \Phi^r(x,t|b,X_r) - r\, \tilde \Phi^r(x,t|b,X_r) + r\,  \tilde \Phi^r(X_r,t|b,X_r)\, ,
\end{equation}
which is exactly (\ref{FPrBmexit}) if $\tilde F(x) =-F(x)$. Finally, we have $ \tilde \Phi^r(a,t|b,X_r)=0$, $ \tilde \Phi^r(b,t|b,X_r)=1$ and the initial conditions $ \tilde \Phi^r(x,t=0|b,X_r)=0$ for $x<b$ and $ \tilde \Phi^r(b,t=0|b,X_r)=1$. We thus once again obtain that the two quantities $E_b^r(x,t,X_r)$ and $\tilde \Phi^r(x,t|b,X_r)$ satisfy the same equations with the same boundary and initial conditions, and thus the identity \eqref{mainRelation_averaged} holds.

\begin{figure}[t]
\centering
    \begin{minipage}[c]{.32\linewidth}
        \centering
        \includegraphics[width=1.\linewidth]{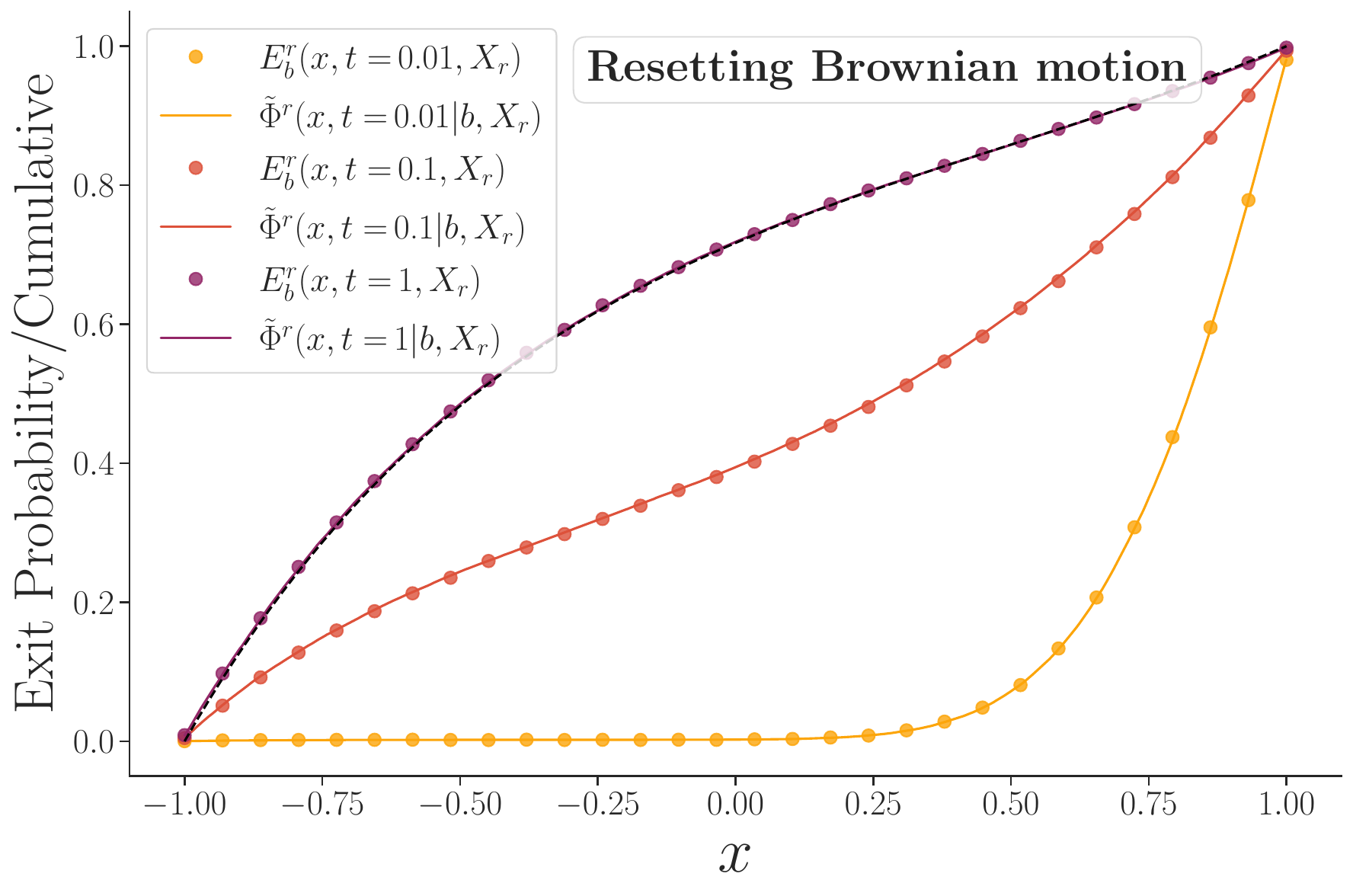}
    \end{minipage}
    \hspace{0.8cm}
    \begin{minipage}[c]{.32\linewidth}
        \centering
        \includegraphics[width=1.\linewidth]{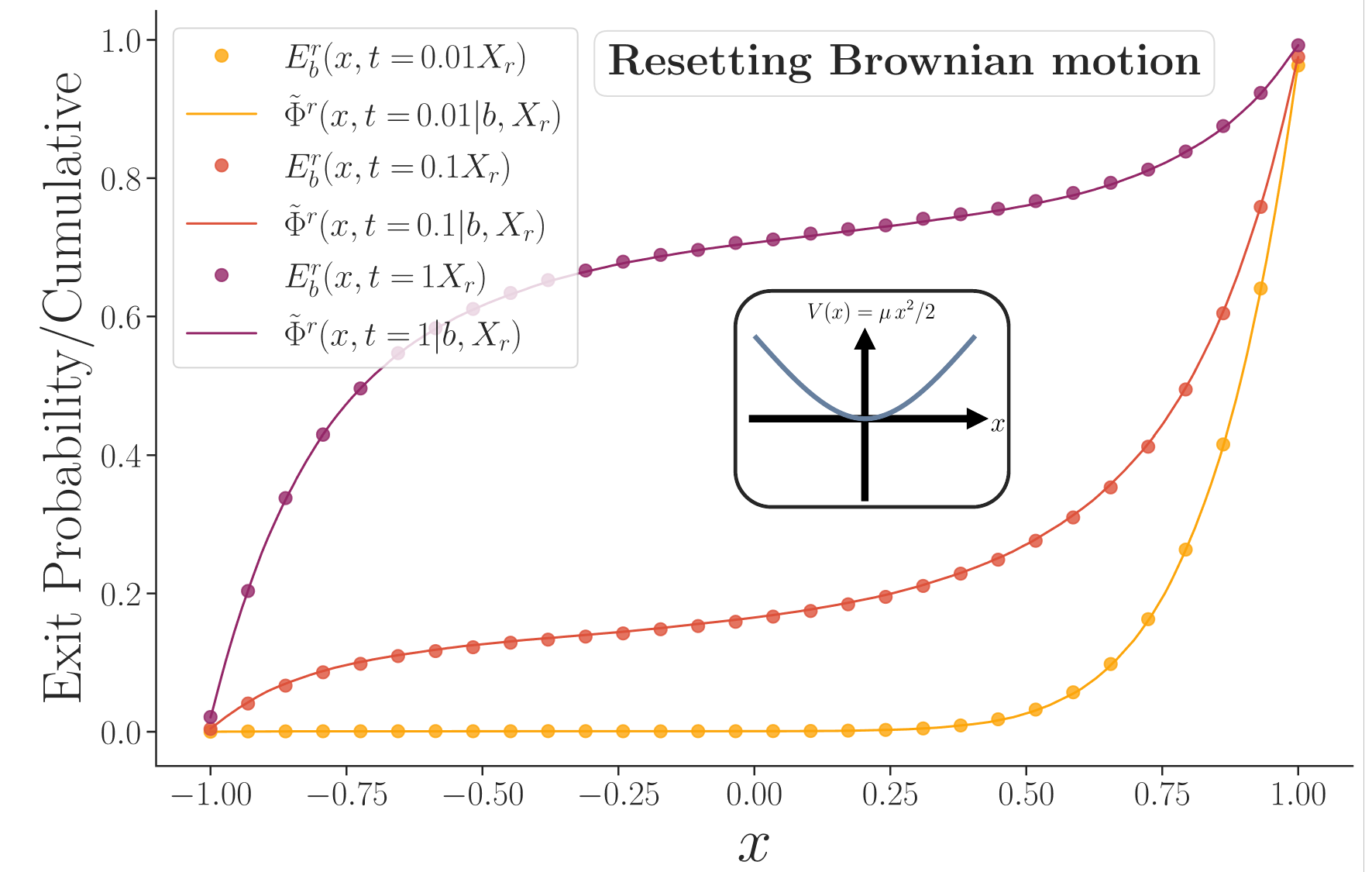}
    \end{minipage}
    \caption{We verify the duality relation (\ref{mainRelation_averaged}) for a Brownian motion subjected to Poissonian resetting. \textbf{Left}: Comparison of simulation results for the exit probability (dots) and cumulative distribution (lines) of the resetting Brownian motion without external potential, with $T=4$, $r=10$ and $X_r=1.5$, at different times. The dashed black curve corresponds to the analytic prediction for the stationary state \eqref{rBm_stationary}. \textbf{Right}: Same plot in the presence of a harmonic potential $V(x)=\frac{\mu}{2}x^2$ with $\mu=20$ for $E_b$, and $-V(x)$ for $\tilde \Phi$. In both cases the overlap is perfect.}
\label{ResettingFig}
\end{figure}

When $F(x)=0$, it is possible to write explicitly the solution to (\ref{FPrBmexit})-(\ref{FPrBmcumu}) in the stationary state  (see \cite{resettingInInterval}). It reads, with $\alpha_r=\sqrt{r/T}$,
\begin{equation} \label{rBm_stationary}
    E_b^r(x,t \to \infty,X_r) = \tilde \Phi^r(x,t\to \infty|b,X_r) = \frac{\sinh(\alpha_r(x-X_r))+\sinh(\alpha_r(X_r-a))}{\sinh(\alpha_r(b-X_r))+\sinh(\alpha_r(X_r-a))} \,.
\end{equation}
In Figure \ref{ResettingFig}, we validate our statements with simulations. 

{\red In App.~\ref{app:resetting}, we instead consider a rBm with hard walls and resetting position $X_r$. We construct its dual which is a Brownian motion with absorbing walls which resets at $a$ if $x(t)<X_r$ and at $b$ if $x(t)>X_r$. Note that in this case, the particle is always absorbed at one of the two walls after the first resetting event.}

\subsection{Generalization for a generic continuous stochastic process subjected to resetting}

 Following the approach used for the rBm, one can include resetting in the general model \eqref{LangevinIntroduction} considered in Section \ref{FPderivationsection}. To be even more general, let us consider a process which can reset at different points $X_r^1,...,X_r^n$ with independent rates $r_1,...,r_n$. One simply has to say that, during the time interval $[t,t+dt]$, the process $x(t)$ either follows the dynamics defined in \eqref{LangevinIntroduction} with probability $(1-\sum_i r_i dt)$, or it resets to $X_r^i$ with probability $r_i dt$. The dual process $y(t)$ then follows the dynamics of \eqref{LangevinIntroductionDUAL} with probability $(1-\sum_i r_i dt)$, and with probability $r_i dt$, it resets to $a$ if $y(t)<X_r^i$ and to $b$ if $y(t)>X_r^i$. In this case, the derivation of Sec. \ref{ContinuousProof} still holds, up to some additional terms in the Fokker-Planck equations to account for the resetting events. These terms can be easily treated in the same way as in the rBm case above, and we can thus conclude that the duality relation \eqref{mainRelation} holds once again. 
 {\red We can also extend in the same way the result of App.~\ref{app:resetting}, where it is instead $y(t)$ (with hard walls) which is subjected to standard resetting, while its dual $x(t)$ (with absorbing walls) is reset at the walls.}

\subsection{Random walks subjected to resetting}\label{discretereset}

Resetting is not restricted to continuous stochastic processes. It can also be introduced for discrete time random walks \cite{resettingReview, randomwalkresetPRL}. In this case, it is again possible to define a dual process such that the identity \eqref{mainRelation} holds. Let us go back to the derivation of Sec.~\ref{DiscreteProof}, but adding the possibility for resetting events. We introduce an additional random variable $R_n$ such that at each time $n$, $R_n=1$ with some probability $0\leq r\leq 1$ and $R_n=0$ otherwise, independently of the past history. Whenever $R_n=1$, the process $X_n$ resets to some value $X_r\in[a,b]$. More precisely, $X_n$ now follows the same dynamics as in Sec.~\ref{DiscreteProof}, given by \eqref{defXt} between times $n-1$ and $n$ if $R_n=0$, but if $R_n=1$, then $X_{n}=X_r$ independently of its previous value (unless it has already been absorbed at $X_n=a^-$ or $b$, in which case it remains there). As for continuous stochastic processes, one could allow for independent resettings at different points $X_r^i$ with different rates $r^i$, but we will restrict to $i=1$ for simplicity.

We now define the dual process $Y_n$ as follows: let $\tilde R_n$ a process with the same law as $R_n$. If $\tilde R_{n}=0$, $Y_{n}$ is given by \eqref{defdual}, but instead if $\tilde R_{n}=1$, one has:
\be
Y_{n} = \begin{cases} a \quad {\rm if} \ Y_{n-1}\leq X_r \;, \\ b \quad {\rm if} \ Y_{n-1} > X_r \;. \end{cases}
\ee
Note the asymmetry in the definition of the resetting events for $Y_n$. This is connected to the shift of the boundary condition from $a$ to $a^-$ in $X_n$. If one is interested in the exit probability at $a$, $E_a(x,n)$, then one should instead set $Y_{n}=a$ when $Y_{n-1}=X_r$. Again, this precision becomes irrelevant if the process is such that the event $Y_n=X_r$ has zero measure.

Finally, given a trajectory of the process $X_n$ and a value of $Y_0=X_0^R$, the dual trajectory $X_n^R$ is the realisation of $Y_n$ with $\tilde W_n= W_{T+1-n}$ and $\tilde R_n= R_{T+1-n}$ for all $i$.

One can check that, with these definitions, the proof of Sec. \ref{DiscreteProof} remains valid. Since the $R_n$'s are independent, the hypothesis \eqref{DBstatement1} is obviously still satisfied when considering the joint probability of $W_n$ and $R_n$. Here we will only give a quick argument as for why the equivalence \eqref{equivalence} still holds. The rest of the derivation can be easily adapted.

We start by noting that the generalised equivalence \eqref{equivgeneral} holds on any interval $\llbracket n_1,n_2-1 \rrbracket$ where $n_1$ and $n_2$ are two successive resetting events. If a resetting event happens at time $n$, one can see that there are only 4 possible situations:

(i) $a^- < X_{n-1} < b$ and $X_r < X^R_{T-n} \leq b$, in which case $X_{n}=X_r<X^R_{T-n}$ and $X^R_{T-n+1}=b>X_{n-1}$,

(ii) $a^- < X_{n-1} < b$ and $a \leq X^R_{T-n} \leq X_r$, in which case $X_{n}=X_r\geq X^R_{T-n}$ and $X^R_{T-n+1}=a \leq X_{n-1}$,

(iii) $X_{n-1}=b \geq X^R_{T-n+1}$, in which case $X_{n}=b \geq X^R_{T-t-\Delta t}$,

(iv) $X_{n-1}=a^- < X^R_{T-n+1}$, in which case $X_{n}=a^- < X^R_{T-n}$.

\noindent Thus the equivalence \eqref{equivgeneral} still holds for all times $n_1,n_2\in\llbracket 0,T \rrbracket$, which implies \eqref{equivalence}.
\\

{\red Let us now consider the reverse case and assume that whenever $R_n=1$, instead of resetting at $X_n=X_r$ one has (if $X_n \in (a,b)$)
\be
X_{n} = \begin{cases} a^- \quad {\rm if} \ X_{n-1}< X_r \;, \\ b \quad {\rm if} \ X_{n-1} \geq X_r\, . \end{cases}
\ee
(note that the strict inequality is now on the first line). On the other hand, if $\tilde R_n = 1$, $Y_n$ undergoes an usual resetting event, i.e. $Y_n=X_r$. We denote $n_1$ the time of the first resetting event for $X_n$. The equivalence \eqref{equivgeneral} holds on the interval $\llbracket 0,n_1-1 \rrbracket$. One should then distinguish two cases:

(i) If $X_{n_1-1} \geq X_r$, then $X_T=b \geq X_0^R$, and $X^R_{T-n_1+1}=X_r \leq X_{n_1-1}$, and thus $X_0 \geq X^R_T$.

(i) If $X_{n_1-1} < X_r$, then $X_T=a^- < X_0^R$, and $X^R_{T-n_1+1}=X_r > X_{n_1-1}$, and thus $X_0 < X^R_T$.

\noindent Thus the equivalence \eqref{equivgeneral} holds once again.
}

\section{How general is the duality ?} \label{sec:generalize}

In the derivation of the duality for continuous stochastic processes (see Section \ref{FPderivationsection}), we have made the crucial assumptions that $x(t)$, whose dynamics is given by~(\ref{LangevinIntroduction}), is driven by an equilibrium Markov process $\bm {\theta}(t)$ that satisfies the detailed balance condition{\red s~\eqref{detailed_balance_Introduction1}-\eqref{detailed_balance_Introduction2}}. This way, it was possible to write the Fokker-Planck equations and derive the duality relation~{\red (\ref{mainRelation})}. However, for the discrete case in Sec.~\ref{discrete_case}, we have used assumptions that are a bit weaker in that regard: the Markovianity and detailed balance assumptions are replaced by a simple time-reversal property {\red \eqref{timereversalmainresults}}. This may suggest that this duality relation is a bit more general in the case of continuous stochastic processes as well. In this section, we consider two processes that do not satisfy these assumptions, but for which the duality still seems to hold. 

\begin{figure}[t]
\centering
    \begin{minipage}[c]{.32\linewidth}
        \centering
        \includegraphics[width=1.\linewidth]{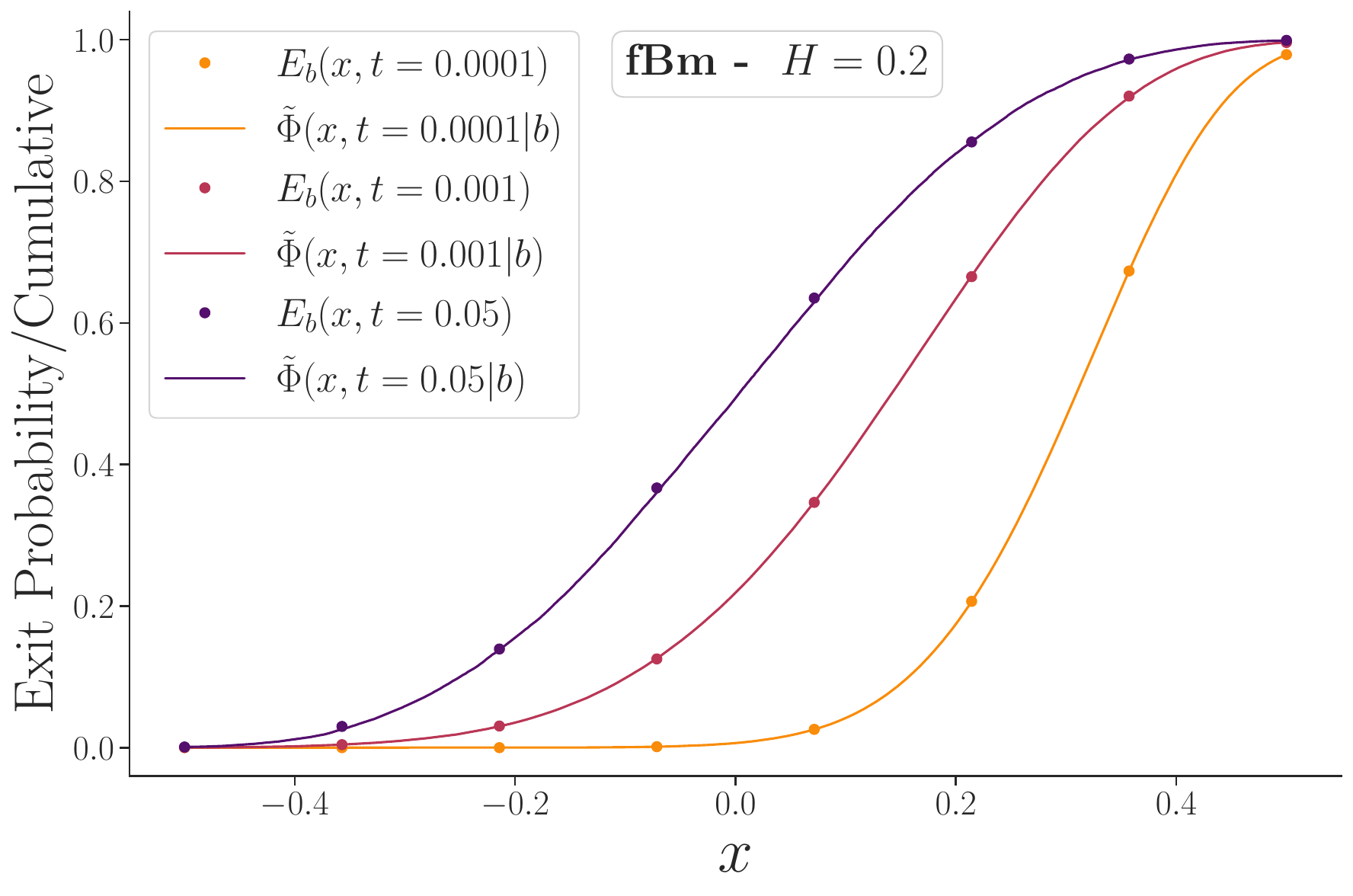}
    \end{minipage}
    \hfill%
    \begin{minipage}[c]{.32\linewidth}
        \centering
        \includegraphics[width=1.\linewidth]{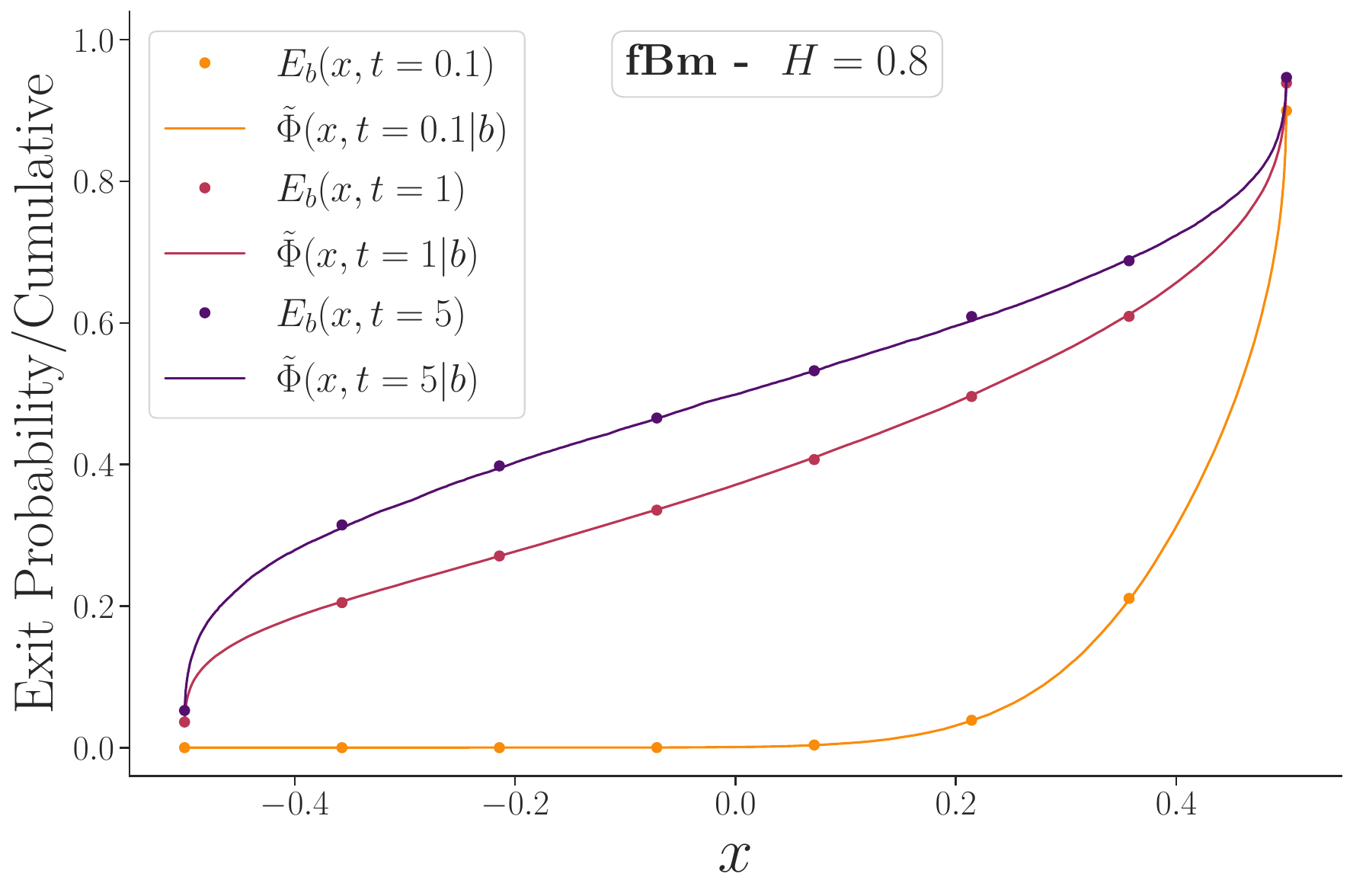}
    \end{minipage}
    \hfill
        \begin{minipage}[c]{.32\linewidth}
        \centering
        \includegraphics[width=1.\linewidth]{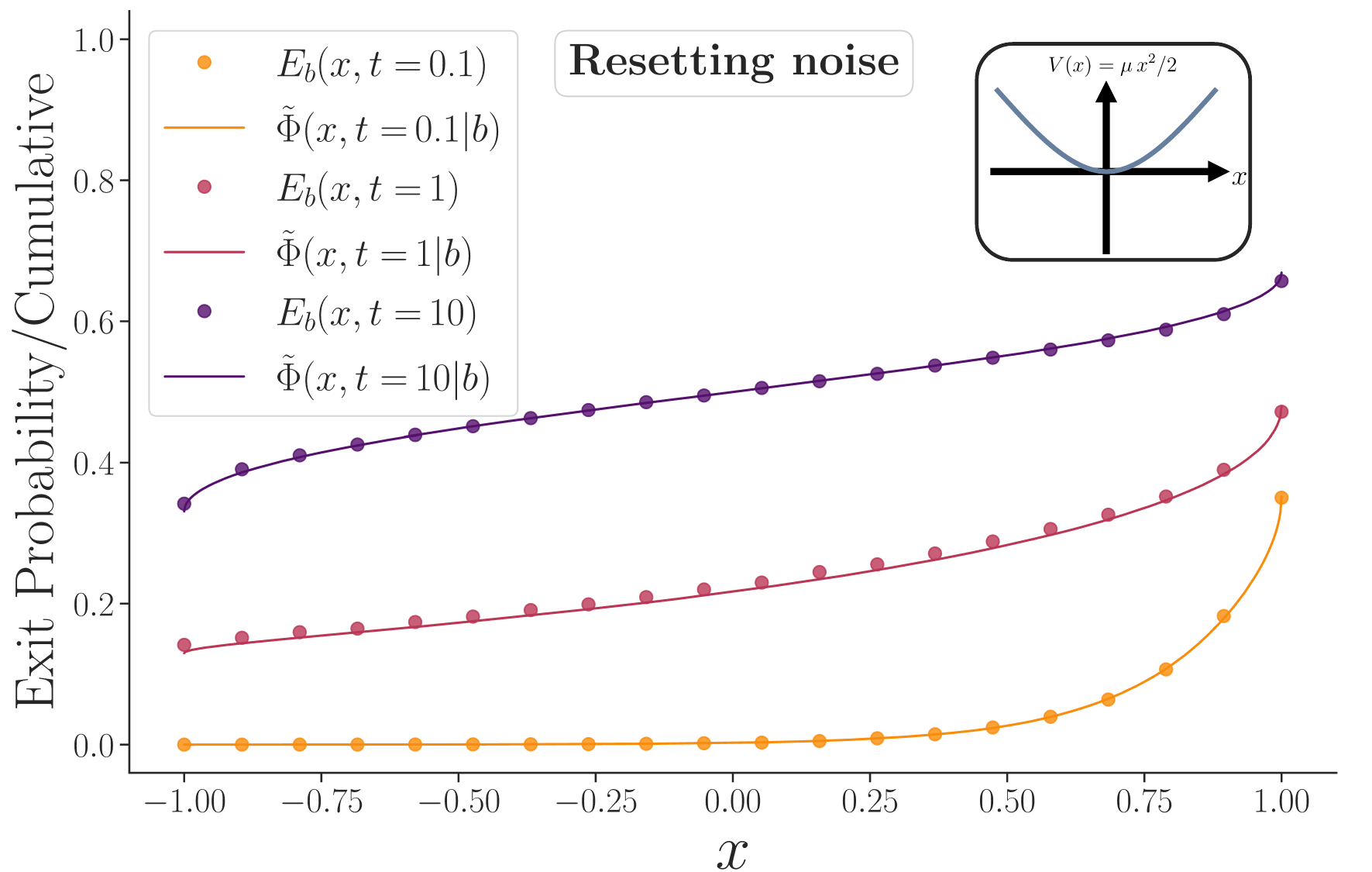}
    \end{minipage}
    \caption{\textbf{Left and central panels}: We show with simulations that the duality~(\ref{mainRelation_averaged}) remains valid even for processes with non-Markovian noise (but stationary increments) such as the fractional Brownian motion. The dots show the value of the exit probability while the solid lines are the cumulative distributions. Both quantities are evaluated numerically by averaging over trajectories. We have tested the duality at different times for Hurst parameters $H = 0.2$ on the left plot, and $H=0.8$ on the central one. \textbf{Right}: Same plot for a process subjected to a resetting noise in a harmonic potential $V(x)=\frac{\mu}{2}x^2$ with $\mu=1$, $r=1$ and $T=4$. This model does not satisfy the detailed balance hypothesis, but the duality \eqref{mainRelation_averaged} seems to hold nonetheless.}
\label{fBmFig}
\end{figure}

\begin{itemize}
    \item \textit{Fractional Brownian motion}:
     We first consider a fractional Brownian motion $x(t)$, i.e. a continuous-time centered Gaussian process with two-time correlations function \cite{fBmMandelbrot}
\begin{equation}
    \langle x(t_1)x(t_2) \rangle = t_1^{2H} + t_2^{2H} - |t_1-t_2|^{2H}\, ,
\end{equation}
where $0<H<1$ the Hurst parameter. The fBm increments are generated by a fractional Gaussian noise (fGn) $\xi_f(t)$ which is a stationary non-Markovian process with correlations
\begin{equation}
     \langle \xi_f(t_1)\xi_f(t_2) \rangle = 2H(2H -1)|t_1-t_2|^{2H-2}\, .
\end{equation}\label{fgn}

Due to the non-Markovian property of the fGn, it is not possible to write the standard Fokker-Planck equation as done in Section \ref{FPderivationsection}. The fBm breaks the assumptions we have made, but simulations show that the duality still holds at any time $t$. In Figure~\ref{fBmFig}, we validate numerically the duality for $H=0.2$, and $H=0.8$.

\item \textit{Resetting noise}:
The last example is a process $x(t)$ driven by a resetting noise as considered in \cite{resettingNoise}. To be more specific, $x(t)$ evolves in a harmonic trap of strength $\mu$ and is subjected to a resetting Brownian motion $y_r(t)$ such that it follows the Langevin dynamics
\begin{equation}
\frac{dx(t)}{dt}=-\mu \, x(t) + r\, y_r(t)\, ,
\label{lange.2}
\end{equation}
and the rBm evolves through
\begin{equation}
y_r(t+dt) = \left\{
    \begin{array}{ll}
        0 \mbox{ with probability } r\, dt \\
        y_r(t) + \sqrt{2T}\, \xi(t)\, dt \mbox{ with probability } (1-r\, dt)
    \end{array}
\right.
,
\end{equation}
where $\xi(t)$ is a Gaussian noise with zero mean and two times correlation function $\langle \xi(t)\xi(t')\rangle=\delta(t-t')$.
The driven process $y_r(t)$ is Markovian but it violates detailed balance due to the resettings. Indeed, while it is possible for $y_r(t)$ to jump from $x\neq 0$ to the origin, the reversed event has zero probability. Thus, this process seems to violate our hypotheses more strongly than the fBm, since the driving process ${\bm \theta}(t)=y_r(t)$ is intrinsically irreversible. In the right panel of Figure {\ref{fBmFig}}, we have tested the duality relation \eqref{mainRelation_averaged} for this model. Although there seems to be a small discrepancy at intermediate times, the agreement is still quite surprising.

\end{itemize}

\section{Conclusion}

In this paper we have shed light on a  relation between the exit probability from an interval of a stochastic process driven by a stationary noise, and the position distribution of its dual in the presence of hard walls. This duality holds in both continuous and discrete time. Such duality relations between absorbing and reflective boundary conditions have a long history in mathematics. Here we provide an explicit and practical formulation of this duality for a variety of models relevant in physics {\red (for which this duality had not yet been applied)}, such as active particles, diffusing diffusivity models or stochastic resetting.
Although this framework includes most of the models studied by physicists, it does not include in theory the fractional Brownian motion or a process driven by a resetting noise that breaks the detailed balance assumption. However, we have shown using simulations that the duality is very likely to hold also in these cases, at least approximately. Therefore, it would be interesting to find a derivation of the duality that includes these models. 
Additionally, a notable assumption in our framework is that the driven noise should have a stationary distribution. As a result, constructing the duality for processes like the random acceleration process, where the acceleration is a Gaussian white noise, remains an open challenge.

Another open question, of particular relevance for applications, is whether a similar connection can be formulated in higher dimensions. Such a duality in multi-dimension has been formulated in an abstract way in other contexts \cite{DualityMultidimensions}, but expressing this duality in a way which is useful to practical computations even for simple models seems much more challenging than in 1d. Finally, a recent work proposed the construction of the Siegmund dual of an $N$-particle system \cite{Nparticleduality}. The study of $N$-particle systems is very relevant when considering active particles where collective motion emerges. {\red In this respect, we have derived a new result for a system of $N$ independent particles with applications to extreme value statistics.}

Understanding both the spatial behavior and first passage properties of stochastic processes is crucial, particularly in the context of active particles. The duality presented here provides an elegant means of linking these aspects together. {\red Computing both the exit probability and the cumulative distribution of the dual is challenging. However, the duality allows us to easily derive one from the other, offering a significant analytical advantage. Furthermore, measuring the exit probability of a stochastic process via simulations or experiments requires many trajectories. In contrast, studying the dual process, if it is ergodic, would involve observing only one long trajectory, thereby simplifying the problem.} 
We hope that this new connection will be useful to the growing number of people working in these fields, and pave the way to new results, both analytical and numerical.

\section*{Acknowledgments}
We thank P. Le Doussal, S. N. Majumdar and G. Schehr for helpful comments and discussions throughout the project. We thank two anonymous reviewers for their detailed feedback and insightful comments.

\section*{Appendices}
\appendix

\section{Illustration of the duality relation for Brownian motion} \label{AppBM}

\subsection{Derivation of the duality relation for Brownian motion using the Fokker-Planck equation} \label{ProofsSimple}

In this Appendix, we give the derivation of Sec.~\ref{ContinuousProof} in the simple case of a diffusing particle subjected to a force $f(x)$, with space-dependent {\red diffusion coefficient} $T(x)$ and hard walls at $a$ and $b$. We start from the Fokker-Planck equation for the particle density (with the It\=o convention)
\begin{equation}
\partial_t \tilde P = - \partial_x [\tilde f(x) \tilde P] + \partial^2_{xx}[T(x)\tilde P] \;.
\end{equation}
Introducing the cumulative distribution $\tilde \Phi(x,t|b)=\int_a^x dy \tilde P(y,t|b)$ and using that $\tilde P(x,t|b)=\partial_x \tilde \Phi(x,t|b)$ we have
\begin{equation}
\partial_x \{-\partial_t \tilde \Phi - \tilde f(x)\partial_x \tilde \Phi + \partial_x [T(x) \partial_x \tilde \Phi] \} = 0 \;.
\label{cumudiffut}
\end{equation}
The zero-flux boundary condition at $x=a$ (as well as at $x=b$) reads
\begin{equation}
0 = \tilde f(a)\tilde P(a,t|b) -\partial_{x}[T(a)\tilde P(a,t|b)] = \tilde f(a)\partial_x \tilde \Phi(a,t|b) -\partial_x [T(x) \partial_x\tilde \Phi(a,t|b)] \;,
\end{equation}
and in addition $\partial_t \tilde \Phi(a,t|b)=\partial_t \tilde \Phi(b,t|b)=0$. Thus we can integrate Eq. \eqref{cumudiffut} to obtain
\begin{equation}
\partial_t \tilde \Phi = [- \tilde f(x)+\partial_x T(x)]\partial_x \tilde \Phi + T(x)\, \partial^2_{xx}\tilde \Phi \;.
\end{equation}
But this is exactly the backward-Fokker Planck equation that the exit probability $E_b(x,t)$ defined in Eq.~\eqref{defExitProba} should satisfy if we take $\tilde f(x) = -f(x)+\partial_x T(x)$. Additionally $\tilde \Phi(x,t|b)$ and $E_b(x,t)$ satisfy the same boundary conditions $E_b(a,t)=\tilde \Phi(a,t|b)=0$ and $E_b(b,t)=\tilde \Phi(b,t|b)=1$ and the same initial condition $E_b(x,0)=\tilde \Phi(x,0|b)=\mathbb{1}_{x\geq b}$. Since this is enough to specify these functions completely, one has
\begin{equation}
E_b(x,t)=\tilde \Phi(x,t|b) \;.
\end{equation}

\subsection{Duality relation for the survival probability of Brownian motion} \label{suvivalBM}
We consider here a $1$d Brownian motion in the interval $]-\infty,b]$ whose position is described by $x(t)$, with $x(0) =x_0 \in ]-\infty,b]$. For the process $x(t)$, $x=b$ is an absorbing boundary condition. We also introduce another $1$d Brownian motion $y(t)$, the dual process of $x(t)$, this time with a hard wall (or reflective) boundary condition at $x=b$. We want to verify the duality relation~\eqref{mainRelation} of section \ref{mainresults}, applied to the computation of the survival probability using (\ref{survivalmainresult}). In this context, it reads
\begin{equation}
    1- Q_b(x_0,t) = \tilde{\mathbb{P}}(y(t) \leq x_0 |y_0=b)\, ,
\label{survivalBM}
\end{equation}
where $Q_b(x_0,t)$ is the probability that the particle $x(t)$ stayed in $]-\infty,b]$ up to time $t$. We will compute the two sides of Eq. (\ref{survivalBM}) and show that they are indeed equal. The survival probability reads
\begin{equation}
    Q_b(x_0,t) = \int_{-\infty}^b dx\, p_b(x,t|x_0)\, ,
\end{equation}
with $p_b(x,t|x_0)$ the Brownian propagator with an absorbing wall at $x=b$. It can be calculated using the method of images. Indeed, the idea is to say there is a Brownian propagator at initial position $x_0$, and we subtract a second Brownian propagator - the image with respect to the wall - at initial position $2b -x_0$ such that $p_b(b,t|x_0) = 0$. Over time, this image propagator subtracts the mass of the particles absorbed by the wall. That way, we have

\begin{equation}
    p_b(x,t|x_0) = \frac{1}{\sqrt{4\pi\, D\, t}}\left(e^{-\frac{(x-x_0)^2}{4D\, t}}-e^{-\frac{(x-(2b-x_0))^2}{4D\, t}}\right)\, ,
\end{equation}
and we directly integrate it form $-\infty$ to $b$ in order to obtain the survival probability
\begin{equation}
    Q_b(x_0,t) = \text{erf}\left(\frac{b - x_0}{\sqrt{4D\, t}}\right)\, .
\end{equation}
On the other hand, for the process $y(t)$ the wall at $x=b$ is reflective and we have to add the image propagator such that the flux at the wall is zero, $\partial_y \tilde{p}_b(y,t|y_0)|_{y=b} = 0$. In addition, to compute the probability on the r.h.s of Eq.~(\ref{survivalBM}) we have to take the initial condition $y_0= b$,
\begin{equation}
    \tilde{p}_b(y,t|b) = \lim_{y_0 \to b}\, \frac{1}{\sqrt{4\pi\, D\, t}}\left(e^{-\frac{(y-y_0)^2}{4D\, t}}+e^{-\frac{(y-(2b-y_0))^2}{4D\, t}}\right)= \frac{1}{\sqrt{\pi\, D\, t}}e^{-\frac{(y-b)^2}{4D\, t}}\, .
\end{equation}
For a Brownian motion, the r.h.s of Eq. (\ref{survivalBM}) is given by
\begin{equation}
  1-  \tilde{\mathbb{P}}(y(t) \leq x_0 |y_0 =b) = 1-\int_{-\infty}^{x_0} dy \, \tilde{p}_b(y,t|b) = \text{erf}\left(\frac{b-x_0}{\sqrt{4D\, t}}\right) = Q_b(x_0,t)\, ,
\end{equation}
and we obtain the expected duality relation 
\begin{equation}
 1- Q_b(x_0,t) =\tilde{\mathbb{P}}(y(t) \leq x_0 |y_0 =b)\, .
\end{equation}

\section{Derivation of the Siegmund duality relation from the Fokker-Planck equation} \label{ContinuousProof_duality}

In this Appendix we show that~(\ref{identity_general}) for continuous time stochastic processes can be generalised to
\begin{equation} \label{identity_general_app1}
     \mathbb{P}(x(t)\geq y_0|x(0)=x_0,{\bm \theta}) = \tilde{\mathbb{P}}(y(t)\leq x_0|{\bm \theta}(t)={\bm \theta};y(0)=y_0,{\bm \theta}(0)^{eq})  \, .
\end{equation}
This is the continuous time equivalent of \eqref{Siegmund_precise}, which relates the cumulative distribution of the process with absorbing walls with the cumulative of its dual {\red with hard walls} initialized at position $y$. It gives back \eqref{mainRelation} when taking $y=b$. Multiplying by $p_{eq}({\bm \theta})$ and integrating over ${\bm \theta}$ yields the symmetric identity
\begin{equation} \label{Siegmund_app}
    \mathbb{P}(x(t)\geq y_0|x(0)=x_0,\bm{\theta}(0)^{eq}) = \tilde{\mathbb{P}}(y(t)\leq x_0|y(0)=y_0,\bm{\theta}(0)^{eq})\, .
\end{equation}

The derivation is very similar to the one presented in Sec.~\ref{ContinuousProof} for the case $y=b$, but the notations are more heavy which is why we present it in the appendix. We consider the same process $x(t)$ as before, defined in \eqref{LangevinIntroduction} and \eqref{SDEtheta_Introduction} with the hypotheses \eqref{detailed_balance_Introduction1}-\eqref{detailed_balance_Introduction2}. The Fokker-Planck equation for the joint probability density $P(x,\bm{\theta},t|x_0,\bm{\theta}_0)$ reads (with the It\=o convention)
\begin{eqnarray}
\partial_t P &=& - \partial_x [f(x,\bm{\theta}) P] + \partial_x^2 [\mathcal{T}(x,\bm{\theta}) P] - \sum_i \partial_{\theta_i}[g_i(\bm{\theta}) P] + \sum_{i,j} \partial_{\theta_i\theta_j}^2 [\mathcal{D}_{ij}(\bm{\theta}) P] \nonumber \\
&&+ \int d\bm{\theta}' \left[\mathcal{W}(\bm{\theta}|\bm{\theta}')P(x,\bm{\theta}',t|x_0,{\bm \theta}_0) - \mathcal{W}(\bm{\theta}'|\bm{\theta})P(x,\bm{\theta},t|x_0,{\bm \theta}_0)\right] \,.
\label{FPgeneral_app}
\end{eqnarray}
One can also write the backwards Fokker-Planck
\begin{eqnarray}
\partial_t P &=& f(x_0,\bm{\theta}_0) \partial_{x_0} P + \mathcal{T}(x_0,\bm{\theta}_0) \partial_{x_0}^2 P + \sum_i g_i(\bm{\theta}_0) \partial_{\theta_{0,i}} P + \sum_{i,j} \mathcal{D}_{ij}(\bm{\theta}_0) \partial_{\theta_{0,i}\theta_{0,j}}^2 P \nonumber \\
&&+ \int d\bm{\theta}'_0\, \mathcal{W}(\bm{\theta}'_0|\bm{\theta}_0)[P(x,\bm{\theta},t|x_0,\bm{\theta}'_0) - P(x,\bm{\theta}_,t|x_0,\bm{\theta}_0)] \;. 
\label{backwardFPgeneral_app}
\end{eqnarray}
As in Sec. \ref{ContinuousProof}, we denote $p(\bm{\theta},t)$ the density probability of $\bm{\theta}$. To simplify the notations, in the rest of this appendix, $P$ without arguments will always denote $P(x,\bm{\theta},t|x_0,{\bm \theta}_0)$, while $P_{\bm \theta}$ will denote $P(x,t|\bm{\theta};x_0,{\bm \theta}_0)$.

Let us first consider the dynamics \eqref{LangevinIntroduction}-\eqref{SDEtheta_Introduction} on an interval $[a,b]$ with absorbing boundary conditions at $x=a$ and $x=b$. We are interested in the probability that a particle starting at position $x_0$ at time $t=0$, with a certain value of $\bm{\theta}$, has a position larger than some value $y_0$ at time $t$. Integrating \eqref{backwardFPgeneral_app} over all possible values of $\bm{\theta}$ and over all $x\geq y_0$ yields
\begin{eqnarray}
\partial_t \mathbb{P}(x(t)\geq y_0|x_0,\bm{\theta}_0) &=& f(x_0,\bm{\theta}_0) \partial_{x_0} \mathbb{P}(x(t)\geq y_0|x_0,\bm{\theta}_0) + \mathcal{T}(x_0,\bm{\theta}_0) \partial_{x_0}^2 \mathbb{P}(x(t)\geq y_0|x_0,\bm{\theta}_0) \nonumber \\
&&+ \sum_i g_i(\bm{\theta}_0) \partial_{\theta_{0,i}} \mathbb{P}(x(t)\geq y_0|x_0,\bm{\theta}_0) + \sum_{i,j} \mathcal{D}_{ij}(\bm{\theta}_0) \partial_{\theta_{0,i}\theta_{0,j}}^2 \mathbb{P}(x(t)\geq y_0|x_0,\bm{\theta}_0) \nonumber\\
&&+ \int d\bm{\theta}'_0\, \mathcal{W}(\bm{\theta}'_0|\bm{\theta}_0)[\mathbb{P}(x(t)\geq y_0|x_0,\bm{\theta}'_0) - \mathbb{P}(x(t)\geq y_0|x_0,\bm{\theta}_0)] \;.
\label{backwardFPgeneral_integrated_app}
\end{eqnarray}
This equation is complemented by the boundary conditions, for $y_0\in(a,b]$,
\begin{eqnarray} \label{exit_bc_app}
&&\mathbb{P}(x(t)\geq y_0|x_0=a^+,\bm{\theta}_0) = 0 \quad \text{for all } f(a^+,\bm{\theta}_0)<0, \text{ or if } \mathcal{T}(x,\bm{\theta})>0 \text{ in the vicinity of } a, \\
&&\mathbb{P}(x(t)\geq y_0|x_0=b^-,\bm{\theta}_0) = 1 \quad \text{for all } f(b^-,\bm{\theta}_0)>0, \text{ or if } \mathcal{T}(x,\bm{\theta})>0 \text{ in the vicinity of } b, \nonumber 
\end{eqnarray}
and the initial condition
\begin{eqnarray}\label{exit_ic_app}
&&\mathbb{P}(x(0)\geq y_0|x_0,\bm{\theta}_0) = \mathbb{1}_{x_0\geq y_0} \;.
\end{eqnarray}

Let us now consider the probability density in the presence of hard walls at $x=a$ and $x=b$. In addition, we replace the force $f(x,{\bm \theta})$ by some $\tilde f(x,{\bm \theta})$. All the probabilities and probability densities related to this new process will be denoted with a tilde (however the distribution $p_{eq}({\bm \theta})$ remains the same). Here we assume that at $t=0$ the parameter $\bm{\theta}$ is initialised in its equilibrium distribution $p_{eq}(\bm{\theta})$ (and thus it keeps the same distribution at all times). Starting from \eqref{FPgeneral_app}, we derive an equation for the conditional density $\tilde P(x,t|{\bm \theta};x_0,{\bm \theta}_0)$ (writing $\tilde P(x,\bm{\theta},t|x_0,{\bm \theta}_0) = p_{eq}(\bm{\theta}) \tilde P(x,t|{\bm \theta};x_0,{\bm \theta}_0)$ and using the detailed balance conditions \eqref{detailed_balance_Introduction1}-\eqref{detailed_balance_Introduction2}, exactly as in Sec. \ref{ContinuousProof}),
\begin{equation}
\partial_t \tilde P_{\bm{\theta}} = -\partial_x [\tilde f(x,\bm{\theta}) \tilde P_{\bm{\theta}}] + \partial_x^2 [\mathcal{T}(x,\bm{\theta}) \tilde P_{\bm{\theta}}] + \sum_{i} g_i(\bm{\theta}) \partial_{\theta_i} \tilde P_{\bm{\theta}} + \sum_{i,j} \mathcal{D}_{ij}(\bm{\theta}) \partial_{\theta_i\theta_j}^2 \tilde P_{\bm{\theta}} + \int d\bm{\theta}' \mathcal{W}(\bm{\theta}'|\bm{\theta}) [\tilde P_{\bm{\theta}'} - \tilde P_{\bm{\theta}}] \;.
\label{FPconditional_app}
\end{equation}

We now introduce the cumulative distribution of the conditional density $\tilde P_{\bm{\theta}}$,
\begin{eqnarray}
    &&\mathbb{\tilde P}(y(t)\leq x_0|\bm{\theta}(t)=\bm{\theta}_0;y_0,\bm{\theta}(0)^{eq}) = \int_{a^-}^{x_0} dy \, \tilde P(y,t|\bm{\theta}(t)={\bm \theta}_0;y_0,\bm{\theta}(0)^{eq}) \, ,\\
    &&\partial_{x_0} \mathbb{\tilde P}(y(t)\leq x_0|\bm{\theta}(t)=\bm{\theta}_0;y_0,\bm{\theta}(0)^{eq}) = \tilde P(x_0,t|{\bm \theta}(t)={\bm \theta}_0;y_0,\bm{\theta}(0)^{eq})
\end{eqnarray}
where the integral starts at $a^-$ to include a potential delta peak at $x=a$. Writing \eqref{FPconditional_app} in terms of $\mathbb{\tilde P}(y(t)\leq x_0|\bm{\theta}(t)=\bm{\theta}_0;y_0,\bm{\theta}(0)^{eq})$ and then integrating the resulting equation between $x=-\infty$ and $x_0$, using that $\mathbb{\tilde P}(y(t)\leq -\infty|\bm{\theta}(t)=\bm{\theta}_0;y_0,\bm{\theta}(0)^{eq})=0$ as well as its derivatives, and rewriting the temperature term as in Sec. \ref{ContinuousProof}, we obtain a differential equation satisfied by $\mathbb{\tilde P}(y(t)\leq x_0|\bm{\theta}(t)=\bm{\theta}_0;y_0,\bm{\theta}(0)^{eq})$,
\begin{eqnarray}
\hspace{-0.5cm}&&\partial_t \mathbb{\tilde P}(y(t)\leq x_0|\bm{\theta}(t)=\bm{\theta}_0;y_0,\bm{\theta}(0)^{eq}) = \nonumber \\
\hspace{-0.5cm}&& \left[ [-\tilde f(x_0,\bm{\theta}_0) + \partial_{x_0} \mathcal{T}(x_0,\bm{\theta}_0)] \partial_{x_0} + \mathcal{T}(x_0,\bm{\theta}_0) \partial_{x_0}^2 + \sum_{i} g_i(\bm{\theta}) \partial_{\theta_{0,i}} + \sum_{i,j} \mathcal{D}_{ij}(\bm{\theta}) \partial_{\theta_{0,i}\theta_{0,j}}^2 \right] \mathbb{\tilde P}(y(t)\leq x_0|\bm{\theta}(t)=\bm{\theta}_0;y_0,\bm{\theta}(0)^{eq}) \nonumber \\
\hspace{-0.5cm}&&+ \int d\bm{\theta}' \mathcal{W}(\bm{\theta}'_0|\bm{\theta}_0) [\mathbb{\tilde P}(y(t)\leq x_0|\bm{\theta}(t)=\bm{\theta}'_0;y_0,\bm{\theta}(0)^{eq}) - \mathbb{\tilde P}(y(t)\leq x_0|\bm{\theta}(t)=\bm{\theta}_0;y_0,\bm{\theta}(0)^{eq})] \;,
\label{FPphi_app}
\end{eqnarray}
Let us now fix $\tilde f(x,{\bm \theta}) = -f(x,{\bm \theta}) + \partial_x \mathcal{T}(x,\bm{\theta})$. Then this is exactly the same as the equation \eqref{backwardFPgeneral_app} for $\mathbb{P}(x(t)\geq y_0|x_0,\bm{\theta}_0)$. The boundary conditions are
\begin{eqnarray}\label{cumul_bc_app}
&&\mathbb{\tilde P}(y(t)\leq a^+|\bm{\theta}(t)=\bm{\theta}_0;y_0,\bm{\theta}(0)^{eq}) = 0 \quad \text{for all } \tilde f(a^+,\bm{\theta}_0)>0, \text{ or if } \mathcal{T}(x,\bm{\theta})>0 \text{ in the vicinity of } a, \\
&&\mathbb{\tilde P}(y(t)\leq b^-|\bm{\theta}(t)=\bm{\theta}_0;y_0,\bm{\theta}(0)^{eq}) = 1 \quad \text{for all } \tilde f(b^-,\bm{\theta}_0)<0, \text{ or if } \mathcal{T}(x,\bm{\theta})>0 \text{ in the vicinity of } b, \nonumber
\end{eqnarray}
and the initial condition
\begin{eqnarray}\label{cumul_ic}
&& \mathbb{\tilde P}(y(0)\leq x_0|\bm{\theta}(t)=\bm{\theta}_0;y_0,\bm{\theta}(0)^{eq}) = \mathbb{1}_{x_0\geq y_0} \;,
\end{eqnarray}
which are also the same as \eqref{exit_bc_app} and \eqref{exit_ic_app} when writing $\tilde f(x,{\bm \theta}) = -f(x,{\bm \theta}) + \partial_x \mathcal{T}(x,\bm{\theta})$ (indeed if $\mathcal{T}(x,\bm{\theta})=0$ near the wall then one simply has $\tilde f(x,{\bm \theta}) = -f(x,{\bm \theta})$ in this region). Thus, the two quantities $\mathbb{P}(x(t)\geq y_0|x_0,\bm{\theta}_0)$ and $\tilde{\mathbb{P}}(y(t)\leq x_0|\bm{\theta}(t)=\bm{\theta}_0;y_0,\bm{\theta}(0)^{eq})$ follow the same differential equation with the same boundary and initial conditions, hence the identity
\begin{equation}
    \mathbb{P}(x(t)\geq y_0|x_0,\bm{\theta}_0) = \tilde{\mathbb{P}}(y(t)\leq x_0|\bm{\theta}(t)=\bm{\theta}_0;y_0,\bm{\theta}(0)^{eq}) \;.
\label{identity_general_app2}
\end{equation}

\section{Siegmund duality for extreme value statistics of $N$ independent particles}\label{NparticlesAppendix}

Consider $N$ independent particles with positions $x_i(t)$ ($i=1,2,\ldots,N$) following the Langevin dynamics (\ref{LangevinIntroduction}), with absorbing walls at $a$ and $b$, where $-\infty \leq a < b \leq \infty$, and their Siegmund duals $y_i(t)$ described by (\ref{LangevinIntroductionDUAL}), with hard walls at $a$ and $b$. All the $x_i$'s and $y_i$'s are driven by a process $\bm{\theta}(t)$ initialized in its equilibrium distribution. We assume that all the $x_i$'s, and all the $y_i$'s, have the same initial conditions $x_i(0)=x\in[a,b]$ and $y_i(0)=y\in[a,b]$. According to the above results, we have for all $i$
\be\label{siegmundappendixN}
\mathbb{P}(x_i(t) \geq y | x_i(0)=x) = \mathbb{\tilde P}(y_i(t) \leq x | y_i(0)=y) \;.
\ee 
Since the particles are independent, the probability that all the $x_i$'s are above $y$ reads
\be
\mathbb{P}\left(\min_{0\leq i\leq N} x_i(t) \geq y\, \Big|\, x_i(0)=x\right) = \prod_i \mathbb{P}(x_i(t) \geq y | x_i(0)=x)\;.
\ee 
Similarly, the probability that all $y_i$'s are bellow the threshold $x$ is given by
\be
\mathbb{P}\left(\max_{0\leq i\leq N} y_i(t) \leq x\, \Big|\, y_i(0)=y\right)= \prod_i \mathbb{P}(y_i(t) \leq x | y_i(0)=y) \;.
\ee 
Hence
\be\label{}
\mathbb{P}\left(\min_{0\leq i\leq N} x_i(t) \geq y\, \Big|\, x_i(0)=x\right) = \mathbb{P}\left(\max_{0\leq i\leq N} y_i(t) \leq x\, \Big|\, y_i(0)=y\right)\;.
\ee 
We now assume that $b$ is finite and that $y_i(0)=b$. Then the exit probability of the minimum of the $x_i$'s is related to the cumulative distribution of the maximum of the $y_i$'s through
\be\label{}
\mathbb{P}\left(\min_{0\leq i\leq N} x_i(t) = b\, \Big|\, x_i(0)=x\right) = \mathbb{P}\left(\max_{0\leq i\leq N} y_i(t) \leq x\, \Big|\, y_i(0)=b\right)\;.
\ee 

We can also derive a similar relation using the complementary event of (\ref{siegmundappendixN})
\be\label{siegmundappendixN}
\mathbb{P}(x_i(t) < y | x_i(0)=x) = \mathbb{\tilde P}(y_i(t) > x | y_i(0)=y) \;,
\ee 
and the same reasoning gives
\be\label{}
\mathbb{P}\left(\max_{0\leq i\leq N} x_i(t) < y\, \Big|\, x_i(0)=x\right) = \mathbb{P}\left(\min_{0\leq i\leq N} y_i(t) > x\, \Big|\, y_i(0)=y\right)\;.
\ee 
As a consequence, taking again the complementary, we also have
\be\label{}
\mathbb{P}\left(\max_{0\leq i\leq N} x_i(t) \geq y\, \Big|\, x_i(0)=x\right) = \mathbb{P}\left(\min_{0\leq i\leq N} y_i(t) \leq x\, \Big|\, y_i(0)=y\right)\;.
\ee
In the presence of a wall located at $b$, then the exit probability of the maximum of the $x_i$'s are related to the cumulative distribution of the $y_i$'s via
\be\label{}
\mathbb{P}\left(\max_{0\leq i\leq N} x_i(t) = b\, \Big|\, x_i(0)=x\right) = \mathbb{P}\left(\min_{0\leq i\leq N} y_i(t) \leq x\, \Big|\, y_i(0)=b\right)\;.
\ee

\section{Exit probability of a persistent random walk}\label{PRW}

In this appendix, we derive the results of Sec. \ref{sec:PRW} for the persistent random walk on a 1d lattice with $L+2$ sites $(i=0,...,L+1)$, in the stationary state. The dynamics of the process $x_n$ and its dual $y_n$ (which in this case are the same up to the boundary conditions) are defined in \eqref{PRWx}, \eqref{PRWsigma} and \eqref{PRWy}. Let us first derive the infinite time results \eqref{PRWEsol} and \eqref{PRWpsol}.

We first consider the case where there are absorbing boundary conditions at sites $0$ and $L+1$, i.e. if the particle reaches one of those two sites, it stays there forever. Let us compute the exit probability defined in \eqref{PRWEdef}, at infinite time. This quantity satisfies the following equations
\begin{eqnarray}
&&E_i^+ = p\, E_{i+1}^+ + (1-p)\, E_{i+1}^- \quad {\rm with} \quad 1 \leq i \leq L\, , \\
&&E_i^- = p\, E_{i-1}^- + (1-p) E_{i-1}^+ \quad {\rm with} \quad 1 \leq i \leq L\, , \\
&&E_1^-= 0\, , \\
&&E_{L}^+ = 1\, .
\end{eqnarray}
To solve this system of equations, one can write the first terms and show by recurrence the two following relations for $1 \leq i \leq L$,
\begin{eqnarray}
    &&E_i^-=(i-1)(1-p)E_1^+\, ,\\
    &&E_i^+-E_i^-= E_1^+\label{eq2exitdiscretepersistent}\, .
\end{eqnarray}
Now we fix $E_1^+$ by imposing $E_L^+=1$ in Eq.~(\ref{eq2exitdiscretepersistent}),
\begin{eqnarray}
    E_L^+=1 = E_1^+ + (L-1)(1-p)E_1^+\, .
\end{eqnarray}
Therefore
\begin{eqnarray}
    E_1^+ =\frac{1}{(1-p)L+p}\, .
\end{eqnarray}
In the end, the solution is, for $1 \leq i \leq L$,
\begin{eqnarray}
    && E_i^+ = \frac{1-p}{(1-p)L+p} (i-1) + \frac{1}{(1-p)L+p} \;, \\
    && E_i^- = \frac{1-p}{(1-p)L+p} (i-1) \;.
\end{eqnarray}

Let us now compute the stationary distribution of positions of the dual process $y_n$. In this case, we consider hard walls at sites $1$ and $L+1$,  i.e. if the particle is on site 1 and jumps to the left it stays at 1, and similarly at the other end of the lattice. We recall that $p_i^\pm$ is the stationary probability that the particle is on site $i$ and that the previous jump was $\pm 1$. In the stationary state, the master equations for this system read
\begin{eqnarray}
&&p_i^+ = p\  p_{i-1}^+ + (1-p)\  p_{i-1}^- \quad {\rm for} \quad 2 \leq i \leq L \, ,\\
&&p_i^- = p\  p_{i+1}^- + (1-p)\  p_{i+1}^+ \quad {\rm for} \quad 2 \leq i \leq L \, ,\\
&&p_1^+ = p_{L+1}^- = 0 \, ,\\
&&p_1^- = p\  p_1^- + p\  p_2^- + (1-p)\ p_2^+ \, ,\\ 
&&p_{L+1}^+ = p\  p_{L+1}^+ + p\  p_L^+ + (1-p)\ p_L^-\, .
\end{eqnarray}
One can solve this system of coupled equations by recurrence. First, we notice that
\begin{eqnarray}
    && p_2^+= (1-p)\, p_1^- \, ,\\
    &&(1-p)\, p_1^- =  p\  p_2^- + (1-p)\ p_2^+ \, , \\
\end{eqnarray} such that $p_2^+=p_2^-$. Now, assume that there exists $2 \leq i \leq L$ such that $p_i^+ = p_i^-$. Let us show that we have $p_{i+1}^+ = p_{i+1}^-$. We have
\begin{eqnarray}
&&p_{i+1}^+ = p\  p_{i}^+ + (1-p)\  p_{i}^- \, ,\\
&&p_i^- = p\  p_{i+1}^- + (1-p)\  p_{i+1}^+ \, .\\
\end{eqnarray} Using $p_i^+ = p_i^-$, we can simplify
\begin{eqnarray}
&&p_{i+1}^+ = p_{i}^+= p_{i}^- \, ,\\
&&p_{i+1}^+ = p\  p_{i+1}^- + (1-p)\  p_{i+1}^+ \, ,\\
\end{eqnarray}
and we conclude $p_{i+1}^+ = p_{i+1}^-$. Therefore $\forall i \in \llbracket 2,L\rrbracket$, $p_i^+ = p_i^-=c$. We need to impose that the sum of the probabilities is $1$, i.e.
\begin{eqnarray}
    \sum_{n=1}^{L+1}\,  \left(p_i^+ + p_i^-\right) =1 \, .
\end{eqnarray}
Combining with equations on the boundaries, we have
\begin{eqnarray}
    &&p_1^- + p_{L+1}^+ +2(L-1)\, c =1 \, ,\\
    &&p_1^- = p_{L+1}^+ = \frac{c}{1-p}\, .\\ 
\end{eqnarray}
Hence, the solution is
\begin{eqnarray}
    && p_i^+ = p_i^- = \frac{1}{2} \frac{1-p}{(1-p)L+p} \quad {\rm for} \quad  2 \leq i \leq L \, ,\\
    && p_1^- = p_{L+1}^+ = \frac{1}{2} \frac{1}{(1-p)L+p} \, , \\
    && p_1^+ = p_{L+1}^- = 0 \, .
\end{eqnarray}

Thus, as predicted by \eqref{mainRelationStationary}, we recover the identity
\begin{equation}
    E_i^\pm = 2 \sum_{j=1}^{i} p_j^\mp = \tilde \Phi_i^{\mp}\, .
\end{equation}

More generally, the master equations of both processes can be written using transfer matrices, which allows for exact numerical computation of the two quantities at all times. Let us denote $q_i^{\pm}(n)$ ($1\leq i\leq L$) the probability that the process $x_n$ is at position $i$ at time $n$ and that the {\it next} jump will be $\pm 1$. We then define the vectors
\be
Q^+(n) = \begin{pmatrix} q_1^+(n) \\ \vdots \\ q_L^+(n) \end{pmatrix} \quad , \quad Q^-(n) = \begin{pmatrix} q_1^-(n) \\ \vdots \\ q_L^-(n) \end{pmatrix} \;.
\ee
For the purpose of the computation, we can consider the states $(x_n=L,\sigma_{n+1}=+1)$ and $(x_n=1,\sigma_{n+1}=-1)$ to be the absorbing states, since a particle in one of these two states is sure to be absorbed at the next step. The time evolution of $Q^{\pm}(n)$ can then be written
\bea \label{PRWabsdyn}
Q^+(n) = A^{++} Q^+(n-1) + A^{+-} Q^-(n-1) \quad , \quad Q^-(n) = A^{-+} Q^+(n-1) + A^{--} Q^-(n-1) \;,
\eea
with (for $1 \leq i,j \leq L$)
\bea
&&A^{++}_{ij} = p\delta_{i,j-1} +\delta_{i,L}\delta_{j,L} \quad , \quad  A^{+-}_{ij} = (1-p)\delta_{i,j+1} \;, \nonumber \\
&&A^{-+}_{ij} = (1-p)\delta_{i,j-1} \quad , \quad A^{--}_{ij} = p\delta_{i,j+1} +\delta_{i,1}\delta_{j,1} \;.
\eea
To compute $E_i^{\sigma}(n)$ (where $\sigma=\pm1$), we initialise the system with $q_i^\sigma(0)=1$ (putting all the other coefficients to $0$), and let the system evolve following the dynamics \eqref{PRWabsdyn} until time $n-1$, at which point one has
\be
E_i^\sigma(n)=q_L^+(n-1|x_0=i,\sigma_0=\sigma) \;,
\ee
where we have written explicitly the initial conditions for the sake of clarity.
Similarly, for the dual process $y_n$ we define (with $p_i^{\pm}(n)$, $1\leq i \leq L+1$, the probability that the process $y_n$ is at position $i$ at time $n$ and that the {\it previous} jump was $\pm 1$)
\be
P^+(n) = \begin{pmatrix} p_2^+(n) \\ \vdots \\ p_{L+1}^+(n) \end{pmatrix} \quad , \quad P^-(n) = \begin{pmatrix} p_1^-(n) \\ \vdots \\ p_L^-(n) \end{pmatrix} \;,
\ee
where we have taken into account the fact that, due to the hard walls at $i=1$ and $i=L+1$, $p_1^+(n)=p_{L+1}^-(n)=0$. The time evolution of $P^{\pm}(n)$ can then be written
\bea \label{PRWhardyn}
P^+(n) = H^{++} P^+(n-1) + H^{+-} P^-(n-1) \quad , \quad P^-(n) = H^{-+} P^+(n-1) + H^{--} P^-(n-1) \;,
\eea
with
\be
H^{++}_{ij} = p\delta_{i,j-1} +\delta_{i,L}\delta_{j,L} \quad , \quad  H^{+-}_{ij} = H^{-+}_{ij} = (1-p)\delta_{i,j} \quad , \quad H^{--}_{ij} = p\delta_{i,j+1} +p\delta_{i,1}\delta_{j,1} \;.
\ee
Here the initialisation is $p_{L+1}^+(1)=p\, p_{L+1}^+(0) +(1-p) p_{L+1}^-(0)=\frac{p}{2}+\frac{1-p}{2}=\frac{1}{2}$ and $p_{L}^-(1)=p\, p_{L+1}^-(0) +(1-p) p_{L+1}^+(0)=\frac{p}{2}+\frac{1-p}{2}=\frac{1}{2}$ (and $0$ for the other coefficients). We let the system evolve according to \eqref{PRWhardyn} until time $n$, and we compute the cumulative $\sum_{j=1}^{i} p_j^\pm$.

Note that one can recover the results for a run-and-tumble particle without potential as a limit of this model by writing $L=\frac{l}{v_0 \Delta t}$ (with $l=b-a$) and $p=1-\gamma \Delta t$ and taking the limit $\Delta t \to 0$. Imposing that $i\, \Delta x = x$, with $\Delta x =v_0 \Delta t$, one can recover the exit probability of a free RTP $E_b(x)$,
\begin{eqnarray}
    && \frac{1}{2}\left(E_i^- + E_i^+\right) = \frac{(1-p)}{(1-p)L+p} (i-1) + \frac{\frac{1}{2}}{(1-p)L+p} \xrightarrow[\Delta t \to 0]{} \frac{\frac{1}{2}+ \frac{\gamma}{v_0}\, x}{1+\frac{\gamma\, l}{v_0}}=E_b(x)\, .
\end{eqnarray}

As a remark, it may sometimes appear more natural to consider instead the exit probability conditioned on the \textit{previous} jump
\begin{equation}
    \hat E_i^\pm(n) = \mathbb{P}(\text{particle exits at } L+1 \text{ before or at time } n| \text{ particle starts at site } i \,\& \text{ previous jump is } \pm 1) \, .
\end{equation}
For this model, it is simply related to $E_i^\pm(n)$ by
\bea
    &&\hat E_i^+(n) = E_{i-1}^+(n) \quad {\rm for} \ 2 \leq i \leq L\, , \\
    &&\hat E_i^-(n) = E_{i+1}^+(n) \quad {\rm for} \ 1 \leq i \leq L-1\, , \\ &&\hat E_1^+(n) = p \hat E_2^+(n) \, , \quad \hat E_L^-(n) = 1-p + p \hat E_{L-1}^-(n) \, .
\eea

{\red
\section{Dual of a resetting Brownian motion with hard walls} \label{app:resetting}

Let us consider a resetting Brownian motion $y(t)$ in the presence of a force $\tilde F(x)$, with hard walls at positions $a$ and $b$. With a fixed rate $r$, it restarts its dynamics at position $X_r$, such that it evolves through
\begin{equation}
y(t+dt) = \left\{
    \begin{array}{ll}
    y(t) + \left(\tilde F(y(t)) + \sqrt{2T}\, \xi(t)\right) dt \mbox{ with probability } (1-r\, dt)\\
        X_r \mbox{ with probability } r\, dt 
    \end{array}
\right.
,\label{rbmlangevin_app}
\end{equation}
where $T$ is a diffusion coefficient, and $\xi(t)$ is a Gaussian white noise with zero mean and correlation function given by $\langle \xi(t)\xi(t')\rangle=\delta(t-t')$. For simplicity we restrict to the initial condition $y(0)=b$, although a more general derivation with an arbitrary initial condition $y(0)=y\in[a,b]$ along the lines of App.~\ref{ContinuousProof_duality} would be possible. The probability distribution $\tilde p(y,t)$ of the position of this process obeys the forward Fokker-Planck equation
\begin{equation}\label{FPdensitydualrbm_app}
    \partial_t \tilde p(y,t) = T\, \partial_y^2 \tilde p(y,t) - \partial_y \left(\tilde F(y) \tilde p(y,t)\right) -r\tilde p(y,t) + r \delta(y-X_r) \, .
\end{equation}
We would like to find the dual process of $y(t)$. For this, we begin by deriving from \eqref{FPdensitydualrbm_app} the differential equation satisfied by the cumulative distribution
\begin{equation}
    \tilde \Phi^r(x,t|b,X_r) = \int_{a^-}^x dy\, \tilde p(y,t)\, .
\end{equation}
Integrating (\ref{FPdensitydualrbm_app}) between $a^-$ and $x$ leads to
\begin{equation} \label{FPrBmcumu_app}
    \partial_t \tilde \Phi^r(x,t|b,X_r) = T\, \partial^2_x \tilde \Phi^r(x,t|b,X_r) -\tilde F(x)\, \partial_x \tilde \Phi^r(x,t|b,X_r) - r\, \tilde \Phi^r(x,t|b,X_r) + r\,  \mathbb{1}_{x>X_r} \, .
\end{equation}
The boundary and initial conditions are $ \tilde \Phi^r(a,t|b,X_r)=0$, $ \tilde \Phi^r(b,t|b,X_r)=1$, and $ \tilde \Phi^r(x,t=0|b,X_r)=0$ for $x<b$ and $ \tilde \Phi^r(b,t=0|b,X_r)=1$.

Let us now consider a process $x(t)$ which evolves through
\begin{equation}\label{dualrbm_app}
x(t+dt) = \left\{
    \begin{array}{ll}
   
        x(t) + \left(F(x(t)) + \sqrt{2T}\, \xi(t)\right)\, dt \mbox{ with probability } (1-r\, dt)\\
        a \mbox{ with probability } r\, dt \mbox{ if } x(t)<X_r\\
              b \mbox{ with probability } r\, dt \mbox{ if } x(t)>X_r 
    \end{array}
\right.
,
\end{equation}
with absorbing walls at $a$ and $b$. We want to compute the exit probability at wall $b$ and time $t$ for this new process. We once again derive the backward Fokker-Planck equation by considering the evolution of the process between times $0$ and $dt$. Starting from $x$, the exit probability at time $t+dt$ reads
\begin{equation}
    E_b^r(x,t+dt,X_r)=(1-rdt)\,  \mathbb{E}_{\xi}\left[E_b^r\left(x+\left(F(x) + \sqrt{2T}\, \xi(t)\right) dt,t,X_r\right)\right] + r dt\, \mathbb{1}_{x>X_r}\, ,
\end{equation}
The last term comes from the fact that, if a resetting event happens between times $0$ and $dt$, the particle will be reset either to $b$ if $x>X_r$, in which case $E_b^r(b,t,X_r)=1$, or to $a$ if $x<X_r$, in which case $E_b^r(a,t,X_r)=0$. Expanding the first term on the right hand-side and averaging over $\xi(t)$ leads to the backward Fokker-Planck equation satisfied by $x(t)$,
\begin{equation}\label{FPrBmexit}
    \partial_t E_b^r(x,t,X_r) = T\, \partial^2_x E_b^r(x,t,X_r) + F(x)\, \partial_x E_b^r(x,t,X_r) - r E_b^r(x,t,X_r) + r \, \mathbb{1}_{x>X_r} \, .
\end{equation}
The boundary conditions are $E_b^r(a,t,X_r)=0$, and $E_b^r(b,t,X_r)=1$, while the initial condition reads $E_b^r(x,t=0,X_r)=0$ for $x<b$, and $E_b^r(b,t=0,X_r)=1$. Once again, we see that, taking $\tilde F(x)=-F(x)$, $E_b^r(x,t,X_r)$ and $\tilde \Phi^r(x,t|b,X_r)$ satisfy the same equations with the same boundary and initial conditions, and thus the identity \eqref{mainRelation_averaged} holds.}


\begin{thebibliography}{26}

\bibitem{redner}
S. Redner, {\it A guide to first-passage processes}, Cambridge university press, (2001).

\bibitem{Metzler_book}
R. Metzler, S. Redner and G. Oshanin, {\it First-passage phenomena and their applications},
Vol. 35, World Scientific, (2014).

\bibitem{EVS1} A. J. Bray, S. N. Majumdar and G. Schehr, {\it Persistence and first-passage properties in nonequilibrium systems}, Adv. Phys. {\bf 62}, 225 (2013).

\bibitem{handbookSM} C. W. Gardiner, {\em Handbook of stochastic methods}, Berlin: springer (1985).

{\red \bibitem{livreSG} S. N. Majumdar, G. Schehr, {\em Statistics of Extremes and Records in Random Sequences}, Oxford University Press, (2024).}

\bibitem{activeintro1} M. C. Marchetti, J. F. Joanny, S. Ramaswamy, T. B. Liverpool, J. Prost,  M. Rao, R. Aditi Simha, {\em Hydrodynamics of soft active matter}, Rev. Mod. Phys. {\bf 85}, 3 1143--1189 (2013). 

\bibitem{activeintro2} C. Bechinger, R. Di Leonardo, H. L\"owen, C. Reichhardt, G. Volpe, G. Volpe, {\em Active particles in complex and crowded environments}, Rev. Mod. Phys. {\bf 88}, 4 045006 (2016). 

\bibitem{activeintro3} S. Ramaswamy, {\em Active matter}, J. Stat. Mech. 054002 (2017). 

\bibitem{activeintro4} E. Fodor, M. Cristina Marchetti, {\em The statistical physics of active matter: From self-catalytic colloids to living cells}, Physica A: Statistical Mechanics and its Applications {\bf 504} 106-120 (2018). 

\bibitem{activeintro5} P. Romanczuk, M. Bär, W. Ebeling, B. Lindner, and L. Schimansky-Geier, {\em Active Brownian particles}, Eur. Phys. J. Spec. Top. {\bf 202}, 1–162 (2012).

\bibitem{colorednoise} P. Hanggi, and P. Jung,  {\em Colored Noise in Dynamical Systems}, Adv. Chem. Phys. {\bf 89}, 239 (1995).

\bibitem{Tailleur_RTP}
J. Tailleur, M. E. Cates, {\em Statistical mechanics of interacting run-and-tumble bacteria}, Phys. Rev. Lett. {\bf 100}, 218103 (2008). 

\bibitem{DKM19} A. Dhar, A. Kundu, S. N. Majumdar, S. Sabhapandit and G. Schehr, {\em Run-and-tumble particle in one-dimensional confining potentials: Steady-state, relaxation, and first-passage properties}, Phys. Rev. E {\bf 99}, 032132 (2019).

\bibitem{Wijland21} D. Martin, J. O'Byrne, M. E. Cates, E. Fodor, C. Nardini, J. Tailleur, F. van Wijland, {\em Statistical mechanics of active Ornstein-Uhlenbeck particles}, Phys. Rev. E {\bf 103}, 032607 (2021).

\bibitem{MalakarRTP} K. Malakar, V. Jemseena, A. Kundu, K. Vijay Kumar, S. Sabhapandit, S. N. Majumdar, S. Redner, A. Dhar, {\em 
Steady state, relaxation and first-passage properties of a run-and-tumble particle in one-dimension}, {\red J. Stat. Mech. 043215 (2018).}

\bibitem{Sevilla}
F. J. Sevilla, A. V. Arzola, and E. P. Cital,{\em Stationary superstatistics distributions of trapped run-and-tumble particles}, Phys. Rev. E {\bf 99}, 012145 (2019).

\bibitem{separation1} J. Schwarz-Linek, C. Valeriani, A. Cacciuto, M. E. Cates, D. Marenduzzo, A. N. Morozov, and W. C. K. Poon, {\em Phase separation and rotor self-assembly in active particle suspensions}, Proc. Natl. Acad. Sci. USA {\bf 109}, 4052 (2012). 


\bibitem{separation2} G. S. Redner, M. F. Hagan, and A. Baskaran, {\em Structure and Dynamics of a Phase-Separating Active Colloidal Fluid}, Phys. Rev. Lett. {\bf 110}, 055701 (2013).


\bibitem{separation3} J. Stenhammar, R. Wittkowski, D. Marenduzzo, and M. E. Cates, {\em Activity-Induced Phase Separation and Self-Assembly in Mixtures of Active and Passive Particles}, Phys. Rev. Lett. {\bf 114}, 018301 (2015).

\bibitem{separation4}
J. O'Byrne, A. Solon, J. Tailleur, and Y. Zhao, {\it An introduction to motility-induced phase separation}, preprint arXiv:2112.03979 (2021).

\bibitem{flocking1} J. Toner, Y. Tu, S. Ramaswamy, {\em Hydrodynamics and phases of flocks}, Ann. of Phys. {\bf 318}, 170 (2005).

\bibitem{flocking2} N. Kumar, H. Soni, S. Ramaswamy, A.~K. Sood, {\em Flocking at a distance in active granular matter}, Nature Comm. {\bf 5},  4688 (2014). 

\bibitem{Berg2004} H. C. Berg, {\em E. Coli in Motion}, (Springer Verlag, Heidelberg, Germany) (2004).

\bibitem{Cates2012} M. E. Cates, {\em Diffusive transport without detailed balance: Does microbiology need statistical physics?}, Rep. Prog. Phys. {\bf 75}, 042601 (2012). 

\bibitem{Kac1974}
M. Kac, Rocky Mountain J. Math. {\bf 4}, 497 (1974).

\bibitem{Ors1990}
E. Orsingher, Stoch. Process. Their Appl. {\bf 34}, 49 (1990)


\bibitem{AOUP} L. L. Bonilla, {\em Active Ornstein-Uhlenbeck particles}, Phys. Rev. E \textbf{100}, 022601 (2019).

\bibitem{ABM} U. Basu, S. N. Majumdar, A. Rosso, and G. Schehr, {\em Active Brownian motion in two dimensions}, Physical Review E, 98(6), 062121 (2018).


\bibitem{benichou1} O. B\'enichou, C. Loverdo, M. Moreau, and R. Voituriez, {\it Intermittent search strategies} Reviews of Modern Physics, {\bf83}(1), 81. (2011).

\bibitem{benichou2} O. B\'enichou, M. Coppey, M. Moreau, P. H. Suet, and R. Voituriez, {\it Optimal search strategies for hidden targets} Physical review letters, {\bf94}(19), 198101 (2005).


\bibitem{SurvivalRTPDriftDeBruyne} B. De Bruyne, S. N. Majumdar and G. Schehr, {\em Survival probability of a run-and-tumble particle in the presence of a drift}, J. Stat. Mech. (2021) 043211.

\bibitem{Singh2020} P. Singh, S. Sabhapandit and A. Kundu, {\em Run-and-tumble particle in inhomogeneous media in one dimension}, J. Stat. Mech. (2020) 083207.

\bibitem{Singh2022} P. Singh, S. Santra and A. Kundu, {\em Extremal statistics of a one-dimensional run and tumble particle with an absorbing wall}, J. Phys. A: Math. Theor. {\bf 55} 465004 (2022).

\bibitem{MFPT1DABP} S. A. Iyaniwura and Z. Peng, {\em Asymptotic analysis and simulation of mean first passage time for active Brownian particles in 1-D},  	arXiv:2310.04446 (2023).

\bibitem{RTPsurvivalMori} F. Mori, P. Le Doussal, S. N. Majumdar and G. Schehr, {\em Universal Survival Probability for a d-Dimensional Run-and-Tumble Particle}, Phys. Rev. Lett. {\bf 124}, 090603 (2020).

\bibitem{TVB12}
V. Tejedor, R. Voituriez, and O. B{\' e}nichou, {\em Optimizing persistent random searches}, Phys. Rev. Lett. {\bf 108}, 088103 (2012).

\bibitem{RBV16}
J.-F. Rupprecht, O. B{\'e}nichou, and R. Voituriez, {\em Optimal search strategies of run-and-tumble walks}, Phys. Rev. E {\bf 94}, 012117 (2016).

\bibitem{MFPT_1D_RTP} M. Gu\'eneau, S. N. Majumdar and G. Schehr, {\em Optimal mean first-passage time of a run-and-tumble particle in a class of one-dimensional confining potential}, EPL {\bf 145} 61002 (2024).

\bibitem{ExitProbaShort} M. Guéneau and L. Touzo, {\em Relating absorbing and hard wall boundary conditions for active particles}, J. Phys. A: Math. Theor. 57 225005 (2024).


\bibitem{BressloffStickyBoundaries} P. C. Bressloff, {\em Encounter-based model of a run-and-tumble particle II: absorption at sticky boundaries}, J. Stat. Mech. (2023) 043208.

\bibitem{AngelaniGenericBC} L. Angelani, {\em One-dimensional run-and-tumble motions with generic boundary conditions}, J. Phys. A: Math. Theor. {\bf 56} 455003 (2023).

\bibitem{RTPpartiallyAbsorbingTarget} E. Jeon, B. Go and Y. W. Kim, {\em Searching for a partially absorbing target by a run-and-tumble particle in a confined space}, arxiv-2310.04016 (2023).


\bibitem{AOUPEscapeLecomte} E. Woillez, Y. Kafri and V. Lecomte, {\em Nonlocal stationary probability distributions and escape rates for an active Ornstein–Uhlenbeck particle}, J. Stat. Mech. (2020) 063204.


\bibitem{Lee2013} C. F. Lee, {\em Active particles under confinement: aggregation at the wall and gradient formation inside a channel}, New J. Phys. {\bf 15} 055007 (2013).

\bibitem{Yang2014} X. Yang, M. L. Manning and M. C. Marchetti, {\em Aggregation and segregation of confined active particles}, Soft Matter, {\bf 10}, 6477 (2014).

\bibitem{Uspal2015} W. E. Uspal, M. N. Popescu, S. Dietrich and M. Tasinkevych, {\em Self-propulsion of a catalytically active particle near a planar wall: from reflection to sliding and hovering}, Soft Matter {\bf 11}, 434 (2015).

\bibitem{Duzgun2018} A. Duzgun and J. V. Selinger, {\em Active Brownian particles near straight or curved walls: Pressure and boundary layers}, Phys. Rev. E {\bf 97}, 032606 (2018).


\bibitem{AngelaniHardWalls} L. Angelani, {\em Confined run-and-tumble swimmers in one
dimension}, J. Phys. A: Math. Theor. {\bf 50}, 325601 (2017).

\bibitem{hardWallsJoanny} C. Sandford, A. Y. Grosberg, and J.-F. Joanny, {\em Pressure and flow of exponentially self-correlated active particles},
Phys. Rev. E {\bf 96}, 052605 (2017).

\bibitem{hardWallsCaprini} L. Caprini and U. M. B. Marconi, {\em Active particles under confinement and effective force generation among surfaces}, Soft matter {\bf 14}, 9044-9054 (2018).



\bibitem{Levy} P. L\'evy, {\em Processus stochastiques et mouvement brownien}, Gauthier-Villars, Paris (1948).

\bibitem{Lindley} D. Lindley, {\em The theory of queues with a single server}, Mathematical Proceedings of the Cambridge Philosophical Society {\bf 48}(2), 277-289 (1952).

\bibitem{Siegmund} D. Siegmund, {\em The Equivalence of Absorbing and Reflecting Barrier Problems for Stochastically Monotone Markov Processes}, Ann. Probab. {\bf 4}(6): 914-924 (1976).


\bibitem{SiegmundDualityClifford} P. Clifford and A. Sudbury, {\em A Sample Path Proof of the Duality for Stochastically Monotone Markov Processes}, Ann. Probab. {\bf13}(2): 558-565 (1985).

\bibitem{KolokoltsovDuality} V. N. Kolokoltsov, {\em Stochastic monotonicity and duality for one-dimensional Markov processes}, Mathematical Notes {\bf 89}, 652–660 (2011).

\bibitem{partiallyOrdered} P. Lorek, {\em Siegmund duality for Markov Chains on partially ordered state spaces}, Probability in the Engineering and Informational Sciences, {\bf 32}(4), 495-521 (2018).

\bibitem{DualityZhao} P. Zhao, {\em Siegmund Duality for Continuous Time Markov Chains on $\mathbb{Z}_+^d$}, Acta Mathematica Sinica. English Series; Heidelberg {\bf 34}, 9: 1460-1472 (2018). 

\bibitem{hittingProbaAnomalous} S. N. Majumdar, A. Rosso, A. Zoia, {\em Hitting probability for anomalous diffusion processes}, Phys. Rev. Lett. {\bf 104}, 020602 (2010).

\bibitem{WidomExitprobaLevyflight} H. Widom  {\em Stable processes and integral equation}, Trans. Amer. Math. Soc. {\bf 98}, 430 (1961).


\bibitem{Levy3} R. M. Blumenthal, R. K. Getoor, and D. B. Ray, {\em On the Distribution of First Hits for the Symmetric Stable Processes}, Trans. Am. Math. Soc. {\bf 99}, 540 (1961).



\bibitem{DenisovLevy} S. I. Denisov, W. Horsthemke, P. Hänggi {\em Steady-state L\'evy flights in a confined domain}, Physical Review E, {\bf77}, 061112 (2008).

\bibitem{AsmussenDiscrete} S. Asmussen and K. Sigman, {\em Monotone Stochastic Recursions and their Duals}, Probability in the Engineering and Informational Sciences {\bf 10}(1), 1-20 (1996).

\bibitem{SigmanContinuous} K. Sigman and R. Ryan, {\em Continuous-time monotone stochastic recursions and duality}, Advances in Applied Probability {\bf 32}(2), 426-445 (2000).

\bibitem{DualityMultidimensions} B. Blaszczyszyn and K. Sigman, {\em Risk and duality in multidimensions}, Stochastic Processes and their Applications {\bf 83} (2), 331-356 (1999).

\bibitem{JansenDualityReview} S. Jansen and N. Kurt, {\em On the notion(s) of duality for Markov processes}, Probab. Surveys {\bf 11}: 59-120 (2014).

\bibitem{DualityGenerators} V. Kolokoltsov and R. Lee, {\em Stochastic duality of Markov processes: a study via generators},  arXiv:1304.1688 (2013).

\bibitem{CoxEntranceExitLaws} J. T. Cox and U. Rösler, {\em A duality relation for entrance and exit laws for Markov processes}, Stochastic Processes and their Applications {\bf 16} 2, 141-156 (1984).

\bibitem{PathwiseDualSturm} A. Sturm and J. M. Swart, {\em Pathwise Duals of Monotone and Additive Markov Processes}, Journal of Theoretical Probability {\bf 31}, 932–983 (2018).


\bibitem{Kolotsovkthorder} V. Kolokoltsov, {\em Stochastic Monotonicity and Duality of kth Order with Application to Put-Call Symmetry of Powered Options}, Journal of Applied Probability, {\bf 52}(1), 82-101 (2015).

\bibitem{DualityLevyProcessesGoffard} P.-O. Goffard and A. Sarantsev, {\em Exponential convergence rate of ruin probabilities for level-dependent Lévy-driven risk processes}, Journal of Applied Probability, {\bf 56}(4), 1244-1268 (2019).

\bibitem{MohleDualityGenetics} M. Möhle, {\em The concept of duality and applications to Markov processes arising in neutral population genetics models}, Bernoulli {\bf 5}(5): 761-777 (1999).

\bibitem{DualityBranchingFoucart} C. Foucart, {\em Local explosions and extinction in continuous-state branching processes with logistic competition}, arXiv:2111.06147 (2021).

\bibitem{LiggettInteractingParticles} T. M. Liggett, {\em Interacting Particle Systems}, Springer-Verlag, Berlin (1985).

\bibitem{DualityBoundaryDriven} G. Carinci, S. Floreani, C. Giardin{\`a} and F. Redig, {\em Boundary driven Markov gas: duality and scaling limits}, Ensaios Matem{\'a}ticos (2021).

\bibitem{Comtet2011} A. Comtet and Y. Tourigny, {\em Excursions of diffusion processes and continued fractions}, Annales de l'Institut Henri Poincar\'e -- Probabilit\'es et Statistiques, {\bf 47}, 850-874 (2011).

\bibitem{Comtet2020} A. Comtet, F. Cornu and G. Schehr, {\em Last-Passage Time for Linear Diffusions and Application to the Emptying Time of a Box}, J. Stat. Phys. {\bf 181}, 1565-1602 (2020).

\bibitem{ThibautDual} T. Arnoulx de Pirey, {\em Extreme value statistics of non-Markovian processes from a new class of integrable nonlinear differential equations}, arXiv:2402.05091 (2024).

{\red
\bibitem{Szabo} A. Szabo, G. Lamm and G. H. Weiss, {\it Localized Partial Traps in Diffusion Processes and Random Walks}, J. Stat. Phys. {\bf 34}, 225 (1984).

\bibitem{Spouge} J.L. Spouge, A. Szabo and G.H. Weiss, {\it Single-particle survival in gated trapping}, Phys. Rev. E, {\bf 54}(3), 2248 (1996).

\bibitem{Scher1} Y. Scher and S. Reuveni, {\it Unified approach to gated reactions on
networks}, Phys. Rev. Lett. {\bf 127}(1), 018301 (2021).

\bibitem{Scher2} Y. Scher, A. Kumar, M.S. Santhanam and S. Reuveni, {\it Continuous gated first-passage processes}, Rep. Prog. Phys. {\bf 87} 108101 (2024).

\bibitem{Guerin} T. Guérin, M. Dolgushev, O. Bénichou and R. Voituriez, {\it Universal kinetics of imperfect reactions in confinement}, Communications chemistry {\bf 4}(1), 157 (2021).

\bibitem{defect1} Elliott W. Montroll and Renfrey B. Potts, {\it Effect of Defects on Lattice Vibrations}, Phys. Rev. {\bf 100}, 525 (1955).

\bibitem{defect2} V. M. Kenkre, {\it Memory Functions, Projection Operators, and the Defect Technique: Some Tools of the Trade for the Condensed Matter Physicist}, Springer Nature (2021).

\bibitem{defect3} T. Kay, T.J. McKetterick and L. Giuggioli, {\it The defect technique for partially absorbing and reflecting boundaries: application to the
Ornstein–Uhlenbeck process}, International Journal of Modern Physics B, {\bf 36}(07-08), 2240011 (2022).
}

\bibitem{DiffDiffChubynsky} M. V. Chubynsky and G. W. Slater, {\em Diffusing Diffusivity: A Model for Anomalous, yet Brownian, Diffusion}, Phys. Rev. Lett. {\bf 113}, 098302 (2014).

\bibitem{DiffDiffChechkin} A. V. Chechkin, F. Seno, R. Metzler and I. M. Sokolov, {\em Brownian yet Non-Gaussian Diffusion: From Superstatistics to Subordination of Diffusing Diffusivities}, Phys. Rev. X {\bf 7}, 021002 (2017).

\bibitem{DiffDiffJain} R. Jain and K. L. Sebastian, {\em Diffusing diffusivity: a new derivation and comparison with simulations}, Journal of Chemical Sciences {\bf 129}, 929–937 (2017).

\bibitem{DiffDiffFPTSposini} V. Sposini, A. Chechkin and R. Metzler, {\em First passage statistics for diffusing diffusivity}, J. Phys. A: Math. Theor. {\bf 52} 04 (2019).

\bibitem{DDsoftmatter1} T. J. Lampo, S. Stylianidou, M. P. Backlund, P. A. Wiggins and A. J. Spakowitz {\em Cytoplasmic RNA-protein particles exhibit non-Gaussian subdiffusive behavior}, Biophysical journal {\bf 112} 3 532-542 (2017). 

\bibitem{DDsoftmatter2} G. Kwon, B. J. Sung and A. Yethiraj, {\em Dynamics in crowded environments: is non-Gaussian Brownian diffusion normal?}, The Journal of Physical Chemistry B, {\bf 118} 28 8128-8134 (2014). 

\bibitem{DDsoftmatter3} B. Wang, J. Kuo, S. C. Bae and S. Granick {\em  When Brownian diffusion is not Gaussian}, Nature materials {\bf 11} 6 481-485 (2012). 

\bibitem{PRWMasoliverReview} J. Masoliver and K. Lindenberg, {\em Continuous time persistent random walk: a review and some generalizations}, Eur. Phys. J. B {\bf 90}: 107 (2017).

\bibitem{PRWSurvivalLacroixMori} B. Lacroix-A-Chez-Toine and F. Mori, {\em Universal survival probability for a correlated random walk and applications to records}, J. Phys. A: Math. Theor. {\bf 53} 495002 (2020).

\bibitem{PRWWeiss} G. H. Weiss, {\em Some applications of persistent random walks and the telegrapher's equation}, Physica A {\bf 311} 381–410 (2002).

\bibitem{resettingPRL} M. R. Evans, and S. N. Majumdar, {\em Diffusion with stochastic resetting}, Physical review letters, 106(16), 160601 (2011).

\bibitem{resettingReview} M. R. Evans, S. N. Majumdar and G. Schehr, {\em Stochastic resetting and applications}, J. Phys. A: Math. Theor. {\bf 53} 193001 (2020).

\bibitem{resettingBriefReview} S. Gupta and A. M. Jayannavar, {\em Stochastic resetting: A (very) brief review}, Frontiers in Physics {\bf 10} 789097 (2022).

\bibitem{Besga20} B. Besga, A. Bovon, A Petrosyan, S. N. Majumdar, S. Ciliberto, {\em Optimal mean 
first-passage time for a Brownian searcher subjected to resetting: experimental and theoretical results},   
Phys. Rev. Research, {\bf 2}, 032029 (R) (2020). 

\bibitem{Faisant21} F. Faisant, B. Besga, A. Petrosyan, S. Ciliberto, S. N. Majumdar, {\em 
Optimal mean first-passage time of a Brownian searcher with resetting in one and two dimensions: 
experiments, theory and numerical tests},
J. Stat. Mech. 113203 (2021).

\bibitem{resettingInInterval} A. Pal, and V. V. Prasad {\em First passage under stochastic resetting in an interval},
Phys. Rev. E {\bf 99} 032123 (2019).

\bibitem{resettingRTP} M. R. Evans, S. N. Majumdar {\em Run and tumble particle under resetting: a renewal approach}, J. Phys. A: Math. Theor. {\bf 51} 475003 (2018). 

\bibitem{riskenBook} H. Risken, {\em The Fokker-Planck equation: Methods of solution and applications}, Springer Series in Synergetics, Springer Berlin, Heidelberg (1996).

\bibitem{DRABP1} I. Santra, U. Basu and S. Sabhapandit, {\em Active Brownian motion with directional reversals}, Phys. Rev. E {\bf 104}, L012601 (2021).


\bibitem{DRABP2} I. Santra, U. Basu and S. Sabhapandit, {\em Direction reversing active Brownian particle in a harmonic potential}, Soft Matter {\bf 17}, 10108 (2021).


\bibitem{EVS2} A. Comtet, and S. N. Majumdar, {\it Precise asymptotics for a random walker’s maximum} Journal of Statistical Mechanics: Theory and Experiment 2005.06 (2005).

\bibitem{reviewEVSPal} S. N. Majumdar, A. Pal, G. Schehr, {\it Extreme value statistics of correlated random variables: a pedagogical review}, Physics Reports {\bf 840} 1-32 (2020).

\bibitem{singletraj} V. Sposini, R. Metzler and G. Oshanin, {\em Single-trajectory spectral analysis of scaled Brownian motion}, New J. Phys. {\bf 21}, 073043  (2019).

\bibitem{CTRW1} E. W. Montroll and G. H. Weiss, {\it Random walks on lattices. II.}, Journal of Mathematical Physics {\bf 6}(2), 167-181 (1965).

\bibitem{CTRW2} H. Scher and M. Lax, {\it Stochastic transport in a disordered solid. I. Theory} Physical Review B  {\bf 7} 10 4491 (1973).


\bibitem{CTRWA} J. P. Bouchaud, A. Georges, {\em Anomalous diffusion in disordered media: Statistical mechanisms, models and physical applications}, Physics Reports, {\bf 195}, 127-293 (1990).

\bibitem{CTRWB} R. Metzler, J. Klafter {\em The random walk's guide to anomalous diffusion: a fractional dynamics approach}, Physics Reports, {\bf339}, 1-77 (2000).

\bibitem{CTRWC} R. Kutner, J. Masoliver {\em The continuous time random walk, still trendy: fifty-year history, state of art and outlook}, Eur. Phys. J. B {\bf90}, 50 (2017).

\bibitem{resettingRTP2} G. Tucci, A. Gambassi, S. N. Majumdar, G. Schehr {\em First-passage time of run-and-tumble particles with noninstantaneous resetting}, Physical Review E, 106(4), 044127 (2022).


\bibitem{resettingNoise} M. Gueneau, S. N. Majumdar, and G. Schehr, {\em Active particle in a harmonic trap driven by a resetting noise: an approach via Kesten variables}, J. Phys. A: Math. Theor. {\bf 56} 475002 (2023). 


\bibitem{IntermittentReset} G. Mercado-V\`asquez, D. Boyer, S. N. Majumdar, and G. Schehr  {\em Intermittent resetting potentials},  Journal of Statistical Mechanics: Theory and Experiment 113203 (2020).

\bibitem{randomwalkresetPRL} L. Kusmierz, S. N. Majumdar, S. Sabhapandit, G. Schehr, {\em First order transition for the optimal search time of Lévy flights with resetting}, Physical review letters, {\bf 113} (22), 220602 (2014). 

\bibitem{fBmMandelbrot} B. Mandelbrot, J. W. Van Ness, {\em Fractional Brownian motions, fractional noises and applications}, SIAM review {\bf10} (4) 422-437 (1968).


\bibitem{Nparticleduality} T. Assiotis, N. O'Connel, J. Warren, {\em Interlacing Diffusions}, Séminaire de Probabilités L, 301-380 (2016).


\end{thebibliography}
\end{document}